\documentclass[a4paper,10pt]{article}
\usepackage{graphicx}
\usepackage{amsmath}
\usepackage{footnote}
\usepackage{multirow}
\usepackage{subfigure}
\usepackage{longtable}

\begin{document}
\title{Polarization study of the P-wave charmonium radiative decay into a light vector meson at $e^{+}e^{-}$ collider experiment} 

\author{Yong-Qing Chen$^{1}$ ~~Peng-Cheng Hong$^{2}$\footnotemark[1] ~~Zhuo Chen$^{1}$ ~~Wei Shan$^{1}$\footnotemark[2]  ~~Wei-Min Song$^{2}$\footnotemark[3] }
\footnotetext[1]{Email:hongpc@ihep.ac.cn}
\footnotetext[2]{Email:shanw@hunnu.edu.cn}
\footnotetext[3]{Email:weiminsong@jlu.edu.cn}
\date{} 
\maketitle

\begin{center}
$^1$School of Physics and Electronics, Hunan Normal University, Changsha, China\\
$^2$College of Physics, Jilin University, Changchun, China\\
\end{center}


\let\thefootnote\relax
\footnotetext{Supported by the National Natural Science Foundation of China (11805064, 11975118)} 

\begin{abstract}
In this work, a formalism is presented for the helicity amplitude analysis of the decays $\psi(2S) \to \gamma_1 \chi_{cJ},~ \chi_{cJ} \to \gamma_2 V (V=\rho^0,~\phi,~\omega)$~(the subscript 1,2 is used to distinguish the two radiative photons), and the polarization expressions of the P-wave charmonia $\chi_{cJ}$ and the vector mesons $\rho^0, \phi, \omega$ for experimental measurements at Electron-Positron Collider. In addition, we derive the formulae of the angular distributions of the $\chi_{c1,2} \to \gamma V$ to extract the degree of transverse polarization $P_T$ of $e^+ e^-$ pairs with symmetric beam energy as well as the ratios of two helicity amplitudes $x$ (in $\chi_{c1}$ decays) and $x,~y$ (in $\chi_{c2}$ decays) representing the relative magnitudes of transverse to longitudinal polarization amplitude, and validate it by performing the Monte Carlo simulation. Finally, the statistical sensitivity of $P_T$,~$x$ and $y$ are estimated based on the large $\psi(2S)$ data samples collected at the current and proposed future $e^+e^-$ collider experiment. 
\end{abstract} 

\bigskip

\noindent 

\section{Introduction}
Over past five decades, heavy quarkonium spectra have been established due to a collective of theorists and experimentalists working. More than 40 heavy quark-antiquark bound states, also known as charmonium and bottomonium, are observed with masses ranging from 2.9 GeV to 4.7 GeV or from 9.3 GeV to 11.1 GeV, respectively~\cite{Workman:2022ynf}. The heavy quarkonia below the open-flavor production mass-threshold are relatively well understood, which provide an ideal laboratory to test perturbative and nonperturbative quantum chromodynamics (QCD)~\cite{Barnes:2005pb, Brambilla:2010cs}.

Charmonium decay is usually a focused research topic that is significant and useful for understanding the fundamental characteristics of charmonium. J/$\psi$ and $\psi(2S)$ are two of the observed charmonium states that have a wealth of experimental decay information, which are recently listed in the Particle Data Group (PDG)~\cite{Workman:2022ynf}. Compared to J/$\psi$ and $\psi(2S)$, the experimental measurement pertinent to the decay of spin singlets, such as the P-wave state $h_c$ and the S-wave state $\eta_c$, as well as spin triplets, the P-wave state $\chi_{cJ}$ are significantly less. Therefore, there are currently more active theoretical and experimental investigations into those charmonia decays. 

Experimental measurements of charmonium radiative decay to light-quark vector mesons would help us understand the QCD and QED mechanism between charmonium and light vector mesons by strong interaction and electromagnetic interaction. 
The previous theoretical study for radiative decays of charmonium into light vector mesons, the processes $\chi_{cJ} \to \gamma V$, is based on numerical calculations for the quark-gluon loop diagrams in perturbative QCD (pQCD) frame and nonrelativistic quantum chromodynamics (NRQCD)~\cite{pQCD, QED_NRQCD}. It provides a useful place to investigate the interatctions between quarks and gluons in OZI suppressed processes. And its predicted branching ratios of the decays $\chi_{cJ} \to \gamma V$ have been tested by later $e^{+}e^{-}$ collider experiments.

The measurement for the branching ratios of the P-wave charmonia $\chi_{cJ} \to \gamma V$ have been presented by CLEO-c and BESIII collaborations in 2008~\cite{CLEO} and 2011~\cite{BESIII:2011ysp}, respectively. However, there are still some significant discrepancy between the experimental results and the theoretical predictions as shown in Table~\ref{t_vs_e}. To resolve the discrepancy, an attempt was made to employ a phenomenological model with hadronic loop mechanism~\cite{Chen:2010re}.

\begin{table*}[htbp]
\centering
\caption{Comparison of experimental measurement results~(CLEO-c, BESIII) and theoretical calculated results~(pQCD, NRQCD, NRQCD+QED) for the branching ratio of $\chi_{cJ} \to \gamma V(V=\rho^0, \phi, \omega)$ [in unit of $10^{-6}$]}
\begin{tabular}{ c c c c c c}
\hline \hline
Decay Mode                         & CLEO-c\cite{CLEO} & BESIII~\cite{BESIII:2011ysp}    & pQCD~\cite{pQCD} & NRQCD~\cite{QED_NRQCD} & NRQCD+QED~\cite{QED_NRQCD}  \\
\hline
$\chi_{c0}\to\gamma\rho^0$     &$<9.6$       &$< 10.5$        &1.2   & 3.2    &2.0         \\

$\chi_{c1}\to\gamma\rho^0$     & $243 \pm 19 \pm 22$        &$ 228 \pm 13 \pm 22$      &14    & 41    &42  \\

$\chi_{c2}\to\gamma\rho^0$     &$<50$         &$<20.8 $        &4.4     & 13    &38   \\
\hline
$\chi_{c0}\to\gamma\phi$     &$<$6.4       &$<$16.2        &0.46   & 1.3    &0.03         \\

$\chi_{c1}\to\gamma\phi$     &$<$26        &$25.8\pm5.2\pm2.3$      &3.6    & 11     &11  \\

$\chi_{c2}\to\gamma\phi$     &$<$13        &$<$8.1         &1.1    & 3.3    &6.5   \\
\hline
$\chi_{c0}\to\gamma\omega$     &$<8.8$       &$< 12.9$        &0.13   & 0.35    &0.22         \\

$\chi_{c1}\to\gamma\omega$     &$83 \pm 15 \pm 12$        &$ 69.7 \pm 7.2 \pm 6.6$       &1.6    & 4.6     &4.7  \\

$\chi_{c2}\to\gamma\omega$     &$<7.0$        &$<6.1$         &0.5    & 1.5    &4.2   \\
\hline\hline

\end{tabular}
\label{t_vs_e}
\end{table*}

The Sokolov-Ternov effect induces self-polarization in high-energy $e^+e^-$ beams, which allows them to naturally become transversely polarized in a storage ring~\cite{Sokolov:1963zn}. Based on the $(2712 \pm 14) \times 10^{6} ~ \psi(2S)$ data samples collected by the BESIII detector in 2009, 2012 and 2021, the precise measurement of polarized parameters and the impact of transversely polarized beams on them will become feasible, enabling a thorough examination of these theoretical models and aiding in a better understanding of the properties of P-wave charmonium radiation decays~\cite{BESIII:2024lks}. Despite the measurement of the branching ratio of $\chi_{cJ} \to \gamma V$, accurate measurement on their polarized parameters is sensitive to validate theoretical calculation. 

In this work, we present a helicity amplitude formula for the process $e^{+}e^{-} \to \psi(2S) \to \gamma_1 \chi_{cJ} \to \gamma_1 \gamma_2 V (V=\rho^0,~\phi,~\omega)$ and construct the spin density matrix for $\chi_{cJ} $ and the light vector mesons. The expressions of joint angular distributions are obtained, and some polarization observables are given to be measured in the $e^{+}e^{-}$ collider. The statistical sensitivities for the relative magnitudes of transverse to longitudinal polarization amplitude of light vector mesons are also discussed with Monte Carlo (MC) simulation results in the paper. By considering the transverse polarization of $e^+e^-$ pairs, the angular distribution parameters will be measured with a high accuracy as well as other decay parameters.

\section{Helicity Amplitude Analysis}
\begin{figure*}[htbp]
\centering
    \subfigure[]{
        \includegraphics[scale=0.25]{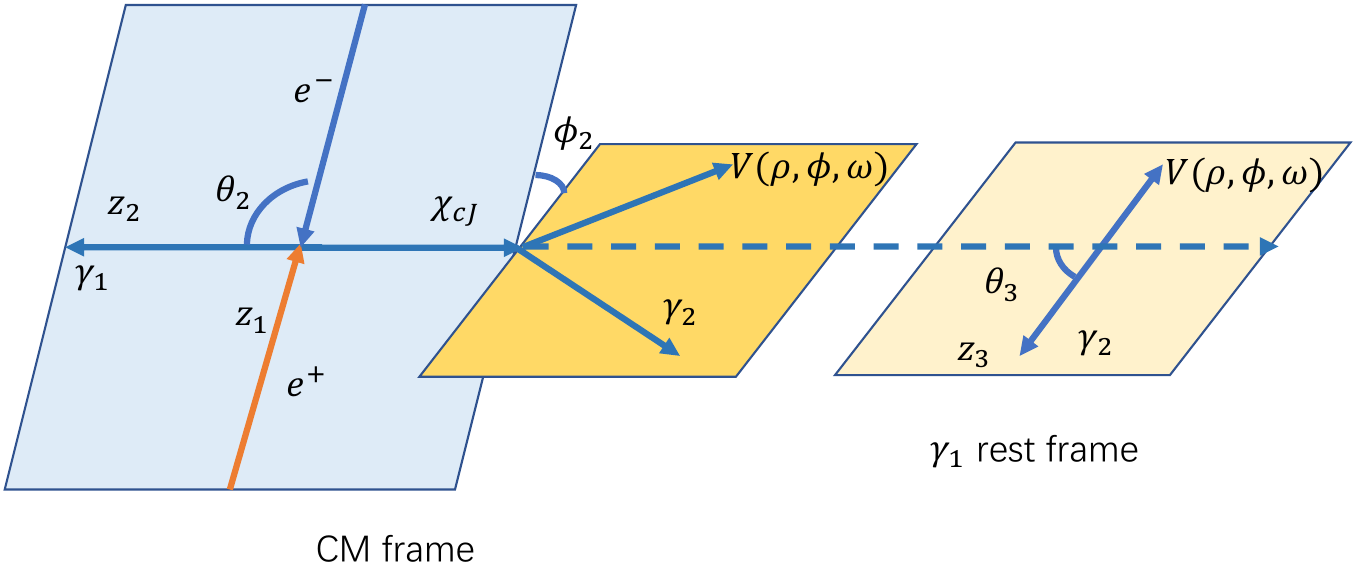}
        \label{decay1}
    }
    \subfigure[]{
        \includegraphics[scale=0.25]{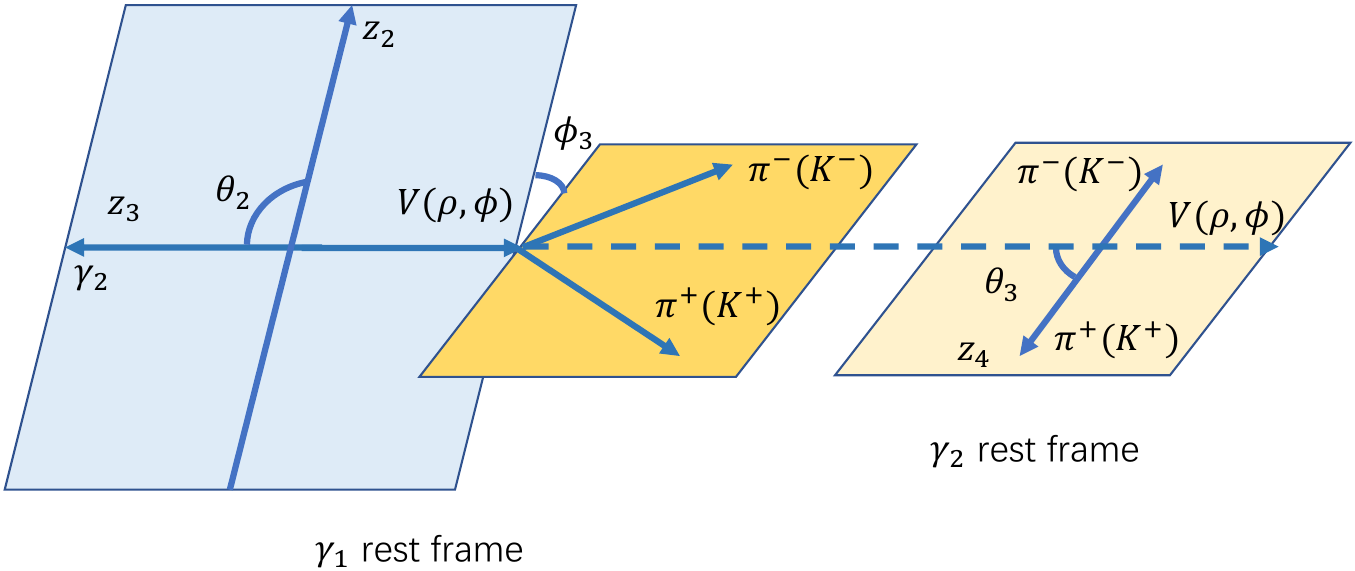}
        \label{decay2}
    }
    \caption{Definition of helicity angles at  $e^{+}e^{-}$ collider experiment.}
\label{decayp}
\end{figure*}

The helicity mechanism is a way to effectively build the dynamic information of the entire decays~\cite{Chung:1993da,Chung:1997jn}. The decay planes as well as helicity angles are visually and clearly depicted in Fig.~\ref{decayp}. In $\psi(2S) \to \gamma_1 \chi_{cJ}$ decay, the helicity angle $\theta_1$ is the polar angle between the direction of the momenta of $e^+$ and $\gamma_1$ in $e^{+}e^{-}$ center-of-mass~(CM) frame. In $\chi_{cJ} \to \gamma_2 V$ decay, the helicity angle $\theta_2$ is choosen as the angle between the direction of momentum of $\gamma_2$ in $\gamma_1$ rest frame and the direction of momentum of $\gamma_1$ from $e^+ e^-$ collision, $\phi_2$ is the azimuthal angle between the $\chi_{cJ}$ production plane and its decay plane. In $\rho^0 \to \pi^+ \pi^-$ and $\phi\to K^+ K^-$ decays, there are two helicity angles, the polar angle $\theta_3$ and the azimuthal angle $\phi_3$, while in a three-body decay $\omega \to  \pi^+ \pi^- \pi^0$, we use the Euler angles $(\alpha,\beta, \gamma)$ to describe its coordinate system rotating process. Specifically, the $\gamma_1$ rest frame is rotated to $\gamma_2$ rest frame by $\gamma$ around $z_3$, $\beta$ around $y_3$ and finally $\alpha$ around $z_3$, where $\beta$ is the angle between the momentum direction of $\gamma_2$ and the cross product direction of the momenta of ${\pi^+}$ and ${\pi^-}$ in $\omega$ rest frame~\cite{Chung:1971ri}.

\begin{table*}[htbp]
\centering
\caption{Definition of helicity angles and amplitudes, where $\lambda_i$ indicates the helicity and energy for the corresponding particles.}
\label{def_a}
\begin{tabular}{lll}
	\hline\hline
	Decay Mode & Solid Angle & Helicity Amplitude \\
	\hline
	$\psi(2S)(\lambda_0) \to \gamma (\lambda_1) \chi_{cJ}(\lambda_2)$ & $\Omega_1$=($\theta_1,\phi_1$) & $B_{\lambda_1,\lambda_2}^J$\\
	\hline
	$\chi_{cJ}(\lambda_2^{\prime})\to\gamma (\lambda_3)V(\lambda_4)$ & $\Omega_2$=($\theta_2,\phi_2$)	&A$_{\lambda_3, \lambda_4}^{Jc}$ \\
	\hline
	$V(\lambda_4^{\prime})\to P  P~ (\rho \to \pi^+ \pi^-,~\phi\to K^+ K^-)$
	& $\Omega_3$=($\theta_3,\phi_3$) & $F^{1}$ \\
	\hline
	$V(\lambda_4') \to P  P  P~(\omega \to \pi^+ \pi^- \pi^0)$  & $  \Omega_3$=($\alpha,\beta, \gamma$) & $F_{0}^{1}$  \\
	\hline\hline
\end{tabular}
\end{table*}

These helicity angles can be constructed by the momentum of final particles. One important point to note is the need to transform the experimentally obtained laboratory-frame momentum to the rest frame of the decaying parent particle for calculation.

The helicity angles and amplitudes for sequential decays are defined in Table~\ref{def_a}. The total amplitude $\mathcal{M}$ for the sequential decay  $\psi(2S) \to \gamma_1 \chi_{cJ} \to \gamma_1 \gamma_2 \rho^0 (\phi) \to \gamma_1 \gamma_2 \pi^+ \pi^- (K^+ K^-)$ can be expressed as
   \begin{eqnarray}
    &\mathcal{M}&( \phi_{1},\theta_{1},\phi_{2}, \theta_{2},\phi_{3}, \theta_{3}; J, M, Jc, \lambda_1 ,\lambda_2,\lambda_3,\lambda_4) \nonumber\\
    &=& B_{\lambda_1 \lambda_2}^{J} D_{M, \lambda_1-\lambda_2}^{J \star}(\phi_{1}, \theta_{1}) A_{\lambda_3 \lambda_4}^{Jc} D_{\lambda_2, \lambda_3-\lambda_4}^{Jc \star}(\phi_{2}, \theta_{2}) \nonumber\\
    &~&\times F^{1} D_{\lambda_4, 0}^{1 \star}(\phi_{3}, \theta_{3}),
   \end{eqnarray}

where $B_{\lambda_1, \lambda_2}^{J}$ ("before" the $\chi_{cJ}$) is the helicity amplitude of $\psi(2S) \to \gamma \chi_{cJ}$ decay, the superscript $J$ is the spin of $\psi(2S)$ with $J=1$, $A_{\lambda_3, \lambda_4}^{Jc}$ ("after" the $\chi_{cJ}$) is the helicity amplitude of the process $\chi_{cJ} \to \gamma \rho^0 (\chi_{cJ} \to \gamma\phi)$, the superscript $Jc$ is the spin of $\chi_{cJ}$ and $F^{1}$ is the helicity amplitude of the decay $\rho^0 \to \pi^+ \pi^-$ or $\phi\to K^+ K^-$ , respectively. The helicity amplitudes are subscripted by the helicity $\lambda_1=\pm 1$ of the radiative photon from $\psi(2S) \to \gamma \chi_{cJ}$ decay, the helicity $\lambda_2$~($\lambda_2=0$ for $\chi_{c0}$, $\lambda_2=0,~\pm 1$ for $\chi_{c1}$ and $\lambda_2=0,~ \pm 1,~\pm 2$ for $\chi_{c2}$) of $\chi_{cJ}$, the helicity $\lambda_3=\pm 1$ of the radiative photon from $\chi_{cJ} \to \gamma V (V=\rho^0, \phi)$, and the helicity  $\lambda_4=0, \pm 1$ of the vector meson ($\rho^0$ or $\phi$). The indices for the helicity of the daughters (psedudoscalar mesons) from $\rho^0$ or $\phi$ decay can be omitted since they are zero. The subscript $M=\pm 1$ for the first Wigner D-function is z-component of spin $J$ of $\psi(2S)$.

The total amplitude $\mathcal{M}$ for the decay
 $\psi(2S) \to \gamma_1 \chi_{cJ} \to \gamma_1 \gamma_2 \omega \to \gamma_1 \gamma_2 \pi^+ \pi^- \pi^0 $  is
\begin{eqnarray}
    &\mathcal{M}&( \phi_{1},\theta_{1},\phi_{2}, \theta_{2},\phi_{3}, \theta_{3}; J, M, Jc,\lambda_1 ,\lambda_2,\lambda_3, \lambda_4) \nonumber\\
    &=& B_{\lambda_1 \lambda_2}^{J} D_{M, \lambda_1-\lambda_2}^{J \star}(\phi_{1}, \theta_{1}) A_{\lambda_3 \lambda_4}^{Jc} D_{\lambda_2, \lambda_3-\lambda_4}^{Jc \star}\left(\phi_{2}, \theta_{2}\right) \nonumber\\
    &~&\times F_{\mu }^{1} D_{\lambda_4, \mu}^{1 \star}(\alpha, \beta, \gamma),
\end{eqnarray}
where $F_{\mu}^{1}$ is the helicity amplitude of the decay $\omega \to \pi^+ \pi^- \pi^0 $ , respectively. And $\mu$ is the z-component of spin angular momentum $\mathrm{J}$ of $\omega$, while the normal to the $\omega$ decay plane are taken as the z-axis. The symmetry relation requires only one amplitude $F_{\mu}^{1}(\mu = 0)$~\cite{AmpF}.

The decay rates $\Gamma$ for the cascade decays $\psi(2S) \to \gamma_1 \chi_{cJ} \to \gamma_1 \gamma_2 \rho^0 (\phi) \to \gamma_1 \gamma_2 \pi^+ \pi^- (K^+ K^-)$  and $\psi(2S) \to \gamma_1 \chi_{cJ} \to \gamma_1 \gamma_2 \omega \to \gamma_1 \gamma_2 \pi^+ \pi^- \pi^0$ are proportional to
\begin{eqnarray}
\label{decay_rates}
\sum_{\substack{M,\lambda_1 ,\lambda_2, \\ \lambda_3,\lambda_4}} \left| \mathcal{M}( \phi_{1},\theta_{1},\phi_{2}, \theta_{2},\phi_{3}, \theta_{3};J, M, Jc, \lambda_1 ,\lambda_2,\lambda_3 \lambda_4) \right|^2 .
\end{eqnarray}

The electromagnetic transitions $\psi(2S) \to \gamma \chi_{cJ}$ are dominant by electric dipole (E1) transitions, and its helicity amplitudes $B_{\lambda_1 \lambda_2}^{J}$ satisfy the E1 transition relations~\cite{Karl:1975jp},
\begin{eqnarray}
\label{B_eqs}        B^{1}_{1,1}&=&B^{1}_{1,0}~~~~~~~~~~~~~~~~~~~\text{for $\psi(2S) \to \gamma_1\chi_{c1}$ decay},\nonumber\\    B^{1}_{1,2}&=&\sqrt{2}B^{1}_{1,1}=\sqrt{6}B^{1}_{1,0}~~~\text{for $\psi(2S) \to \gamma_1\chi_{c2}$ decay.}
\end{eqnarray}
Parity conservation gives the relation $B^{1}_{\lambda_1,\lambda_2}=B^{1}_{-\lambda_1,-\lambda_2}\eta_{\psi(2S)}\eta_{\gamma}\eta_{\chi_{cJ}}(-1)^{s_{\psi(2S)}-s_{\gamma}-s_{\chi_{cJ}}}$, where $\eta$ and $s$ represents the parity and spin of the particle. So we have $B^{1}_{-1,0}=B^{1}_{1,0}$ for the decay $\psi(2S) \to \gamma \chi_{c0}$, $B^{1}_{-1,-1}=-B^{1}_{1,1}$, $B^{1}_{-1,0}=-B^{1}_{1,0}$ for the decay $\psi(2S) \to \gamma \chi_{c1}$, and $B^{1}_{-1,-2}=B^{1}_{1,2}$, $B^{1}_{-1,-1}=-B^{1}_{1,1}$, $B^{1}_{-1,0}=-B^{1}_{1,0}$ for the decay $\psi(2S) \to \gamma \chi_{c2}$. Angular momentum conservation $|\lambda_1-\lambda_2|\leq 1$ requires that these amplitudes($B^{1}_{1,-1}, B^{1}_{-1,1}$) do not exist for the decay $\psi(2S) \to \gamma \chi_{c1}$, and $B^{1}_{1,-1},B^{1}_{-1,1},B^{1}_{-1,2},B^{1}_{1,-2}$ do not exist for the decay $\psi(2S) \to \gamma \chi_{c2}$. The matrix of helicity amplitude can be written as
\begin{eqnarray}
  B=
  \begin{bmatrix}
   B^{1}_{-1,0}\\
   B^{1}_{0,0}\\
   B^{1}_{1,0}
   \end{bmatrix}
   =
  \begin{bmatrix}
   B^{1}_{1,0}\\
   0\\
   B^{1}_{1,0}
   \end{bmatrix}
   ~~~\text{for $\psi(2S) \to \gamma \chi_{c0}$ decay,}
\end{eqnarray}
\begin{eqnarray}
  B&=&
  \begin{bmatrix}
   B^{1}_{-1,-1}& B^{1}_{-1,0} &0 \\
   B^{1}_{0,-1} & B^{1}_{0,0} & B^{1}_{0,1}\\
   0 & B^{1}_{1,0} & B^{1}_{1,1}
   \end{bmatrix}
   =
  \begin{bmatrix}
   -B^{1}_{1,1} & -B^{1}_{1,0} & 0 \\
   0 & 0 & 0\\
   0 & B^{1}_{1,0} &	B^{1}_{1,1}
   \end{bmatrix} \nonumber\\
   &=&
  \begin{bmatrix}
   -B^{1}_{1,0} & -B^{1}_{1,0} & 0 \\
   0 & 0 & 0\\
   0 & B^{1}_{1,0} &	B^{1}_{1,0}
   \end{bmatrix}
   ~\text{for $\psi(2S) \to \gamma \chi_{c1}$ decay,}
\end{eqnarray}
\begin{eqnarray}
  B&=&
  \begin{bmatrix}
   B^{1}_{-1,-2}& B^{1}_{-1,-1} & B^{1}_{-1,0} & B^{1}_{-1,1} &0\\
   B^{1}_{0,-2} & B^{1}_{0,-1} & B^{1}_{0,0} & B^{1}_{0,1} & B^{1}_{0,2}\\
   0 & B^{1}_{1,-1} & B^{1}_{1,0} & B^{1}_{1,1}	& B^{1}_{1,2}
   \end{bmatrix}\nonumber\\
   &=&
  \begin{bmatrix}
   B^{1}_{1,2} & B^{1}_{1,1} & B^{1}_{1,0} & 0 & 0\\
   0 & 0 & 0 & 0 & 0\\
   0 & 0 & B^{1}_{1,0} &	B^{1}_{1,1} & B^{1}_{1,2}
   \end{bmatrix} ~\text{for $\psi(2S) \to \gamma \chi_{c2}$ decay.} \nonumber\\
   &=&
  \begin{bmatrix}
   \sqrt{6}B^{1}_{1,0} & \sqrt{3}B^{1}_{1,0} & B^{1}_{1,0} & 0 & 0 \\
   0 & 0 & 0 & 0 & 0\\
   0 & 0 & B^{1}_{1,0} & \sqrt{3}B^{1}_{1,0} & \sqrt{6}B^{1}_{1,0}
   \end{bmatrix}
\end{eqnarray}
As for $\chi_{cJ} \to \gamma V(\rho^0, \phi, \omega)$ decays, only one independent helicity amplitude for $\chi_{c0}$ decays ($A^{0}_{1,1}$ or $A^{0}_{-1,-1}$, "so-called" transverse polarization amplitude $A^{\perp}_{0V}$ in Ref.~\cite{c2gV}), there are more than one independent helicity amplitudes for $\chi_{c1}$ and $\chi_{c2}$ decays. The helicity amplitudes can be expressed as $A^{Jc}_{\lambda_3, \lambda_4} = a_{\lambda_3, \lambda_4}*e^{i\zeta_{\lambda_3, \lambda_4}}$, where $Jc$ is the spin of the mother particle, $\lambda_i$ is the helicity value of daughter particles, $a$ and $\zeta$ are the magnitude and phase angle of helicity amplitude $A$. Parity conservation leads to the following symmetry relations for helicity amplitudes as
\begin{eqnarray}
\label{A_eqs}    A^{0}_{1,1}&=&A^{0}_{-1,-1}~~~~~~~~~~~~~~~~~~~~\text{for $\chi_{c0}$ decay},\nonumber\\
    A^{1}_{-1,-1}&=&-A^{1}_{1,1},~~A^{1}_{-1,0}=-A^{1}_{1,0}~\text{for $\chi_{c1}$ decay},\nonumber\\    A^{2}_{-1,-1}&=&A^{2}_{1,1},~~A^{2}_{-1,0}=A^{2}_{1,0},~~A^{2}_{-1,1}=A^{2}_{1,-1}\nonumber\\
    &~&~~~~~~~~~~~~~~~~~~~~~~~~~~~~~\text{for $\chi_{c2}$ decay}.
\end{eqnarray}
We define a parameter for polarization observable to describe the relative magnitudes of two helicity amplitudes
\begin{eqnarray}
\label{x_chic1}
	x &\equiv& \frac{|A^1_{1,1}|}{|A^1_{1,0}|} = \frac{|A^{\perp}_{1V}|}{|A^{||}_{1V}|}
\end{eqnarray}
for $\chi_{c1} \to \gamma V(\rho^0, \phi, \omega)$ decays, and two independent parameters
\begin{eqnarray}
	x &\equiv& \frac{|A^2_{1,1}|}{|A^2_{1,0}|}=\frac{|A^{\perp}_{2V}|}{|A^{||}_{2V}|}, ~~~y \equiv \frac{|A^2_{1,-1}|}{|A^2_{1,0}|}=\frac{|T^{\perp}_{2V}|}{|A^{||}_{2V}|}
\end{eqnarray}
for $\chi_{c2} \to \gamma V(\rho, \phi, \omega)$ decays, where $A$ or $T$ with subscript $\perp$ (transverse polarization) or $||$ (longitudinal polarization) is the amplitude defined in Ref.~\cite{c2gV}.
They have the significance of describing the ratio of amplitudes that characterize transverse polarization to longitudinal polarization.

The phase differences between independent amplitudes are defined as
\begin{eqnarray}
	\Delta _1=\zeta _{1,0}-\zeta _{1,1}
\end{eqnarray}
for $\chi_{c1}$, and
\begin{eqnarray}
\label{deltas}
	\Delta_1&=&\zeta _{1,0}-\zeta _{1,1},~~\Delta _2=\zeta _{1,-1}-\zeta _{1,1},
\end{eqnarray}
for $\chi_{c2}$.

\section{Spin Density Matrix}
As one of the method to get the decay angular distributions,  the spin density matrix (SDM) carries the dynamical information of particle decay, and its different parameterization can clearly explain various physical phenomena. For example, when expressed in multipole parameter form $r^L_M$ (The $L$-rank index ranges from 1 to 2J, $M$ is taken from $-L$ to $L$), it can provide information about particle polarization~\cite{Doncel:1973sg,a_chi}.

The spin density matrix of $\psi(2S)$ from polarized $e^+ e^-$ beams can is written to be
\begin{eqnarray}
    \rho^{\psi(2S)} = \frac{1}{2}
    \begin{bmatrix}
        (1-\mathcal{P}_z)(1+\bar{\mathcal{P}}_z) &  0 & P_T^2\\
        0 &  0 & 0\\
        P_T^2 &  0 & (1+\mathcal{P}_z)(1-\bar{\mathcal{P}}_z)
    \end{bmatrix},
\end{eqnarray}
where $\mathcal{P}_z/\bar{\mathcal{P}}_z$ is the degree of longitudinal polarization of $e^+/e^-$ and $P_T$ is the degree of transverse polarization of $e^+/e^-$~\cite{sdm_psi}.

By taking the direction of the photon's momentum as the positive z-axis direction as shown in Fig.~\ref{decay1}, the SDM of $\chi_{c0}$ can be written as
\begin{eqnarray}
	\rho^{\chi_{c0}} &=& \frac{1}{2} b_{1,0}^2 \left( 1 + \cos ^2\left(\theta _1\right)+ P_T^2 \sin ^2\left(\theta _1\right) \cos \left(2 \phi _1\right) \right),
\end{eqnarray}
while the matrix elements of $\rho^{\chi_{c1}}$ and $\rho^{\chi_{c2}}$ are listed in Append.~\ref{ele_chic12}, where $b_{1, 0}$ is the magnitude of helicity amplitude $B_{1, 0}$ in $\psi(2S) \to \gamma  \chi_{cJ}$ decays.
In these SDM calculations, we have used the E1 transition relations in the $\psi(2S) \to \gamma \chi_{cJ}$ decay~\cite{E1}.

In our paper, we take the multipole parameters $r^L_M$ to describe the SDM of vectors($V =\rho^0, \phi, \omega$), it is expressed as

\begin{eqnarray}
		\rho^{V}=\frac{r^0_0}{3}
			\begin{bmatrix}
			r^2_0+\sqrt{3} r^1_0+1 & \sqrt{\frac{3}{2}}(-i r^1_{-1}+r^1_1-i r^2_{-1}+r^2_1)& \sqrt{3}(r^2_2-i r^2_{-2})\\
			\sqrt{\frac{3}{2}}(i r^1_{-1}+r^1_1+i r^2_{-1}+r^2_1)& 1-2 r^2_0 &\sqrt{\frac{3}{2}}(-i r^1_{-1}+r^1_1+i r^2_{-1}-r^2_1)\\
			\sqrt{3}(r^2_2+i r^2_{-2}) &\sqrt{\frac{3}{2}}(i r^1_{-1}+r^1_1-i r^2_{-1}-r^2_1) &r^2_0-\sqrt{3} r^1_0+1
			\end{bmatrix}.
\end{eqnarray}

In $\chi_{cJ} \to \gamma V(\rho, \phi, \omega)$ decays, the real multipole parameters $r^L_M$ have different expressions for $J = 0, 1, 2$ as listed in Appendix~\ref{rLM_exp}.

\section{Joint Angular Distribution}
In order to compare with experimental results from electron-positron collider, the joint angular distribution between the production and decay of $\chi_{cJ}$ is needed.
The expressions of joint angular distribution can be obtained easily using the SDMs of decay particles. Here we construct the joint angular distribution expressions for each decay level, which can be used experimentally to verify the measured results. The joint angular distributions for the processes $\psi(2S) \to \gamma_1 \chi_{cJ}$, $\psi(2S) \to \gamma_1 \chi_{cJ}, ~\chi_{cJ}\to \gamma_2 V (V=\rho^0,~\phi,~\omega)$ and $\psi(2S) \to \gamma_1 \chi_{cJ},~\chi_{cJ}\to \gamma_2 V (V=\rho^0,~\phi,~\omega),~V \to \text{final~states}$ read

\begin{eqnarray}
	&\mathcal{W}(\Omega_1)&\propto \mathrm{Tr}[\rho^{\chi_{cJ}}],\\
	&\mathcal{W}(\Omega_1, \Omega_2)&\propto \mathrm{Tr}[\rho^{V}]=r^0_0,\\
	&\mathcal{W}(\Omega_1, \Omega_2, \Omega_3) &\propto \sum_{\lambda_4,\lambda_4'} \rho^{V}_{\lambda_4,\lambda_4'} D^{1*}_{\lambda_4,0}(\Omega_3)D^{1}_{\lambda_4',0}(\Omega_3)|F^{1}_0|^2\nonumber\\
	& &\propto -\frac{1}{6} r^0_0 [2 \sqrt{3} \sin ^2\theta _3 (r^2_{-2} \sin 2 \phi _3\nonumber\\
	& &+r^2_2 \cos 2 \phi _3) +2 \sqrt{3} \sin 2 \theta _3 (r^2_{-1} \sin \phi _3\nonumber\\
	& &+r^2_1 \cos \phi _3)+3 r^2_0 \cos 2 \theta _3+r^2_0-2],
\end{eqnarray}
respectively.

Taking into account the polarization observables used in experiment, we provide the distribution formulae for polar angle $\theta_i~(i=1,2,3)$ after integrating other polar angles and azimuthal angles.

For $\chi_{c0}$,
\begin{eqnarray}
\label{chic0ang}
	\frac{\mathrm{d}N}{\mathrm{d}\cos\theta_1} &\propto& 1+ \cos^2\theta_1,\\
	\frac{\mathrm{d}N}{\mathrm{d}\cos\theta_3} &\propto& 1 -\cos^2\theta_3, \\
	 \frac{\mathrm{d}{N}}{\mathrm{d} \cos \theta_1\mathrm{d}\phi_1}& \propto &1+\cos ^2\left(\theta _1\right)\nonumber\\
	 &+&{Pt}^2 \left(1-\cos ^2\left(\theta _1\right)\right) \cos \left(2 \phi _1\right)
\label{chic0ang1}
\end{eqnarray}
The angular distribution ${\mathrm{d}N}/{\mathrm{d}\cos\theta_2}$ is trivial due to the projection of the cosine polar angle $\cos\theta_2$ is flat.

For $\chi_{c1}$,
\begin{eqnarray}
\label{chic1ang}
	\frac{\mathrm{d}N}{\mathrm{d}\cos\theta_1} &\propto& 1- \frac{1}{3}\cos^2\theta_1,\\
\label{chic1ang1}
	\frac{\mathrm{d}N}{\mathrm{d}\cos\theta_2} &\propto& 1+ \frac{2x^2-1}{2x^2+3} \cos^2\theta_2,\\
	\frac{\mathrm{d}N}{\mathrm{d}\cos\theta_3} &\propto& 1+ \frac{2-x^2}{x^2} \cos^2\theta_3,  \\
	\frac{\mathrm{d}{N}}{\mathrm{d} \cos \theta_1\mathrm{d}\phi_1}& \propto &1-\frac{1}{3} \cos ^2\left(\theta _1\right)\nonumber\\
	&-&\frac{1}{3} {Pt}^2 \left(1-\cos ^2\left(\theta _1\right)\right) \cos \left(2 \phi _1\right).
\label{chic1ang2}
\end{eqnarray}

For $\chi_{c2}$,
\begin{eqnarray}
\label{chic2ang}
	\frac{\mathrm{d}N}{\mathrm{d}\cos\theta_1} &\propto 1&+ \frac{1}{13}\cos^2\theta_1,\\
\label{chic2ang1}
	\frac{\mathrm{d}N}{\mathrm{d}\cos\theta_2} &\propto 1&+\frac{-6x^2+6y^2-3}{10x^2+6y^2+9}\cos^2\theta_2,\\
	\frac{\mathrm{d}N}{\mathrm{d}\cos\theta_3} &\propto 1& + \frac{2-x^2-y^2}{x^2+y^2}\cos^2\theta_3,\\
	\frac{\mathrm{d}{N}}{\mathrm{d} \cos \theta_1\mathrm{d}\phi_1}& \propto &
 1+\frac{1}{13} \cos^2\theta_1\nonumber\\
 &+& \frac{1}{13}{Pt}^2 \left(1-\cos ^2\left(\theta _1\right)\right) \cos \left(2 \phi _1\right)
\label{chic2ang2}
\end{eqnarray}

To validate the above angular distribution functions, MC simulation is performed with modeling by amplitude sampling of phase space events in Eqs.~\ref{decay_rates}-\ref{deltas}.
The phase differences are naively set to be $\Delta_1= \frac{\pi}{3}$ in $\chi_{c1}$ decays, and $\Delta_1=\frac{\pi}{3}$ and $\Delta_2=\frac{\pi}{4}$ in $\chi_{c2}$ decays.
The degree of beam polarization is simply set to be $P_T=0.24$ for all $\chi_{cJ}$ decays.
The parameter $x$ is set to be 0.43 for $\chi_{c1}\to \gamma \rho^0$ decay, 0.63 for $\chi_{c1}\to \gamma \phi$ decay, and 0.57 for $\chi_{c1}\to \gamma \omega$ decay, which is obtained from BESIII measurement in 2011~\cite{BESIII:2011ysp}.
Referring to Ref.~\cite{E1}, the parameters for $\chi_{c2}\to \gamma V$ decays are arbitrarily chosen to be $x=1.55$, $y=2.06$ for $\chi_{c2}\to \gamma \rho^0$ decay, $x=1.55$, $y=2.13$ for $\chi_{c2}\to \gamma \phi$ decay, and $x=0$, $y=1$ for $\chi_{c2}\to \gamma \omega$ decay, respectively.
500,000 pseudoexperiments for each decay mode are generated and fitted using a probability density function derived from the full angular distributions shown in Eqs~\ref{chic0ang}-\ref{chic2ang2}.

\begin{figure*}[htbp]
	\centering
	\subfigure[]{
		\includegraphics[scale=0.20]{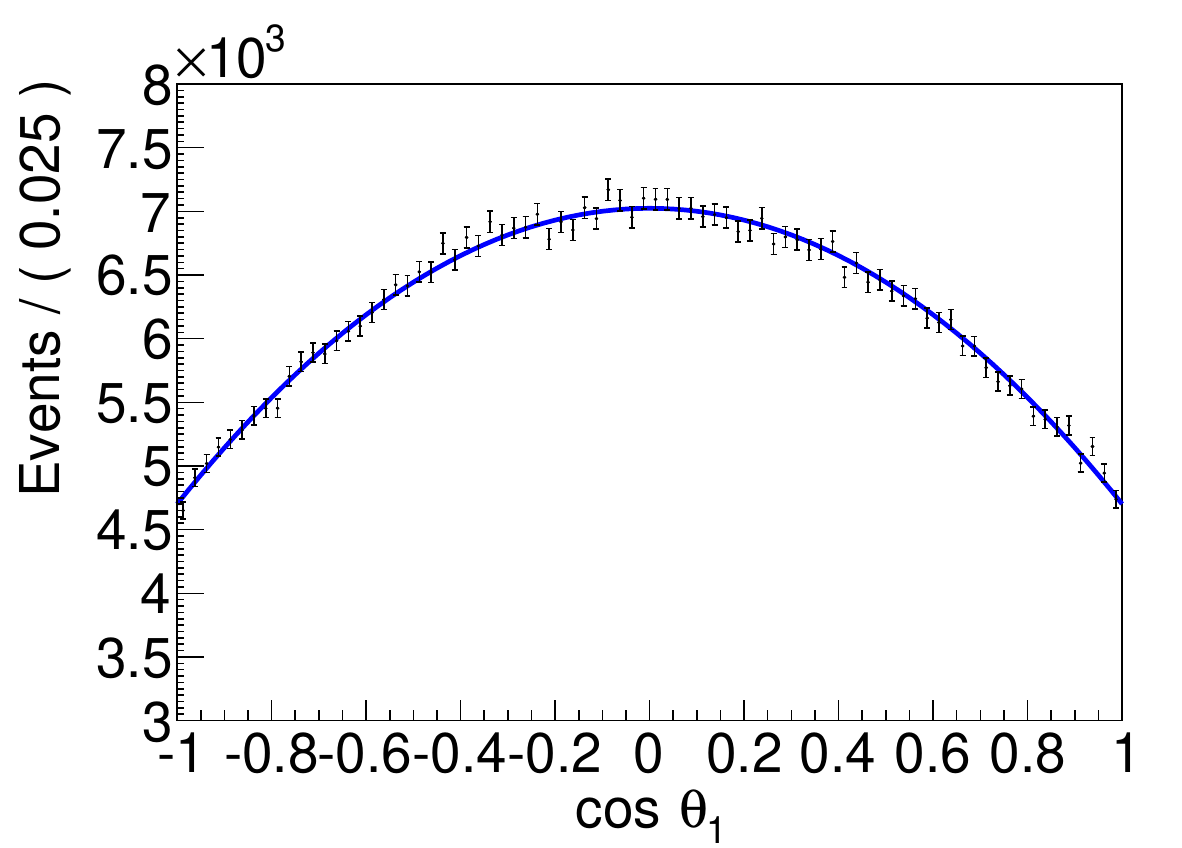}
		\label{c1_p_c1}}
	\subfigure[]{
		\includegraphics[scale=0.20]{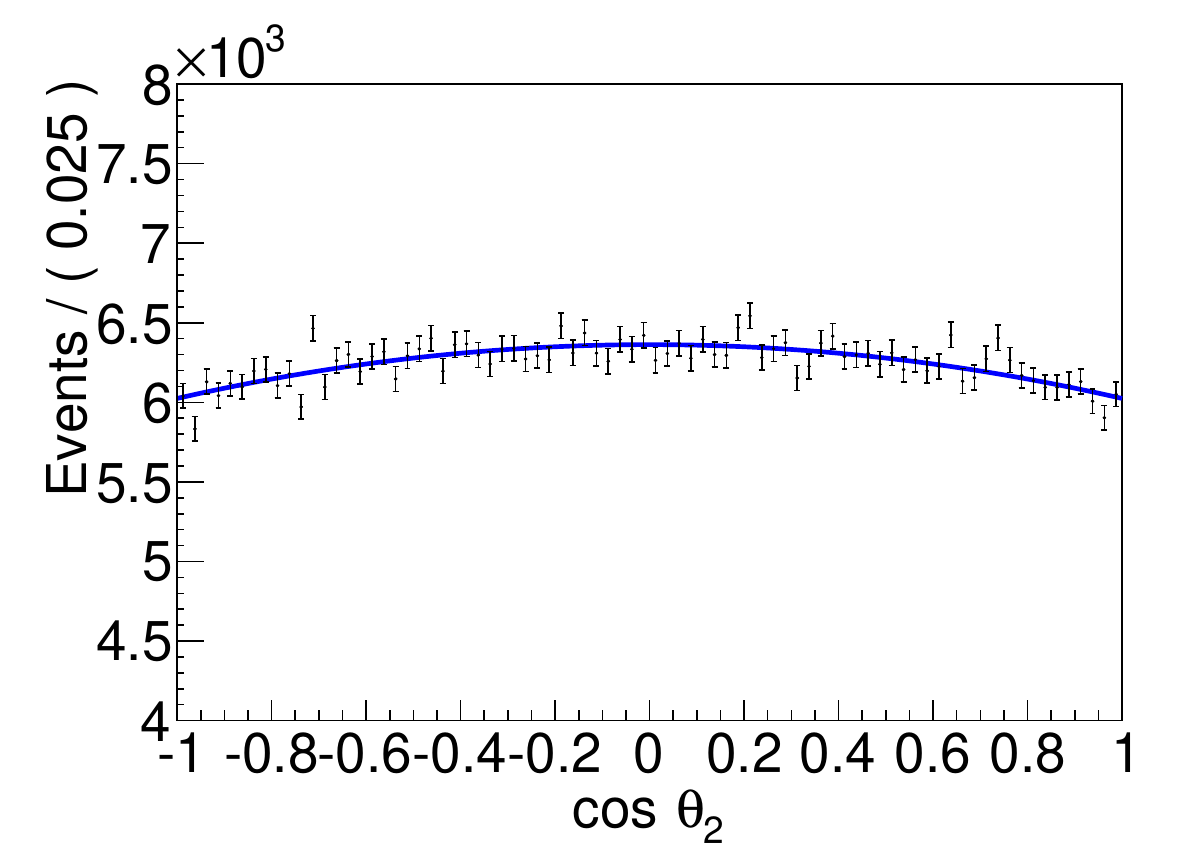}
		\label{c1_p_c2}	}
	\subfigure[]{
		\includegraphics[scale=0.20]{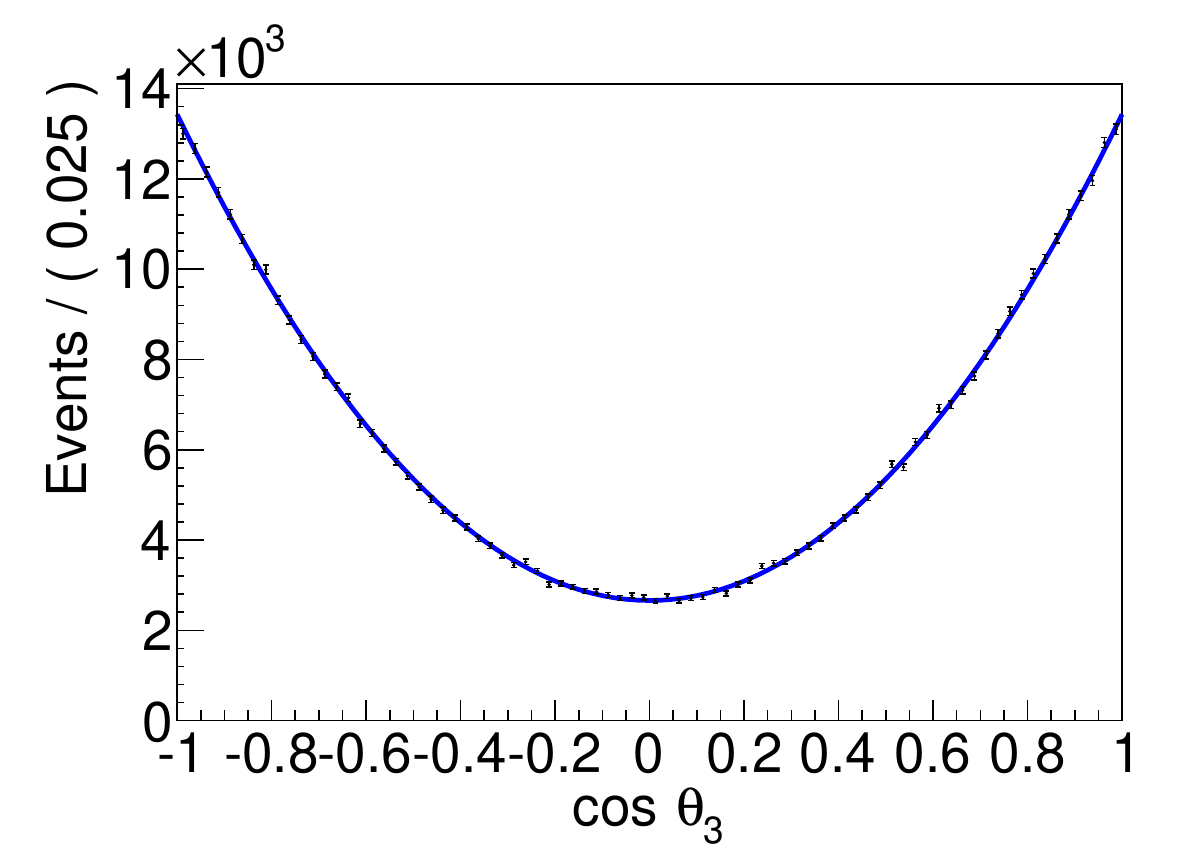}
		\label{c1_p_c3}}
    \subfigure[]{
        \includegraphics[scale=0.20]{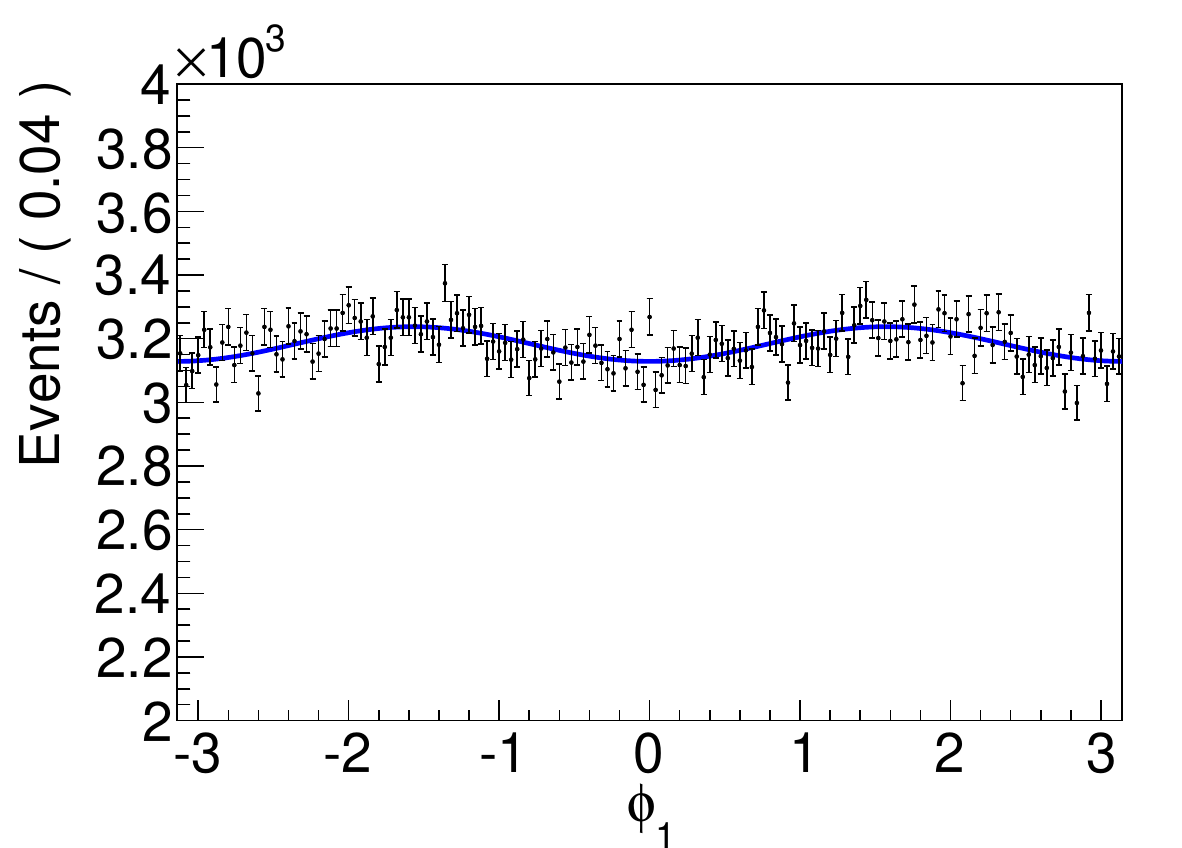}
        \label{c1_p_phi}
    }
	\caption{Fits to the angular distributions of $\cos\theta_i (i=1,2,3)$ and $\phi_1$ in $\psi(2S)\to \gamma \chi_{c1}$, $\chi_{c1} \to \gamma \phi$ and $\phi\to K^+K^-$ decays. Dots with error bars represent MC events, and the blue solid curve denotes the fit.}
\end{figure*}

\begin{figure*}[htbp]
	\centering
	\subfigure[]{
		\includegraphics[scale=0.20]{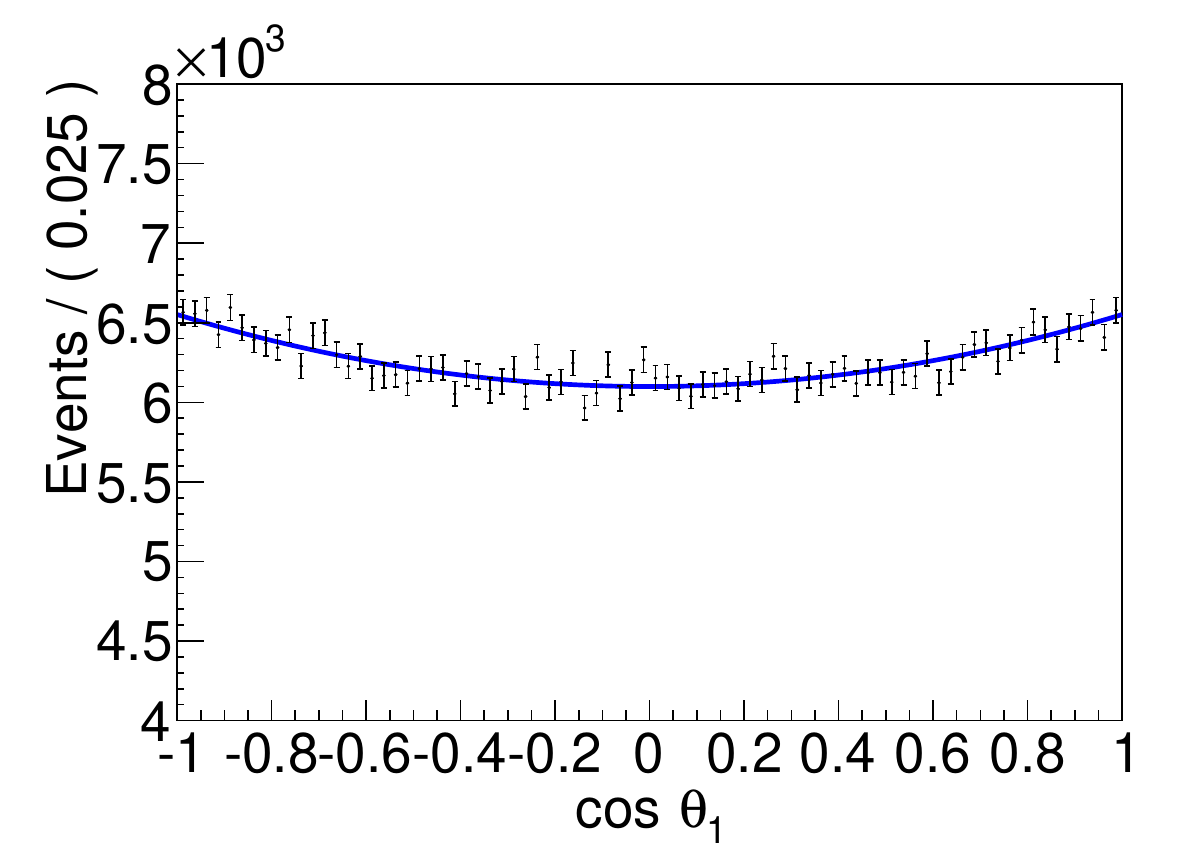}
		\label{c2_p_c1}}
	\subfigure[]{
		\includegraphics[scale=0.20]{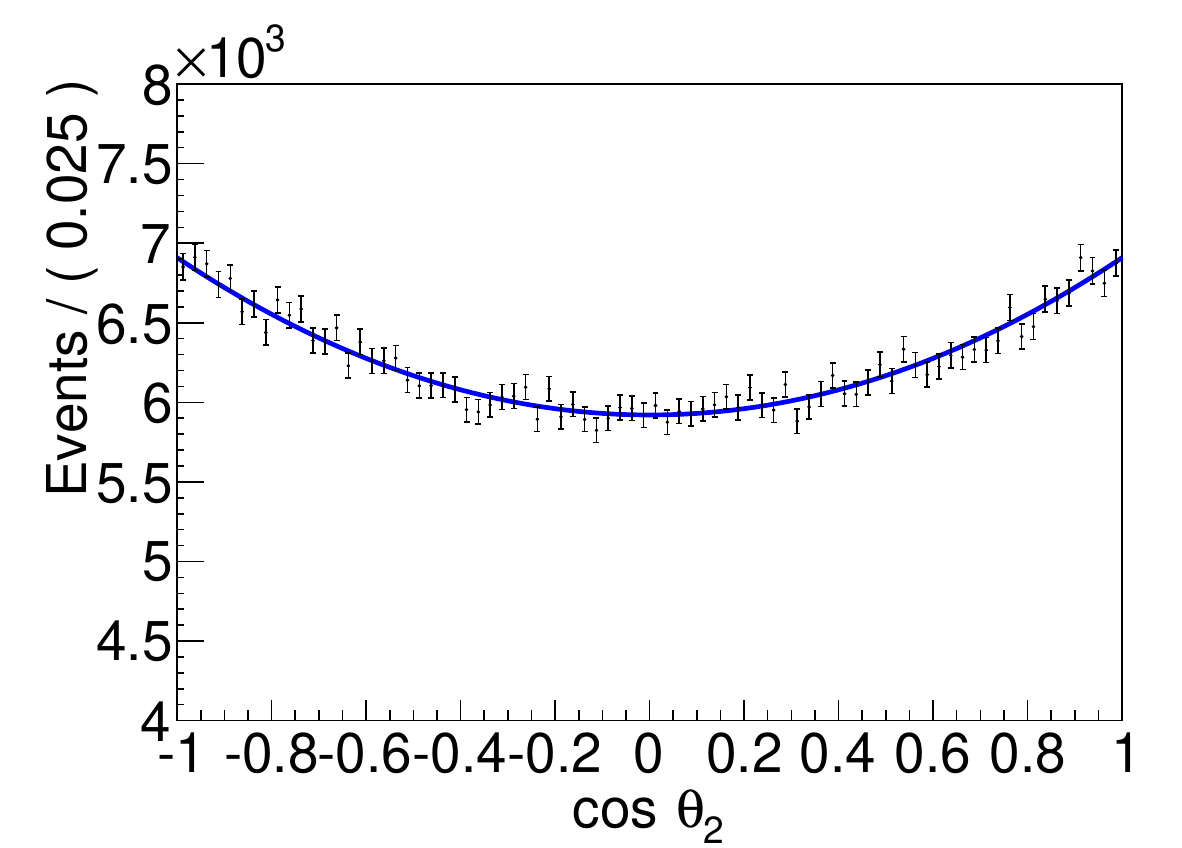}
		\label{c2_p_c2}	}
	\subfigure[]{
		\includegraphics[scale=0.20]{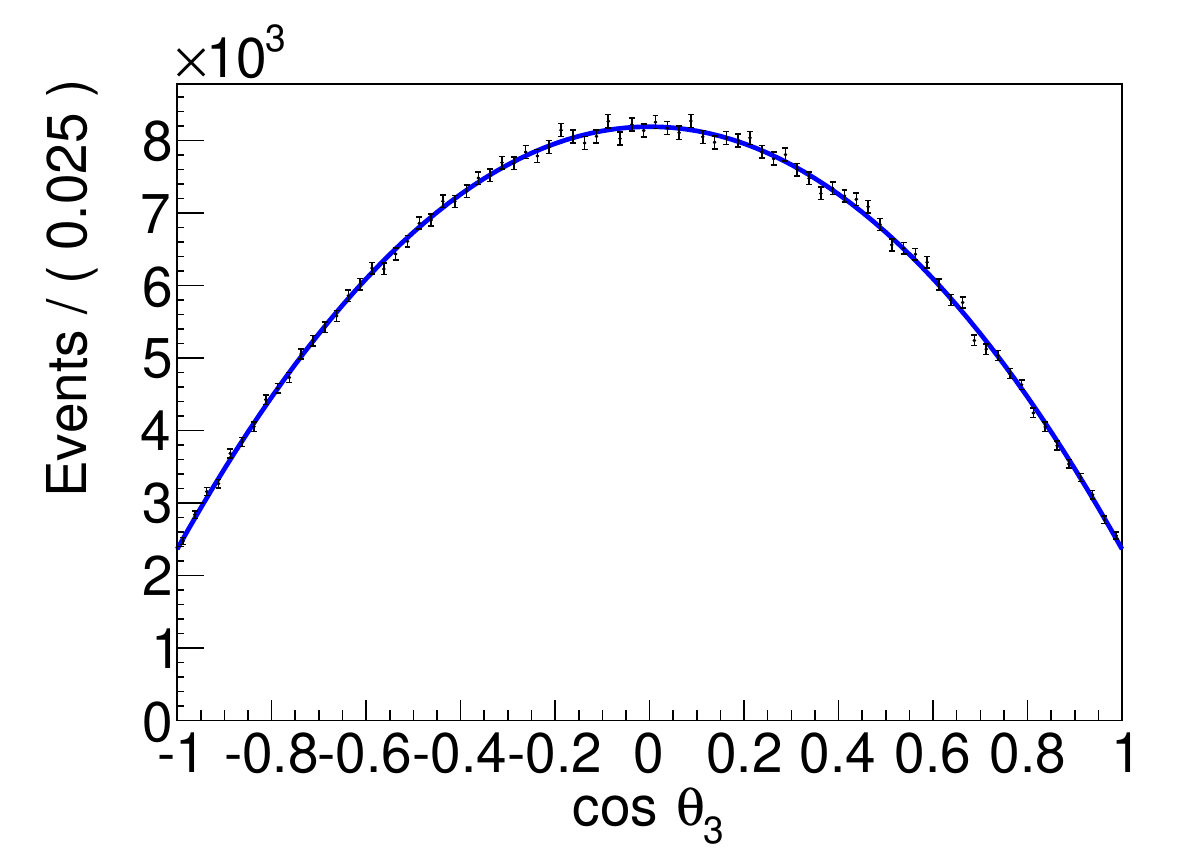}
		\label{c2_p_c3}}
    \subfigure[]{
        \includegraphics[scale=0.20]{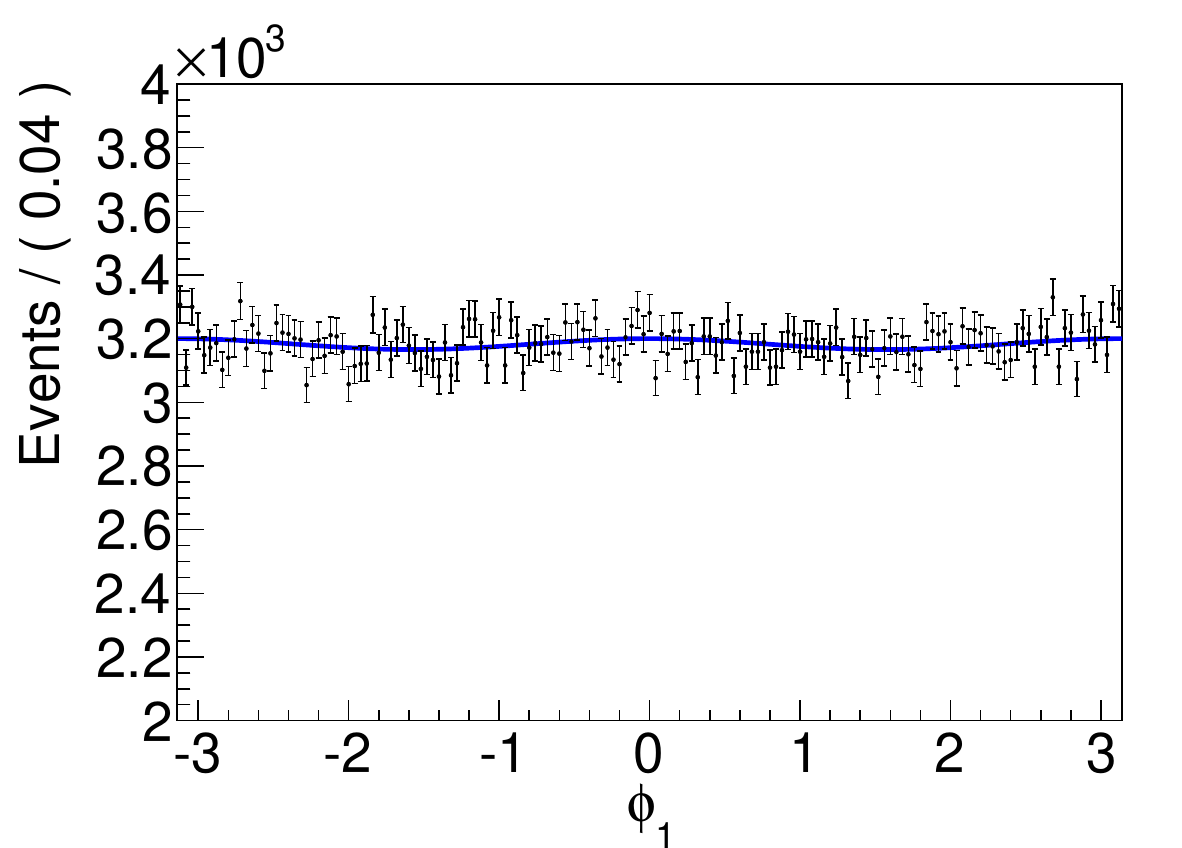}
        \label{c2_p_phi}}
	\caption{Fits to the angular distributions of $\cos\theta_i (i=1,2,3)$ and $\phi_1$ in $\psi(2S)\to \gamma \chi_{c2}$, $\chi_{c2} \to \gamma \phi$ and $\phi\to K^+K^-$ decays. Dots with error bars represent MC events, and the blue solid curve denotes the fit.}
\end{figure*}

Here we list some fit results, and the others can be found in Appendix.~\ref{fit_results}. We use the function $1+\alpha\cos^2\theta$ to fit the angular distributions of $\psi(2S)\to \gamma \chi_{cJ} (J=0,1,2)$ in Eq.~\ref{chic0ang},~\ref{chic1ang},~\ref{chic2ang} and  $\rho^0 \to \pi^+ \pi^-$, $\phi\to K^+K^-$, $\omega \to \pi^+ \pi^- \pi^0$ from $\chi_{c0} \to \gamma V (V=\rho^0,~\phi,~\omega) $ decays in Eq.~\ref{chic0ang1}. In the case, the fitted $\alpha$ values are $0.992 \pm 0.009$ for the decay $\psi(2S)\to \gamma \chi_{c0}$ and $-0.999 \pm 0.001$ for the decay $\phi\to K^+K^-$ from $\chi_{c0} \to \gamma \phi$ decay in Fig.~\ref{c0_o_c1} and Fig.~\ref{c0_o_c3}, respectively, which are consistent with the default values in Eq.~\ref{chic0ang} and Eq.~\ref{chic0ang1} within the standard deviation.
With the measured angular distributions of $\chi_{cJ} \to \gamma V~(J=0,1,2)$ , we can extract the parameter $\hat{x}(x, y, P_T)$ in $\chi_{cJ}$ decays by fitting these equations of joint angular distributions to data. Given an example for fitting the angular distributions of $\chi_{cJ} \to \gamma \phi~(J=1,2)$ based on MC simulation samples, the fitted $x$ and $P_T$ value in $\chi_{c1} \to \gamma \phi$ decays are $0.630 \pm 0.002$ and $0.26\pm0.02$ in Fig.~\ref{c1_p_c2}, Fig.~\ref{c1_p_c3} and Fig.~\ref{c1_p_phi}, which are in agreement with the input $x=0.63$ and $P_T=0.24$. In $\chi_{c2} \to \gamma \phi$ decays, the fitted parameters of $x$ and $y$ are $x=1.558 \pm0.344,~y =2.137 \pm 0.251$ in Fig.~\ref{c2_p_c2} and Fig.~\ref{c2_p_c3}, respectively, which are consistant with the input $x=1.55,~y=2.13$.

\begin{figure*}[htbp]
	\centering
	\subfigure[]{
		\includegraphics[scale=0.25]{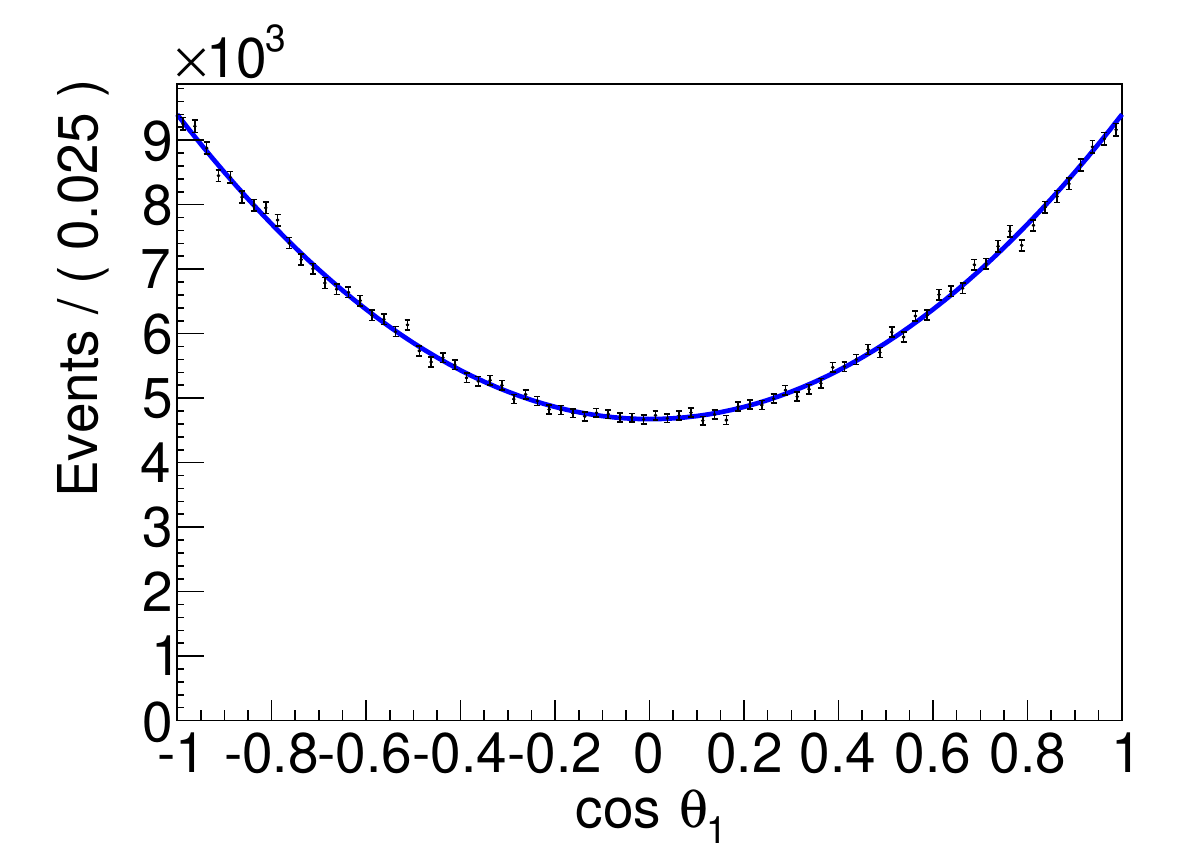}
		\label{c0_p_c1}}
	\subfigure[]{
		\includegraphics[scale=0.25]{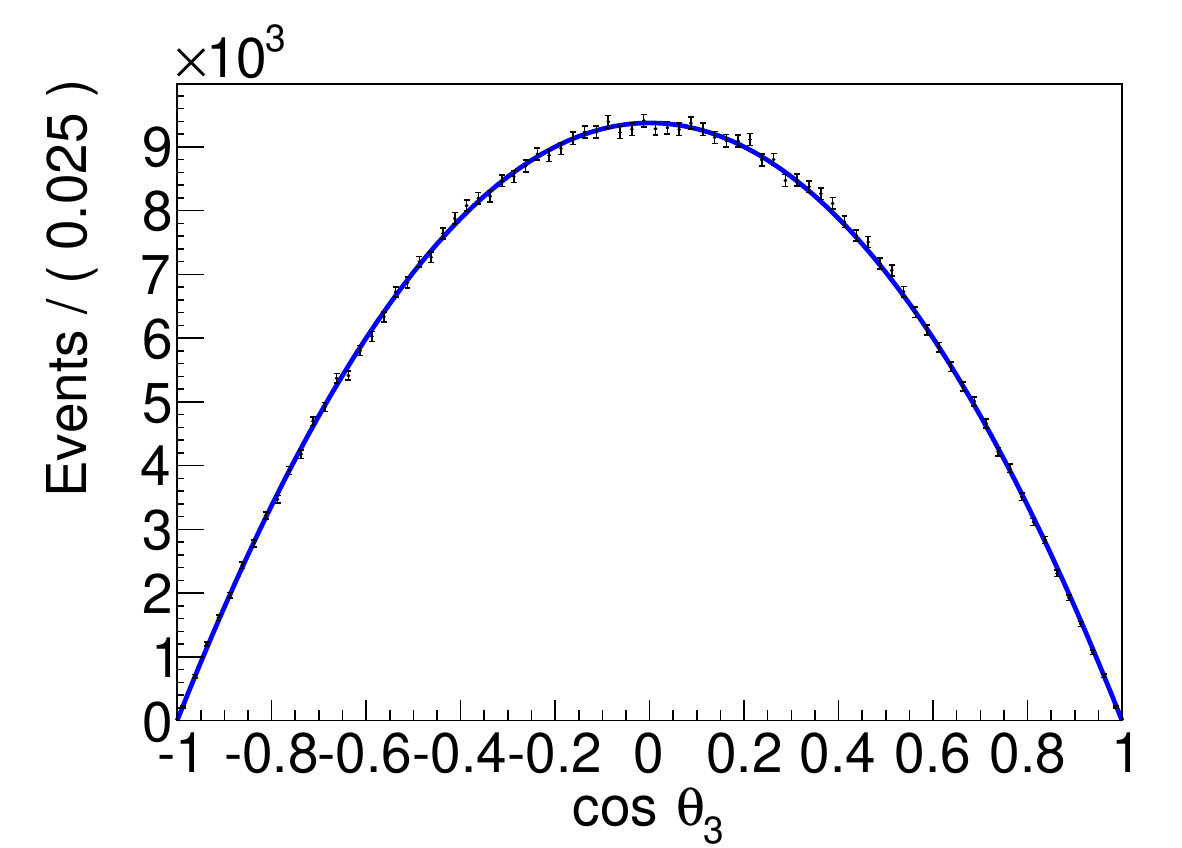}
		\label{c0_p_c3}}
    \subfigure[]{
        \includegraphics[scale=0.25]{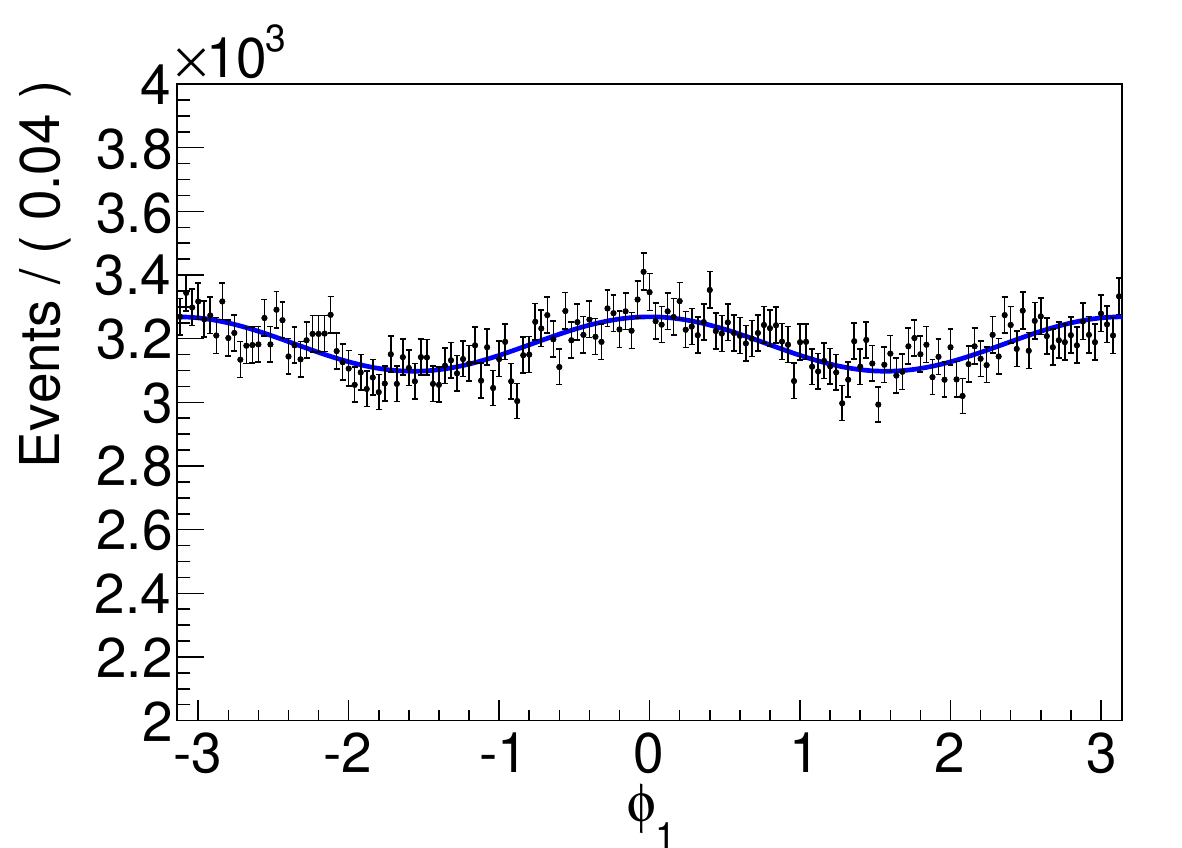}
        \label{c0_p_phi}
    }
	\caption{Fits to the angular distributions of $\cos\theta_1$, $\phi_1$ in $\psi(2S)\to \gamma \chi_{c0}$ and $\cos\theta_3$ in $\phi\to K^+K^-$ from $\chi_{c0} \to \gamma \phi$ decays. Dots with error bars represent MC events, and the blue solid curve denotes the fit.}
\end{figure*}

\section{Sensitivity Estimation}

By investigating the statistical sensitivity of specific parameters, the precise measurement is beneficial from evaluating the potential impact of uncertainties on the final results. The estimation aims to enhance the understanding of the experimental uncertainties associated with these parameters and observed signal yields. It is necessary to estimate the sensitivity in order to guide the data acquisition plan for the large-scale $e^+e^-$ experimental devices, which are operating currently like BEPCII, and to be built in the future, such as Super-Tau Charm Facility (STCF)~\cite{BESIII:2020nme,Achasov:2023gey}.Through rigorous statistical analysis and simulations, the research provides valuable insights into the sensitivity of key parameters, enabling researchers to make informed decisions and draw meaningful conclusions from the experimental data.

Assuming that the parameters will be obtained by fitting to the maximum likelihood function event by event, then the normalized joint angular distribution is defined as
\begin{eqnarray}
	\widetilde{\mathcal{W}}(\Omega_1, \Omega_2, \Omega_3,\hat{x})=\frac{\mathcal{W}(\Omega_1, \Omega_2, \Omega_3,\hat{x})}{\int\cdot\cdot\cdot\int\mathcal{W}(\cdot\cdot\cdot) \prod_{i=1}^{3} \mathrm{d}\mathrm{cos}\theta_i \prod_{j=1}^{3}\mathrm{d}\phi_j },
\end{eqnarray}
where $\hat{x}$ is a set of parameters which contains $P_T$ for $\chi_{c0}$, $x, P_T$ for $\chi_{c1}$ and $x, y, P_T$ for $\chi_{c2}$.
And the maximum likelihood function is
\begin{eqnarray}
	L=\prod_{i=1}^{N}\widetilde{\mathcal{W}}(\Omega_1,\Omega_2, \Omega_3, \hat{x}),
\end{eqnarray}
where $N$ is the number of observed events~\cite{L_function}.
The estimated statistical sensitivity $\delta_{x_i}$ for parameter $\hat{x}(x_i)$ is obtained by
\begin{eqnarray}
    \delta_{x_i} = \frac{\sqrt{V_{x_i, x_i}}}{|x_i|} \times 100\%
\end{eqnarray}
where the covariance matrix gives
\begin{eqnarray}
   V^{-1}_{x_i,x_j} &=& E[-\frac{\partial^2{lnL}}{\partial{x_i} \partial{x_j}}]\nonumber
   \\&=&
   N \int -\widetilde{\mathcal{W}} \cdot (\frac{\partial ^2 \ln \widetilde{{\mathcal{W}}}}{ \partial x_i \partial x_j})  \prod_{k=1}^{3} \mathrm{d}\mathrm{cos}\theta_k \prod_{l=1}^{3}\mathrm{d}\phi_l  .
\end{eqnarray}

\begin{figure*}[htbp]
\centering
    \subfigure[]{
	  \includegraphics[scale=0.12]{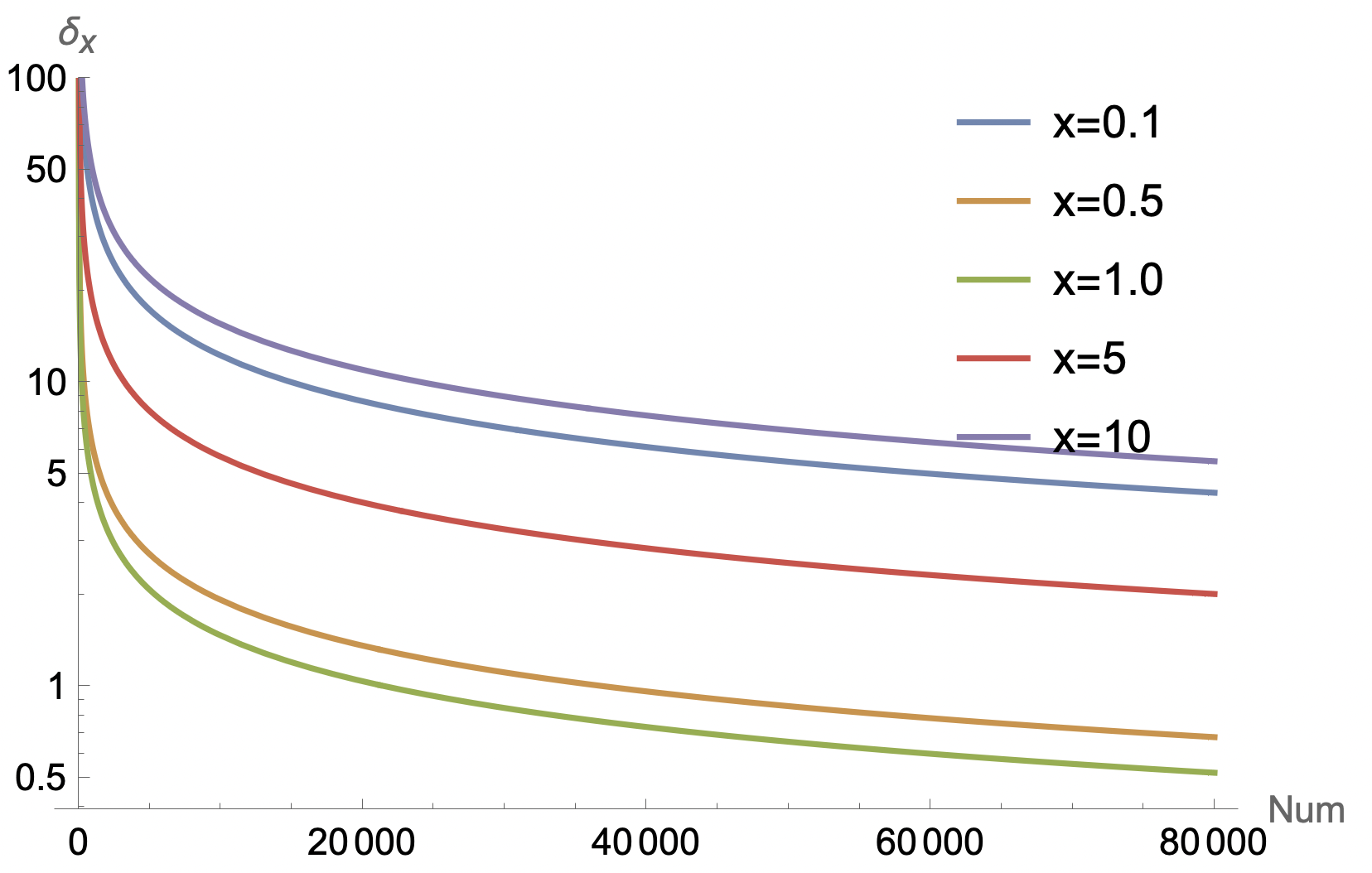}
        \label{chi_c1_x}
    }
    \subfigure[]{
        \includegraphics[scale=0.12]{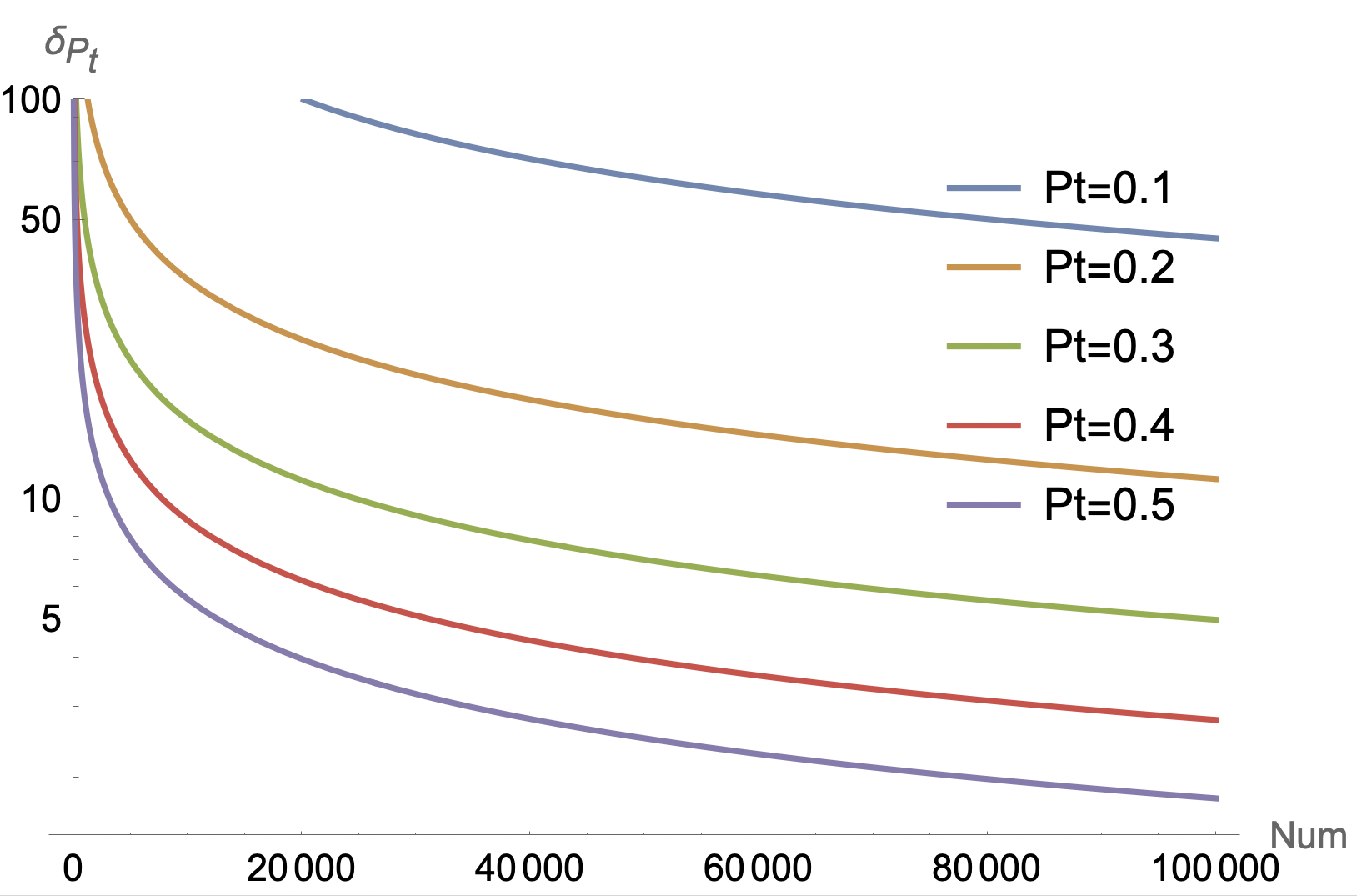}
        \label{chi_c1_Pt}
    }
	\caption{(a)~The sensitivity of $x$ (in $\chi_{c1}$ decay) for different $x$ values relative to the observed events N. (b)~The sensitivity of $P_T$ (in $\chi_{c1}$ decay) for different $P_T$ values relative to the observed events N}
\end{figure*}

\begin{figure*}[htbp]
\centering
	\subfigure[]{
		\includegraphics[scale=0.08]{c1x.png}
		\label{chi_c2_x}}
	\subfigure[]{
		\includegraphics[scale=0.08]{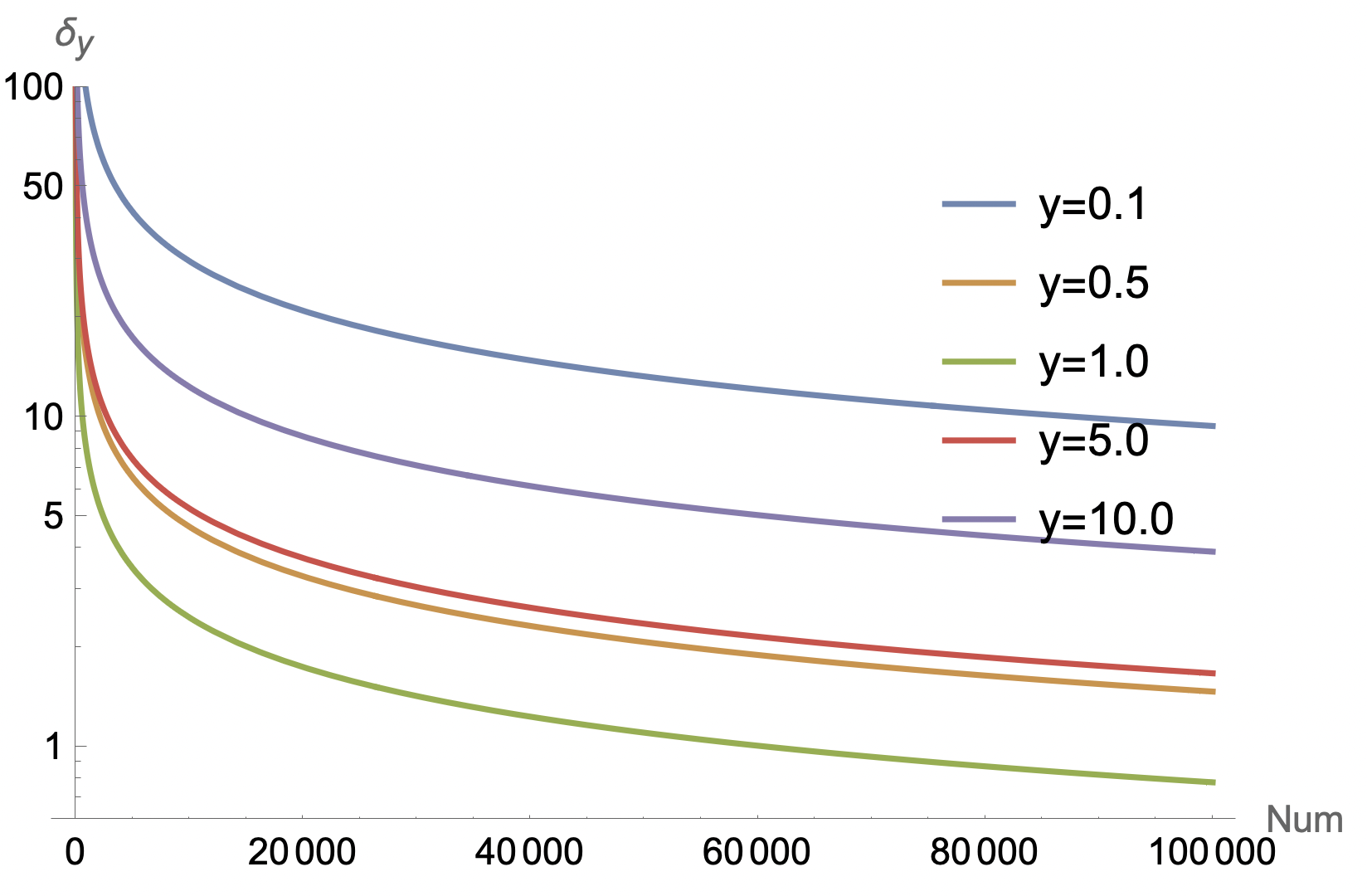}
		\label{chi_c2_y}}
    \subfigure[]{
        \includegraphics[scale=0.08]{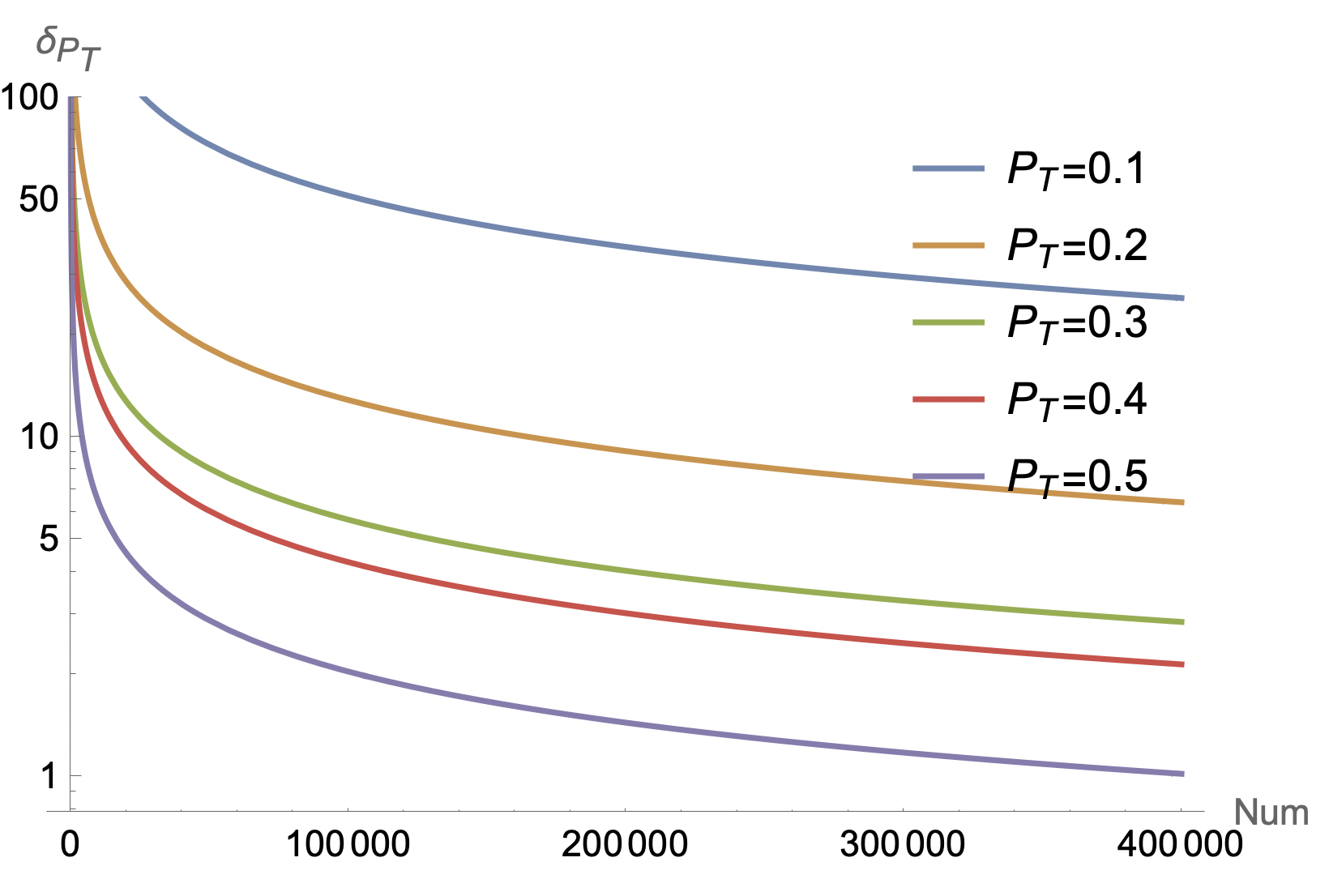}
        \label{chi_c2_Pt}
    }
	\caption{(a)~The sensitivity of $x$ (in $\chi_{c2}$ decay) for different $x$ values relative to the observed events N. (b)~The sensitivity of $y$ (in $\chi_{c2}$ decay) for different $y$ values relative to the observed events N. (c)~The sensitivity of $P_T$ (in $\chi_{c2}$ decay) for different $P_T$ values relative to the observed events N.}
\end{figure*}

By taking the phase difference $\Delta_1= \frac{\pi}{3}$ in $\chi_{c1}$ decays, the dependence of the statistical sensitivity for a set of different $x$ value is plotting in Fig.~\ref{chi_c1_x}. In $\chi_{c2}$ decays, the three phase differences $\Delta_1$ and $\Delta_2$ are set to be $\frac{\pi}{3}$ and $\frac{\pi}{4}$, respectively. The parameter $y(x)$ is set to be $1$ in plotting the dependence of the statistical sensitivity for a set of different $x(y)$ in Fig.~\ref{chi_c2_x} and Fig.~\ref{chi_c2_y}.
\begin{figure*}[htbp]
\centering
	\includegraphics[width=8cm,height=5cm]{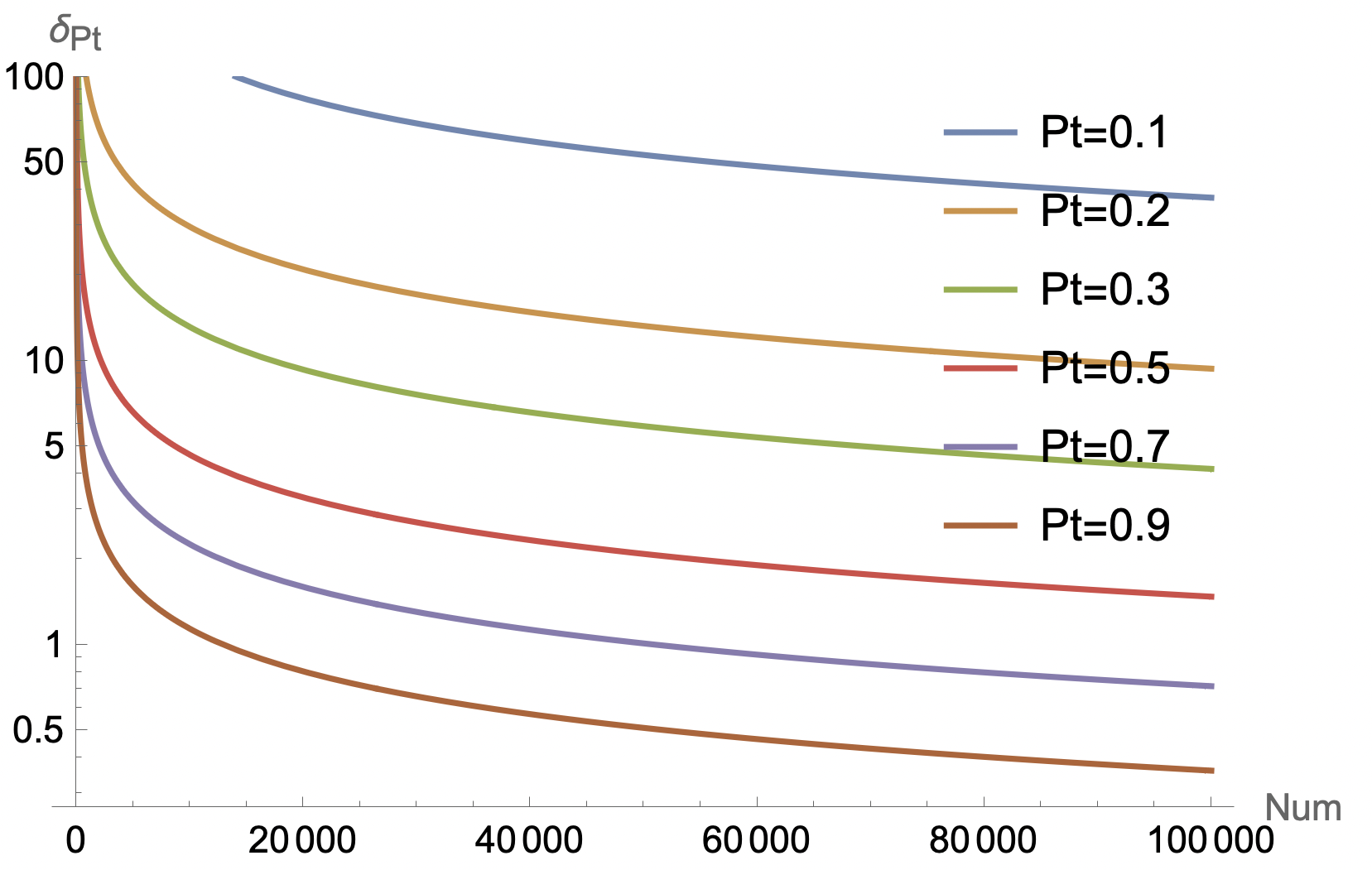}
	\caption{The sensitivity of $P_T$ (in $\chi_{c0}$ decay) for different $P_T$ values relative to the observed events N}
	\label{chi_c0_Pt}
\end{figure*}

From Fig.~\ref{chi_c0_Pt}, Fig.~\ref{chi_c1_Pt} and Fig.~\ref{chi_c2_Pt}, we can see that with the increase in polarization, the number of events required under the same statistical significance decreases.
In Fig.~\ref{chi_c1_x}, we can see that the ratio of the two independent helicity amplitude modulus $x$ in $\chi_{c1}$ decays can be measured at a statistical sensitivity to an order of $1\%$ with at least 20,000 observed signal yields, where background, detector acceptance, and other experimental effects are not taken into account, if assuming $x=1$. In comparison to the decay of $\chi_{c1}$, the decay of $\chi_{c2}$ involves two independent parameters $x$ and $y$, necessitating a higher number of observed events to achieve the same statistical sensitivity in the measurement of these two parameters.

The expected number of observed signal yields for the processes $\psi(2S) \to \gamma_1 \chi_{c1,2} \to \gamma_1 \gamma_2 \rho^0 \to \gamma_1 \gamma_2 \pi^+ \pi^-$, $\psi(2S) \to \gamma_1 \chi_{c1,2} \to \gamma_1 \gamma_2 \phi \to \gamma_1 \gamma_2 K^+ K^-$ and $\psi(2S) \to \gamma_1 \chi_{cJ} \to \gamma_1 \gamma_2 \omega \to \gamma_1 \gamma_2 \pi^+ \pi^- \pi^0 $ are calculated by using the following equation:
\begin{eqnarray}
  N_{sig} &=& N_{\Psi(2S)} \times  Br_{\psi(2S) \to \gamma_1 \chi_{cJ}} \times Br_{\chi_{cJ}\to\gamma_2 V}  \nonumber \\
  &~&\times Br_{V \to final~states}\times \epsilon ,
\end{eqnarray}
where $N_{sig}$ represents the expected number of observed signal events, $N_{\Psi(2S)}$ represents the total number of $\psi(2S)$ data samples at BESIII or
STCF, and  $\epsilon$ is the expected experimental reconstruction efficiency. Additinally, $Br_{\psi(2S) \to \gamma_1 \chi_{cJ}}$, $Br_{\chi_{cJ}\to\gamma_2 V}$ and $Br_{V \to final~states}$ denote the branching ratio of $\psi(2S) \to \gamma_1 \chi_{cJ}~(J=0,1,2)$, $\chi_{cJ}\to\gamma_2 V (V=\rho^0,~\phi,~\omega)$ and $\rho^0 \to \pi^+ \pi^-$~|~$\phi\to K^+ K^-$~|~$\omega \to \pi^+ \pi^- \pi^0$, respectively. BESIII has collected about 2.71 billion~(B) $\psi(2S)$ events in 2009, 2012 and 2021, and plan to collect 3 billion~(B) $\psi(2S)$ events~\cite{BESIII:2020nme}. The future high-luminosity $e^+e^-$ collider STCF will collect approximately 640 billion~(B) $\psi(2S)$ data samples each year~\cite{Achasov:2023gey}. In Table~\ref{calc_para}, we obtain the expected number of observed signal yields for the processes $\psi(2S) \to \gamma_1 \chi_{c0,1,2},~ \chi_{c0,1,2} \to \gamma_2 V (V=\rho^0,~\phi,~\omega)$ based on 2.71 B $\psi(2S)$ data samples at BESIII and 640 B $\psi(2S)$ data samples at STCF. With the expected signal yields, we can estimate the statistical sensitivity of the relative magnitudes of transverse to longitudinal polarization amplitude $\delta_x$ ($\delta_y, \delta_{P_T}$) for the processes $\psi(2S) \to \gamma_1 \chi_{c0,1,2} \to \gamma_1 \gamma_2 V (V=\rho^0,~\phi,~\omega)$ based on 2.71 B and 640 B $\psi(2S)$ events in Table~\ref{sensi_x_y}. The results of $\delta_{P_T}$ indicate that the degree of beam transverse polarization has a large statistical uncertainty based on current data.
The statistical sensitivity $\delta_x$ is $1.4\%-4.3\%$ for $\chi_{c1} \to \gamma V$ decays with 2.71 B $\psi(2S)$ data samples in current BESIII experiment, which reaches at least 5-fold improvement over BESIII measurement in 2011 with $(1.06\pm 0.04)\times10^8~\psi(2S)$ data samples~\cite{BESIII:2011ysp}. The processes $\chi_{c2} \to \gamma V$ is promising to be observed at BESIII, and its statistical sensitivity $\delta_x$ and $\delta_y$ are conservatively estimated to be up to the levels of $10-20\%$ based on 2.71 B $\psi(2S)$ data samples. The STCF experiment is expected to further
improve the sensitivity $\delta_x$ for $\chi_{c1}$ decays and $\delta_x$, $\delta_y$ for $\chi_{c2}$ decays with an impressive precision less than or equal to $1\%$ based the 640 B $\psi(2S)$ data sample,  presenting the improvement of 1 order of magnitude compare to BESIII experiment with 2.71 B $\psi(2S)$ data samples. The sensitivity $\delta_{P_T}$ for $\chi_{c1}$ and $\chi_{c2}$ decays ranges from $2\%$ to $20\%$ at STCF experiment.

\begin{table}[htbp]
\centering
\caption{The expected number of observed signal events and the related parameters for calculating the expected signal yields for the processes $\psi(2S) \to \gamma_1 \chi_{c0,1,2} \to \gamma_1 \gamma_2 V (V=\rho^0,~\phi,~\omega)$ at BESIII and STCF experiment.The upper limits for $\chi_{c2}\to\gamma V$ are at 90\% C.L..}
\label{calc_para}
\resizebox{\textwidth}{!}{
\begin{tabular}{cccccc}
	\hline\hline
	$N_{sig}$ & $N_{\Psi(2S)}$ & $Br_{\psi(2S) \to \gamma_1 \chi_{cJ}} (\%)$~\cite{Workman:2022ynf} & $Br_{\chi_{cJ}\to\gamma_2 V}$~($10^{-5}$)~\cite{Workman:2022ynf}& $Br_{V \to final~states}$~(\%)~\cite{Workman:2022ynf} & $\epsilon$~(\%)~\cite{BESIII:2011ysp}\\
    \hline
	537 & \multirow{9}{*}{$2.71\times 10^9$~(BESIII)} & \multirow{3}{*}{ 9.77~($\psi(2S) \to \gamma \chi_{c0})$} & \textless0.9~($\chi_{c0}\to\gamma \rho^0$) & 100~($\rho^0 \to \pi^+ \pi^-$) & 22.6\\
	253 & &  & \textless0.6~($\chi_{c0}\to\gamma \phi$) & 49.1~($\phi\to K^+ K^-$) & 32.4 \\
	337 & &  & \textless0.8~($\chi_{c0}\to\gamma \omega$) & 89.2~($\omega \to \pi^+ \pi^- \pi^0$) & 18.6 \\
	11072 &   & \multirow{3}{*}{ 9.75~($\psi(2S) \to \gamma \chi_{c1})$} & 21.6~($\chi_{c1}\to\gamma \rho^0$) & 100~($\rho^0 \to \pi^+ \pi^-$) & 19.4\\
	1080 & &  & 2.4~($\chi_{c1}\to\gamma \phi$) & 49.1~($\phi\to K^+ K^-$) & 34.6 \\
	3493 & &  & 6.8~($\chi_{c1}\to\gamma \omega$) & 89.2~($\omega \to \pi^+ \pi^- \pi^0$) & 22.7 \\
	788 &  & \multirow{3}{*}{9.36~($\psi(2S) \to \gamma \chi_{c2}$)} &  \textless1.9~($\chi_{c2}\to\gamma \rho^0$) & 100~($\rho^0 \to \pi^+ \pi^-$)& 15.7 \\
    339 &&  & \textless0.8~($\chi_{c2}\to\gamma \phi$) & 49.1~($\phi\to K^+ K^-$) & 32.6 \\
	261 & &  & \textless0.6~($\chi_{c2}\to\gamma \omega$) & 89.2~($\omega \to \pi^+ \pi^- \pi^0$) & 19.2 \\
    \hline
    $1.27\times10^5$ & \multirow{9}{*}{$6.4\times 10^{11}$~(STCF)} & \multirow{3}{*}{ 9.77~($\psi(2S) \to \gamma \chi_{c0})$} & \textless0.9~($\chi_{c0}\to\gamma \rho^0$) & 100~($\rho^0 \to \pi^+ \pi^-$) & 22.6\\
	$5.97\times10^4$ & &  & \textless0.6~($\chi_{c0}\to\gamma \phi$) & 49.1~($\phi\to K^+ K^-$) & 32.4 \\
	$7.95\times10^4$ & &  & \textless0.8~($\chi_{c0}\to\gamma \omega$) & 89.2~($\omega \to \pi^+ \pi^- \pi^0$) & 18.6 \\

    $2.61\times10^6$ &  & \multirow{3}{*}{ 9.75~($\psi(2S) \to \gamma \chi_{c1})$} & 21.6~($\chi_{c1}\to\gamma \rho^0$) & 100~($\rho^0 \to \pi^+ \pi^-$) & 19.4\\
	$2.55\times10^5$ &  &  & 2.4~($\chi_{c1}\to\gamma \phi$) & 49.1~($\phi\to K^+ K^-$) & 34.6 \\
	$8.25\times10^5$ &  &  & 6.8~($\chi_{c1}\to\gamma \omega$) & 89.2~($\omega \to \pi^+ \pi^- \pi^0$) & 22.7 \\
	$1.86\times10^5$ &  & \multirow{3}{*}{9.36~($\psi(2S) \to \gamma \chi_{c2}$) } &  \textless1.9~($\chi_{c2}\to\gamma \rho^0$) & 100~($\rho^0 \to \pi^+ \pi^-$) & 15.7 \\
        $8.01\times10^4$ & &  & \textless0.8~($\chi_{c2}\to\gamma \phi$) & 49.1~($\phi\to K^+ K^-$) & 32.6 \\
	$6.16\times10^4$ & &  & \textless0.6~($\chi_{c2}\to\gamma \omega$) & 89.2~($\omega \to \pi^+ \pi^- \pi^0$) & 19.2 \\
	\hline\hline
\end{tabular}
}
\end{table}

\begin{table*}[!htbp]
\centering
\caption{The statistical sensitivity $\delta_x$ ($\delta_y$, $\delta_{P_T}$) for the processes $\psi(2S) \to \gamma_1 \chi_{c0,1,2} \to \gamma_1 \gamma_2 V (V=\rho^0,~\phi,~\omega)$ based on 2.71 billion $\Psi(2S)$ events at BESIII and 640 billion $\Psi(2S)$ events at STCF.}
\resizebox{\textwidth}{!}{
\begin{tabular}{ c c c c c c c c c}
\hline \hline
Decay Mode                         &  Parameter & Input value    & $\delta_x$~(2.71B) & $\delta_y$~(2.71B) & $\delta_{P_T}$~(2.71B) & $\delta_x$~(640B) & $\delta_y$~(640B) & $\delta_{P_T}$~(640B) \\
\hline
{$\chi_{c0}\to\gamma \rho^0$}     &    $P_T$     &  0.2 & {$\backslash$}   & -     & {-} &  7.69\% & - & -  \\
\hline
{$\chi_{c0}\to\gamma \phi$}     &    $P_T$     &  0.2 & {$\backslash$}   & -     & {-} &  7.27\% & - & -  \\
\hline
{$\chi_{c0}\to\gamma \omega$}     &    $P_T$     &  0.2 & {$\backslash$}   & -     & {-} &  8.27\% & - & -  \\
\hline
\multirow{3}{*}{$\chi_{c1}\to\gamma \rho^0$}     &    $x$     &  1 & \multirow{3}{*}{1.4\%}    & \multirow{3}{*}{-} & \multirow{3}{*}{32.5\%} & \multirow{3}{*}{0.1\%}     & \multirow{3}{*}{-} & \multirow{3}{*}{2.1\%}  \\
     &   $P_T$          &    0.2 (fixed)  &     &   &  &  &   & \\
     &    $\Delta_1$    &   $\frac{\pi}{3}$~(fixed)   &    &      &  &  \\
\hline
\multirow{3}{*}{$\chi_{c1}\to\gamma \phi$}     &    $x$     &  1 & \multirow{3}{*}{4.3\%}   & \multirow{3}{*}{-} & \multirow{3}{*}{$\backslash$} &   \multirow{3}{*}{0.3\%}   & \multirow{3}{*}{-} & \multirow{3}{*}{6.8\%} \\
    &   $P_T$          &    0.2 (fixed)  &     &   &  &  &   & \\
     &    $\Delta_1$    &   $\frac{\pi}{3}$~(fixed)   &    &      &  &  \\
\hline
\multirow{3}{*}{$\chi_{c1}\to\gamma \omega$}     &    $x$     &  1 & \multirow{3}{*}{2.4\%}     & \multirow{3}{*}{-} & \multirow{3}{*}{59.1\%}&  \multirow{3}{*}{0.2\%}    & \multirow{3}{*}{-} & \multirow{3}{*}{3.8\%}\\
    &   $P_T$          &    0.2 (fixed)  &     &   &  &  &   & \\
     &    $\Delta_1$    &   $\frac{\pi}{3}$~(fixed)   &    &      &  &  \\
\hline
\multirow{5}{*}{$\chi_{c2}\to\gamma \rho^0$}     &  $x$   &  1      &  \multirow{5}{*}{8.6\%}   & \multirow{5}{*}{8.4\%} & \multirow{5}{*}{$\backslash$} &  \multirow{5}{*}{0.6\%} &  \multirow{5}{*}{0.5\%}   & \multirow{5}{*}{9.0\%} \\
    &  $y$  &    1     &     &      &   & \\
    &   $P_T$          &    0.2 (fixed)  &     &   &  &  &   & \\
     &   $\Delta_1$    &   $\frac{\pi}{3}$~(fixed)   &   &      &   &\\
    &   $\Delta_2$    &   $\frac{\pi}{4}$~(fixed)   &   &      &   &\\
\hline
\multirow{5}{*}{$\chi_{c2}\to\gamma \phi$}     &  $x$   &  1      &  \multirow{5}{*}{13.7\%}   &  \multirow{5}{*}{13.3\%}   & \multirow{5}{*}{$\backslash$} & \multirow{5}{*}{0.9\%} & \multirow{5}{*}{0.9\%}& \multirow{5}{*}{14.2\%}\\
    &  $y$  &   1      &     &      &   & \\
    &   $P_T$          &    0.2 (fixed)  &     &   &  &  &   & \\
     &   $\Delta_1$    &   $\frac{\pi}{3}$~(fixed)   &   &      &   &\\
    &   $\Delta_2$    &   $\frac{\pi}{4}$~(fixed)   &   &      &   &\\
\hline
\multirow{5}{*}{$\chi_{c2}\to\gamma \omega$}     &  $x$   &    1    &  \multirow{5}{*}{15.6\%}   &  \multirow{5}{*}{15.1\%}   & \multirow{5}{*}{$\backslash$} & \multirow{5}{*}{1.0\%} &\multirow{5}{*}{1.0\%}&  \multirow{5}{*}{16.2\%} \\
    &  $y$  &     1    &     &      &   & \\
    &   $P_T$          &    0.2 (fixed)  &     &   &  &  &   & \\
     &   $\Delta_1$    &   $\frac{\pi}{3}$~(fixed)   &   &      &   &\\
    &   $\Delta_2$    &   $\frac{\pi}{4}$~(fixed)   &   &      &   &\\
\hline\hline
\end{tabular}
}
\label{sensi_x_y}
\end{table*}

\section{Summary and Outlook}
To better understand the radiative decays of $\psi(2S) \to \gamma \chi_{cJ}, \chi_{cJ} \to \gamma V(\rho^0, \phi, \omega)$, we present formulae of helicity amplitude analysis and derive the joint angular distribution for these decay chains. Furthermore, we provide observables for experimentally measuring the polarization of the vector mesons in $\chi_{cJ}$ decays , perform the Monte Carlo simulation and fit the angular distributions to validate the theoretical calculations. Finally, we investigate the statistical sensitivity of the degree of transverse polarization $P_T$ of $e^+ e^-$ beams and the modulus ratio of helicity amplitudes in the decays of $\chi_{c1}$ and $\chi_{c2}$, and predict the expected number of signal events required to achieve the relevant statistical precision in experimental measurements.
Based on 3 billion $\psi(2S)$ planned data samples at BESIII experiment, the ratio of transverse to longitudinal polarization amplitude for the process $\chi_{cJ} \to \gamma V(\rho, \phi, \omega)$ can be measured in the near future~\cite{BESIII:2020nme}.
The fomulism in the work can be used for near future research in high-energy physics experiments like STCF with 640 billion $\psi(2S)$ data samples per year~\cite{Achasov:2023gey}. Analogous to the radiative decay of a charmonium state with the same spin-parity quantum number, it also provides a reference for future measurements on the super-B factory to study the polarization effect of P-wave bottonia $\chi_{bJ} \to \gamma V(\rho^0, \phi, \omega)$~\cite{Aushev:2010bq}.

$\,$

$\,$

\clearpage

\appendix

\section{The spin density matrix elements of $\chi_{c1}$ and $\chi_{c2}$}\label{ele_chic12}
\subsection{The spin density matrix elements of $\chi_{c1}$}
\begin{eqnarray}\label{ele_chic1}
    \rho^{\chi_{c1}}_{1,1} &=&-\frac{1}{2} b_{1,0}^2 \sin ^2\left(\theta _1\right) \left(P_T^2 \cos \left(2 \phi _1\right)-1\right),\nonumber\\
    \rho^{\chi_{c1}}_{1,0} &=& \frac{b_{1,0}^2 \sin \left(\theta _1\right) \left(\cos \left(\theta _1\right) \left(P_T^2 \cos \left(2 \phi _1\right)-1\right)-i P_T^2 \sin \left(2 \phi _1\right)\right)}{2 \sqrt{2}},\nonumber\\
    \rho^{\chi_{c1}}_{0,1} &=& \frac{b_{1,0}^2 \sin \left(\theta _1\right) \left(\cos \left(\theta _1\right) \left(P_T^2 \cos \left(2 \phi _1\right)-1\right)+i P_T^2 \sin \left(2 \phi _1\right)\right)}{2 \sqrt{2}},\nonumber\\
    \rho^{\chi_{c1}}_{0,0} &=& \frac{1}{4} b_{1,0}^2 \left(\cos \left(2 \theta _1\right)+2 P_T^2 \sin ^2\left(\theta _1\right) \cos \left(2 \phi _1\right)+3\right),\nonumber\\
    \rho^{\chi_{c1}}_{0,-1} &=& \frac{b_{1,0}^2 \sin \left(\theta _1\right) \left(\cos \left(\theta _1\right) \left(1-P_T^2 \cos \left(2 \phi _1\right)\right)+i P_T^2 \sin \left(2 \phi _1\right)\right)}{2 \sqrt{2}},\nonumber\\
    \rho^{\chi_{c1}}_{-1,0} &=& -\frac{b_{1,0}^2 \sin \left(\theta _1\right) \left(\cos \left(\theta _1\right) \left(P_T^2 \cos \left(2 \phi _1\right)-1\right)+i P_T^2 \sin \left(2 \phi _1\right)\right)}{2 \sqrt{2}},\nonumber\\
    \rho^{\chi_{c1}}_{-1,-1} &=& -\frac{1}{2} b_{1,0}^2 \sin ^2\left(\theta _1\right) \left(P_T^2 \cos \left(2 \phi _1\right)-1\right),\nonumber\\
    \rho^{\chi_{c1}}_{1,-1} &=& \rho^{\chi_{c1}}_{-1,1} = 0.
\end{eqnarray}

\subsection{The spin density matrix elements of $\chi_{c2}$}
\begin{eqnarray}\label{ele_chic2}
	\rho^{\chi_{c2}}_{2,2} &=& \frac{3}{4} b_{1,0}^2 \left(\cos \left(2 \theta _1\right)+2 P_T^2 \sin ^2\left(\theta _1\right) \cos \left(2 \phi _1\right)+3\right), \nonumber\\
	\rho^{\chi_{c2}}_{2,1} &=& \frac{3}{2} b_{1,0}^2 \sin \left(\theta _1\right) \left(\cos \left(\theta _1\right) \left(1-P_T^2 \cos \left(2 \phi _1\right)\right)+i P_T^2 \sin \left(2 \phi _1\right)\right),\nonumber\\
	\rho^{\chi_{c2}}_{2,0} &=& \frac{1}{4} \sqrt{\frac{3}{2}} b_{1,0}^2 \left(2 \sin ^2\left(\theta _1\right)+P_T^2 \left(\left(\cos \left(2 \theta _1\right)+3\right) \cos \left(2 \phi _1\right)-4 i \cos \left(\theta _1\right) \sin \left(2 \phi _1\right)\right)\right),\nonumber\\
	\rho^{\chi_{c2}}_{1,2} &=& \frac{1}{2} (-3) b_{1,0}^2 \sin \left(\theta _1\right) \left(\cos \left(\theta _1\right) \left(P_T^2 \cos \left(2 \phi _1\right)-1\right)+i P_T^2 \sin \left(2 \phi _1\right)\right),\nonumber\\
	\rho^{\chi_{c2}}_{1,1} &=& \frac{1}{2} (-3) b_{1,0}^2 \sin ^2\left(\theta _1\right) \left(P_T^2 \cos \left(2 \phi _1\right)-1\right), \nonumber\\
	\rho^{\chi_{c2}}_{1,0} &=& \frac{1}{2} \sqrt{\frac{3}{2}} b_{1,0}^2 \sin \left(\theta _1\right) \left(\cos \left(\theta _1\right) \left(P_T^2 \cos \left(2 \phi _1\right)-1\right)-i P_T^2 \sin \left(2 \phi _1\right)\right),\nonumber\\
	\rho^{\chi_{c2}}_{0,2} &=& \frac{1}{4} \sqrt{\frac{3}{2}} b_{1,0}^2 \left(2 \sin ^2\left(\theta _1\right)+P_T^2 \left(\left(\cos \left(2 \theta _1\right)+3\right) \cos \left(2 \phi _1\right)+4 i \cos \left(\theta _1\right) \sin \left(2 \phi _1\right)\right)\right),\nonumber\\
	\rho^{\chi_{c2}}_{0,1} &=& \frac{1}{2} \sqrt{\frac{3}{2}} b_{1,0}^2 \sin \left(\theta _1\right) \left(\cos \left(\theta _1\right) \left(P_T^2 \cos \left(2 \phi _1\right)-1\right)+i P_T^2 \sin \left(2 \phi _1\right)\right),\nonumber\\
	\rho^{\chi_{c2}}_{0,0} &=& \frac{1}{4} b_{1,0}^2 \left(\cos \left(2 \theta _1\right)+2 P_T^2 \sin ^2\left(\theta _1\right) \cos \left(2 \phi _1\right)+3\right),\nonumber\\
	\rho^{\chi_{c2}}_{0,-1} &=& \frac{1}{2} \sqrt{\frac{3}{2}} b_{1,0}^2 \sin \left(\theta _1\right) \left(\cos \left(\theta _1\right) \left(1-P_T^2 \cos \left(2 \phi _1\right)\right)+i P_T^2 \sin \left(2 \phi _1\right)\right),\nonumber\\
	\rho^{\chi_{c2}}_{0,-2} &=& \frac{1}{4} \sqrt{\frac{3}{2}} b_{1,0}^2 \left(2 \sin ^2\left(\theta _1\right)+P_T^2 \left(\left(\cos \left(2 \theta _1\right)+3\right) \cos \left(2 \phi _1\right)-4 i \cos \left(\theta _1\right) \sin \left(2 \phi _1\right)\right)\right),\nonumber\\
	\rho^{\chi_{c2}}_{-1,0} &=& -\frac{1}{2} \sqrt{\frac{3}{2}} b_{1,0}^2 \sin \left(\theta _1\right) \left(\cos \left(\theta _1\right) \left(P_T^2 \cos \left(2 \phi _1\right)-1\right)+i P_T^2 \sin \left(2 \phi _1\right)\right),\nonumber\\
	\rho^{\chi_{c2}}_{-1,-1} &=& \frac{1}{2} (-3) b_{1,0}^2 \sin ^2\left(\theta _1\right) \left(P_T^2 \cos \left(2 \phi _1\right)-1\right),\nonumber\\
	\rho^{\chi_{c2}}_{-1,-2} &=& \frac{3}{2} b_{1,0}^2 \sin \left(\theta _1\right) \left(\cos \left(\theta _1\right) \left(P_T^2 \cos \left(2 \phi _1\right)-1\right)-i P_T^2 \sin \left(2 \phi _1\right)\right),\nonumber\\
	\rho^{\chi_{c2}}_{-2,0} &=& \frac{1}{4} \sqrt{\frac{3}{2}} b_{1,0}^2 \left(2 \sin ^2\left(\theta _1\right)+P_T^2 \left(\left(\cos \left(2 \theta _1\right)+3\right) \cos \left(2 \phi _1\right)+4 i \cos \left(\theta _1\right) \sin \left(2 \phi _1\right)\right)\right),\nonumber\\
	\rho^{\chi_{c2}}_{-2,-1} &=& \frac{3}{2} b_{1,0}^2 \sin \left(\theta _1\right) \left(\cos \left(\theta _1\right) \left(P_T^2 \cos \left(2 \phi _1\right)-1\right)+i P_T^2 \sin \left(2 \phi _1\right)\right),\nonumber\\
	\rho^{\chi_{c2}}_{-2,-2} &=& \frac{3}{4} b_{1,0}^2 \left(\cos \left(2 \theta _1\right)+2 P_T^2 \sin ^2\left(\theta _1\right) \cos \left(2 \phi _1\right)+3\right),\nonumber\\
    \rho^{\chi_{c2}}_{2,-1} &=& ~\rho^{\chi_{c2}}_{2,-2} =~\rho^{\chi_{c2}}_{1,-1} =~\rho^{\chi_{c2}}_{1,-2} = ~\rho^{\chi_{c2}}_{-1,2} =~\rho^{\chi_{c2}}_{-1,1} =~\rho^{\chi_{c2}}_{-2,1} =~\rho^{\chi_{c2}}_{-2,1}=0
\end{eqnarray}

\section{$r^L_M$ expressions}\label{rLM_exp}

\subsection{$\chi_{c0}$}
The multipole parameters $r^L_M$ for $\chi_{c0}$ are expressed as
\begin{eqnarray}
	r^0_0 =a_{1,1}^2 b_{1,0}^2 \left(1+\cos ^2\left(\theta _1\right)+P_T^2 \sin ^2\left(\theta _1\right) \cos \left(2 \phi _1\right)\right),\nonumber\\
	r^0_0 r^2_0 = \frac{1}{2} a_{1,1}^2 b_{1,0}^2 \left(1+\cos ^2\left(\theta _1\right)+P_T^2 \sin ^2\left(\theta _1\right) \cos \left(2 \phi _1\right)\right).
\end{eqnarray}
The other $r^L_M$ unlisted are equal to zero.

\subsection{$\chi_{c1}$}
The multipole parameters $r^L_M$ for $\chi_{c1}$ are expressed as
\begin{eqnarray}
	r^0_0 &=& \frac{1}{12} a_{1,0}^2 b_{1,0}^2 (-3 \sin \left(2 \theta _1\right) \sin \left(2 \theta _2\right) \cos \left(\phi _2\right)
	+6 P_T^2 \left(1-2 x^2\right) \sin \left(\theta _1\right) \sin \left(2 \theta _2\right) \sin \left(2 \phi _1\right) \sin \left(\phi _2\right) \nonumber\\
	&+& 6 P_T^2 \sin ^2\left(\theta _1\right) \cos \left(2 \phi _1\right) \left(\left(2 x^2-1\right) \cos \left(2 \theta _2\right)-1\right)
	+3 \sin \left(2 \theta _1\right) \sin \left(2 \theta _2\right) \cos \left(\phi _2\right) \left(P_T^2 \left(1-2 x^2\right) \cos \left(2 \phi _1\right)+2 x^2\right) \nonumber\\
	&+&2 \left(2 x^2-1\right) \cos ^2\left(\theta _1\right) \left(3 \cos \left(2 \theta _2\right)+1\right)-2 \left(x^2+1\right) \cos \left(2 \theta _1\right)+10 x^2+10),\nonumber\\
	r^0_0 r^1_{-1} &=& \frac{1}{4} \sqrt{3} x a_{1,0}^2 b_{1,0}^2 \sin \left(\Delta _1\right) (2 \sin \left(2 \theta _2\right) \left(\cos ^2\left(\theta _1\right)+P_T^2 \sin ^2\left(\theta _1\right) \cos \left(2 \phi _1\right)\right) \nonumber\\
	&+&\cos \left(2 \theta _2\right) \left(2 P_T^2 \sin \left(\theta _1\right) \sin \left(2 \phi _1\right) \sin \left(\phi _2\right) +\sin \left(2 \theta _1\right) \cos \left(\phi _2\right) \left(P_T^2 \cos \left(2 \phi _1\right)-1\right)\right)),\nonumber\\
	r^0_0 r^1_1 &=&\frac{1}{4} \sqrt{3} x a_{1,0}^2 b_{1,0}^2 \sin \left(\Delta _1\right) \cos \left(\theta _2\right) (2 P_T^2 \sin \left(\theta _1\right) \sin \left(2 \phi _1\right) \cos \left(\phi _2\right) +\sin \left(2 \theta _1\right) \sin \left(\phi _2\right) \left(1-P_T^2 \cos \left(2 \phi _1\right)\right)),\nonumber\\
	r^0_0 r^2_{-1} &=& \frac{1}{4} \sqrt{3} x a_{1,0}^2 b_{1,0}^2 \cos \left(\Delta _1\right) \cos \left(\theta _2\right) (\sin \left(2 \theta _1\right) \sin \left(\phi _2\right) \left(P_T^2 \cos \left(2 \phi _1\right)-1\right)	-2 P_T^2 \sin \left(\theta _1\right) \sin \left(2 \phi _1\right) \cos \left(\phi _2\right)),\nonumber\\
	r^0_0 r^2_{0}& =& \frac{1}{12} a_{1,0}^2 b_{1,0}^2 (3 \sin \left(2 \theta _1\right) \sin \left(2 \theta _2\right) \cos \left(\phi _2\right)
	-6 P_T^2 \left(x^2+1\right) \sin \left(\theta _1\right) \sin \left(2 \theta _2\right) \sin \left(2 \phi _1\right) \sin \left(\phi _2\right) \nonumber\\
	&+&6 P_T^2 \sin ^2\left(\theta _1\right) \cos \left(2 \phi _1\right) \left(\left(x^2+1\right) \cos \left(2 \theta _2\right)+1\right)
	-3 \sin \left(2 \theta _1\right) \sin \left(2 \theta _2\right) \cos \left(\phi _2\right) \left(P_T^2 \left(x^2+1\right) \cos \left(2 \phi _1\right)-x^2\right) \nonumber\\
	&+&2 \left(x^2+1\right) \cos ^2\left(\theta _1\right) \left(3 \cos \left(2 \theta _2\right)+1\right)-\left(x^2-2\right) \cos \left(2 \theta _1\right)+5 x^2-10),\nonumber\\
	r^0_0 r^2_{1}& =& \frac{1}{4} \sqrt{3} x a_{1,0}^2 b_{1,0}^2 \cos \left(\Delta _1\right) (2 \sin \left(2 \theta _2\right) \left(\cos ^2\left(\theta _1\right)+P_T^2 \sin ^2\left(\theta _1\right) \cos \left(2 \phi _1\right)\right) \nonumber\\
	&+&\cos \left(2 \theta _2\right) (2 P_T^2 \sin \left(\theta _1\right) \sin \left(2 \phi _1\right) \sin \left(\phi _2\right)
	+\sin \left(2 \theta _1\right) \cos \left(\phi _2\right) \left(P_T^2 \cos \left(2 \phi _1\right)-1\right)))
	.
\end{eqnarray}

The other $r^L_M$ unlisted are equal to zero.

\subsection{$\chi_{c2}$}
The multipole parameters $r^L_M$ for $\chi_{c2}$ are expressed as

\begin{eqnarray}
	r^0_0 &=& \frac{1}{64} \{4 x^2 \cos \left(2 \phi _1\right) \sin ^2\left(\theta _1\right) P_T^2+54 y^2 \cos \left(2 \phi _1\right) \sin ^2\left(\theta _1\right) P_T^2
	+108 x^2 \cos \left(4 \theta _2\right) \cos \left(2 \phi _1\right) \sin ^2\left(\theta _1\right) P_T^2 \nonumber\\
    &+& 18 y^2 \cos \left(4 \theta _2\right) \cos \left(2 \phi _1\right) \sin ^2\left(\theta _1\right) P_T^2
	-72 \cos \left(4 \theta _2\right) \cos \left(2 \phi _1\right) \sin ^2\left(\theta _1\right) P_T^2+72 x^2 \cos \left(2 \phi _1\right) \cos \left(2 \phi _2\right) \sin ^2\left(\theta _2\right) P_T^2 \nonumber\\
    &+&108 y^2 \cos \left(2 \phi _1\right) \cos \left(2 \phi _2\right) \sin ^2\left(\theta _2\right) P_T^2-144 \cos \left(2 \phi _1\right) \cos \left(2 \phi _2\right) \sin ^2\left(\theta _2\right) P_T^2 \nonumber\\
	&+&48 x^2 \cos \left(2 \phi _1\right) \cos \left(\phi _2\right) \sin \left(2 \theta _1\right) \sin \left(2 \theta _2\right) P_T^2-72 y^2 \cos \left(2 \phi _1\right) \cos \left(\phi _2\right) \sin \left(2 \theta _1\right) \sin \left(2 \theta _2\right) P_T^2 \nonumber\\
	&+&48 \cos \left(2 \phi _1\right) \cos \left(\phi _2\right) \sin \left(2 \theta _1\right) \sin \left(2 \theta _2\right) P_T^2-72 x^2 \cos \left(2 \phi _1\right) \cos \left(\phi _2\right) \sin \left(2 \theta _1\right) \sin \left(4 \theta _2\right) P_T^2\nonumber\\
	&-&12 y^2 \cos \left(2 \phi _1\right) \cos \left(\phi _2\right) \sin \left(2 \theta _1\right) \sin \left(4 \theta _2\right) P_T^2+48 \cos \left(2 \phi _1\right) \cos \left(\phi _2\right) \sin \left(2 \theta _1\right) \sin \left(4 \theta _2\right) P_T^2\nonumber\\
	&+&96 x^2 \sin \left(\theta _1\right) \sin \left(2 \theta _2\right) \sin \left(2 \phi _1\right) \sin \left(\phi _2\right) P_T^2-144 y^2 \sin \left(\theta _1\right) \sin \left(2 \theta _2\right) \sin \left(2 \phi _1\right) \sin \left(\phi _2\right) P_T^2 \nonumber\\
	&+&96 \sin \left(\theta _1\right) \sin \left(2 \theta _2\right) \sin \left(2 \phi _1\right) \sin \left(\phi _2\right) P_T^2-144 x^2 \sin \left(\theta _1\right) \sin \left(4 \theta _2\right) \sin \left(2 \phi _1\right) \sin \left(\phi _2\right) P_T^2 \nonumber\\
	&-&24 y^2 \sin \left(\theta _1\right) \sin \left(4 \theta _2\right) \sin \left(2 \phi _1\right) \sin \left(\phi _2\right) P_T^2+96 \sin \left(\theta _1\right) \sin \left(4 \theta _2\right) \sin \left(2 \phi _1\right) \sin \left(\phi _2\right) P_T^2 \nonumber\\
	&+&48 \left(2 x^2+3 y^2-4\right) \cos \left(\theta _1\right) \sin ^2\left(\theta _2\right) \sin \left(2 \phi _1\right) \sin \left(2 \phi _2\right) P_T^2+48 x^2 \cos \left(2 \phi _2\right) \sin ^2\left(\theta _1\right) \sin ^2\left(\theta _2\right) \nonumber\\
	&+&72 y^2 \cos \left(2 \phi _2\right) \sin ^2\left(\theta _1\right) \sin ^2\left(\theta _2\right)-96 \cos \left(2 \phi _2\right) \sin ^2\left(\theta _1\right) \sin ^2\left(\theta _2\right)+18 x^2 \cos \left(4 \theta _2\right)
	+3 y^2 \cos \left(4 \theta _2\right) \nonumber\\
    &+&\cos \left(2 \theta _1\right) \{2 x^2+27 y^2+12 P_T^2 \left(2 x^2+3 y^2-4\right) \cos \left(2 \phi _1\right) \cos \left(2 \phi _2\right) \sin ^2\left(\theta _2\right)
	+9 \left(6 x^2+y^2-4\right) \cos \left(4 \theta _2\right) \nonumber\\
    &+& 12 \cos \left(2 \theta _2\right) [-2 x^2+5 y^2
	+P_T^2 \left(6 x^2+y^2-4\right) \cos \left(2 \phi _1\right) \cos \left(2 \phi _2\right) \sin ^2\left(\theta _2\right)-4]-12\}-48 x^2 \cos \left(\phi _2\right) \sin \left(2 \theta _1\right) \sin \left(2 \theta _2\right) \nonumber\\
	&+&72 y^2 \cos \left(\phi _2\right) \sin \left(2 \theta _1\right) \sin \left(2 \theta _2\right)+3 [50 x^2+67 y^2-8 P_T^2 \cos \left(2 \phi _1\right) \sin ^2\left(\theta _1\right)-4 \cos \left(4 \theta _2\right)
	-16 \cos \left(\phi _2\right) \sin \left(2 \theta _1\right) \sin \left(2 \theta _2\right)+52] \nonumber\\
    &+&72 x^2 \cos \left(\phi _2\right) \sin \left(2 \theta _1\right) \sin \left(4 \theta _2\right)
	+12 y^2 \cos \left(\phi _2\right) \sin \left(2 \theta _1\right) \sin \left(4 \theta _2\right)-48 \cos \left(\phi _2\right) \sin \left(2 \theta _1\right) \sin \left(4 \theta _2\right)\nonumber\\
	&+&12 \cos \left(2 \theta _2\right) [\cos \left(2 \phi _1\right) \left(3 \left(6 x^2+y^2-4\right) \cos \left(2 \phi _2\right) \sin ^2\left(\theta _2\right)-2 \left(2 x^2-5 y^2+4\right) \sin ^2\left(\theta _1\right)\right) P_T^2
	-6 x^2+7 y^2\nonumber\\
    &+&2 \left(6 x^2+y^2-4\right) \sin ^2\left(\theta _2\right) (2 \cos \left(\theta _1\right) \sin \left(2 \phi _1\right) \sin \left(2 \phi _2\right) P_T^2
	+\cos \left(2 \phi _2\right) \sin ^2\left(\theta _1\right))-4]\} a_{1,0}^2 b_{1,0}^2,\nonumber\\
	r^0_0 r^1_{-1} &=& \frac{3}{64} a_{1,0}^2 b_{1,0}^2 \{2 x \sin \Delta _1 \{\sin \left(2 \theta _2\right) [4 \sin ^2\left(\theta _1\right) \cos \left(2 \phi _2\right)+\cos \left(2 \theta _1\right) \left(4 P_T^2 \cos \left(2 \phi _1\right) \cos ^2\left(\phi _2\right)-2\right) \nonumber \\
    &+&2 P_T^2 \cos \left(2 \phi _1\right) \left(3 \cos \left(2 \phi _2\right)-1\right)-6]
	+18 P_T^2 \sin \left(4 \theta _2\right) \sin ^2\left(\phi _2\right) \cos \left(2 \phi _1\right)-3 \sin \left(4 \theta _2\right) [2 \sin ^2\left(\theta _1\right) \cos \left(2 \phi _2\right)+\cos \left(2 \theta _1\right) \nonumber\\
    &\times &\left(P_T^2 \cos \left(2 \phi _1\right) \left(\cos \left(2 \phi _2\right)+3\right)-3\right)-1]-8 P_T^2 \sin \left(\theta _1\right) [\cos \left(2 \theta _2\right)
	-3 \cos \left(4 \theta _2\right)] \sin \left(2 \phi _1\right) \sin \left(\phi _2\right)\}\nonumber\\
    &+& 8 \cos \left(\theta _1\right) \{x \sin \Delta _1 [P_T^2 (2 \sin \left(2 \theta _2\right)-3 \sin \left(4 \theta _2\right)) \sin \left(2 \phi _1\right) \sin \left(2 \phi _2\right)\nonumber\\
    &-&2 \sin \left(\theta _1\right) (\cos \left(2 \theta _2\right)
	-3 \cos \left(4 \theta _2\right)) \cos \left(\phi _2\right) \left(P_T^2 \cos \left(2 \phi _1\right)-1\right)]\nonumber\\
    &+&\sqrt{6} y \sin \left(\Delta _1-\Delta _2\right)  [\sin \left(\theta _1\right) \left(3 \cos \left(2 \theta _2\right)+\cos \left(4 \theta _2\right)\right) \cos \left(\phi _2\right) \left(P_T^2 \cos \left(2 \phi _1\right)-1\right)\nonumber\\
	&-&4 P_T^2 \sin \left(\theta _2\right) \cos ^3\left(\theta _2\right) \sin \left(2 \phi _1\right) \sin \left(2 \phi _2\right)]\} \nonumber\\
    &+&\sqrt{6} y \sin \left(\Delta _1-\Delta _2\right) \{6 P_T^2 \sin \left(4 \theta _2\right) \sin ^2\left(\phi _2\right) \cos \left(2 \phi _1\right)-2 \sin \left(2 \theta _2\right) [2 \sin ^2\left(\theta _1\right) \cos \left(2 \phi _2\right)\nonumber\\
	&+&\cos \left(2 \theta _1\right) \left(P_T^2 \cos \left(2 \phi _1\right) \left(\cos \left(2 \phi _2\right)+5\right)-5\right)+P_T^2 \cos \left(2 \phi _1\right) \left(3 \cos \left(2 \phi _2\right)-5\right)-7]
	-\sin \left(4 \theta _2\right) (2 \sin ^2\left(\theta _1\right) \cos \left(2 \phi _2\right)\nonumber\\
    &+&\cos \left(2 \theta _1\right) \left(P_T^2 \cos \left(2 \phi _1\right) \left(\cos \left(2 \phi _2\right)+3\right)-3\right)-1)
	+8 P_T^2 \sin \left(\theta _1\right) \left(3 \cos \left(2 \theta _2\right)+\cos \left(4 \theta _2\right)\right) \sin \left(2 \phi _1\right) \sin \left(\phi _2\right)\}\},\nonumber\\
	r^0_0 r^1_1 &=&\frac{3}{224} a_{1,0}^2 b_{1,0}^2 \{7 \sin \left(\theta _2\right) [\sin \left(2 \phi _2\right) \left(2 \sin ^2\left(\theta _1\right)+P_T^2 \left(\cos \left(2 \theta _1\right)+3\right) \cos \left(2 \phi _1\right)\right)\nonumber\\
	&-&4 P_T^2 \cos \left(\theta _1\right) \sin \left(2 \phi _1\right) \cos \left(2 \phi _2\right)] [2 x \sin \Delta _1 \left(3 \cos \left(2 \theta _2\right)+1\right)
	\nonumber\\
    &+&\sqrt{6} y \sin \left(\Delta _1-\Delta _2\right) \left(\cos \left(2 \theta _2\right)+3\right)]
	+28 \sin \left(\theta _1\right) \cos \left(\theta _2\right) [\cos \left(\theta _1\right) \sin \left(\phi _2\right) \left(1-P_T^2 \cos \left(2 \phi _1\right)\right)
	\nonumber\\
    &+&P_T^2 \sin \left(2 \phi _1\right) \cos \left(\phi _2\right)] [2 x \sin \Delta _1 \left(3 \cos \left(2 \theta _2\right)-1\right)
	+\sqrt{6} y \sin \left(\Delta _1-\Delta _2\right) \left(\cos \left(2 \theta _2\right)+3\right)]\},\nonumber\\
	r^0_0 r^2_{-2} &=& -\frac{3 x y }{8 \sqrt{2}}a_{1,0}^2 b_{1,0}^2 \cos \left(\Delta _2\right) \{-\cos \left(\theta _2\right) \left(3 \cos \left(2 \theta _2\right)+1\right) \sin \left(2 \phi _2\right) [2 \sin ^2\left(\theta _1\right)
	+P_T^2 \left(\cos \left(2 \theta _1\right)+3\right) \cos \left(2 \phi _1\right)] \nonumber\\
    &+&2 P_T^2 \sin \left(2 \phi _1\right) [\cos \left(\theta _1\right) \left(5 \cos \left(\theta _2\right)+3 \cos \left(3 \theta _2\right)\right) \cos \left(2 \phi _2\right)
	+\sin \left(\theta _1\right) \left(7 \sin \left(\theta _2\right)+3 \sin \left(3 \theta _2\right)\right) \cos \left(\phi _2\right)] \nonumber\\
	&-&2 \sin \left(2 \theta _1\right) \sin \left(\theta _2\right) \left(3 \cos \left(2 \theta _2\right)+5\right) \sin \left(\phi _2\right)\left(P_T^2 \cos \left(2 \phi _1\right)-1\right)\}\nonumber\\
	r^0_0 r^2_{-1} &=&-\frac{3}{224}  a_{1,0}^2 b_{1,0}^2 \{2 x \cos \Delta _1 \{14 \cos \left(\theta _2\right) \left(3 \cos \left(2 \theta _2\right)-1\right)\nonumber (2 P_T^2 \sin \left(\theta _1\right) \sin \left(2 \phi _1\right) \cos \left(\phi _2\right) \nonumber\\
    &+&\sin \left(2 \theta _1\right) \sin \left(\phi _2\right) \left(1-P_T^2 \cos \left(2 \phi _1\right)\right))
	-7 \sin \left(\theta _2\right) \left(3 \cos \left(2 \theta _2\right)+1\right) [4 P_T^2 \cos \left(\theta _1\right) \sin \left(2 \phi _1\right) \cos \left(2 \phi _2\right) \nonumber\\
    &-&\sin \left(2 \phi _2\right) \left(2 \sin ^2\left(\theta _1\right)+P_T^2 \left(\cos \left(2 \theta _1\right)+3\right) \cos \left(2 \phi _1\right)\right)]\} \nonumber\\
	&+&7 \sqrt{6} y \cos \left(\Delta _1-\Delta _2\right) \{\sin \left(\theta _2\right) \left(\cos \left(2 \theta _2\right)+3\right) \sin \left(2 \phi _2\right) \left(2 \sin ^2\left(\theta _1\right)+P_T^2 \left(\cos \left(2 \theta _1\right)+3\right) \cos \left(2 \phi _1\right)\right)\nonumber\\
	&-&2 P_T^2 \left(5 \sin \left(\theta _2\right)+\sin \left(3 \theta _2\right)\right) \cos \left(\theta _1\right) \sin \left(2 \phi _1\right) \cos \left(2 \phi _2\right)
	+\sin \left(2 \theta _1\right) \cos \left(3 \theta _2\right) \sin \left(\phi _2\right) \left(1-P_T^2 \cos \left(2 \phi _1\right)\right) \nonumber\\
    &+&\cos \left(\theta _2\right) \left(4 P_T^2 \sin \left(\theta _1\right) \left(\cos \left(2 \theta _2\right)+3\right) \sin \left(2 \phi _1\right) \cos \left(\phi _2\right)-7 \sin \left(2 \theta _1\right) \sin \left(\phi _2\right) \left(P_T^2 \cos \left(2 \phi _1\right)-1\right)\right)\}\},\nonumber\\
 \end{eqnarray}
 \begin{eqnarray}
	r^0_0 r^2_0 &=& \frac{1}{128} \{4 x^2 \cos \left(2 \phi _1\right) \sin ^2\left(\theta _1\right) P_T^2+54 y^2 \cos \left(2 \phi _1\right) \sin ^2\left(\theta _1\right) P_T^2
	+ 108 x^2 \cos \left(4 \theta _2\right) \cos \left(2 \phi _1\right) \sin ^2\left(\theta _1\right) P_T^2 \nonumber\\
    &+&18 y^2 \cos \left(4 \theta _2\right) \cos \left(2 \phi _1\right) \sin ^2\left(\theta _1\right) P_T^2 +144 \cos \left(4 \theta _2\right) \cos \left(2 \phi _1\right) \sin ^2\left(\theta _1\right) P_T^2+48 \cos \left(2 \phi _1\right) \sin ^2\left(\theta _1\right) P_T^2 \nonumber\\
	&+&72 x^2 \cos \left(2 \phi _1\right) \cos \left(2 \phi _2\right) \sin ^2\left(\theta _2\right) P_T^2+108 y^2 \cos \left(2 \phi _1\right) \cos \left(2 \phi _2\right) \sin ^2\left(\theta _2\right) P_T^2
	+288 \cos \left(2 \phi _1\right) \cos \left(2 \phi _2\right) \sin ^2\left(\theta _2\right) P_T^2 \nonumber\\
    &+&48 x^2 \cos \left(2 \phi _1\right) \cos \left(\phi _2\right) \sin \left(2 \theta _1\right) \sin \left(2 \theta _2\right) P_T^2
	-72 y^2 \cos \left(2 \phi _1\right) \cos \left(\phi _2\right) \sin \left(2 \theta _1\right) \sin \left(2 \theta _2\right) P_T^2\nonumber\\
    &-&96 \cos \left(2 \phi _1\right) \cos \left(\phi _2\right) \sin \left(2 \theta _1\right) \sin \left(2 \theta _2\right) P_T^2
	-72 x^2 \cos \left(2 \phi _1\right) \cos \left(\phi _2\right) \sin \left(2 \theta _1\right) \sin \left(4 \theta _2\right) P_T^2\nonumber\\
    &-&12 y^2 \cos \left(2 \phi _1\right) \cos \left(\phi _2\right) \sin \left(2 \theta _1\right) \sin \left(4 \theta _2\right) P_T^2
	-96 \cos \left(2 \phi _1\right) \cos \left(\phi _2\right) \sin \left(2 \theta _1\right) \sin \left(4 \theta _2\right) P_T^2 \nonumber\\
    &+&96 x^2 \sin \left(\theta _1\right) \sin \left(2 \theta _2\right) \sin \left(2 \phi _1\right) \sin \left(\phi _2\right) P_T^2
	-144 y^2 \sin \left(\theta _1\right) \sin \left(2 \theta _2\right) \sin \left(2 \phi _1\right) \sin \left(\phi _2\right) P_T^2\nonumber\\
    &-&192 \sin \left(\theta _1\right) \sin \left(2 \theta _2\right) \sin \left(2 \phi _1\right) \sin \left(\phi _2\right) P_T^2
	-144 x^2 \sin \left(\theta _1\right) \sin \left(4 \theta _2\right) \sin \left(2 \phi _1\right) \sin \left(\phi _2\right) P_T^2\nonumber\\
    &-&24 y^2 \sin \left(\theta _1\right) \sin \left(4 \theta _2\right) \sin \left(2 \phi _1\right) \sin \left(\phi _2\right) P_T^2
	-192 \sin \left(\theta _1\right) \sin \left(4 \theta _2\right) \sin \left(2 \phi _1\right) \sin \left(\phi _2\right) P_T^2
	\nonumber\\
    &+&48 \left(2 x^2+3 y^2+8\right) \cos \left(\theta _1\right) \sin ^2\left(\theta _2\right) \sin \left(2 \phi _1\right) \sin \left(2 \phi _2\right) P_T^2+150 x^2+201 y^2
	+48 x^2 \cos \left(2 \phi _2\right) \sin ^2\left(\theta _1\right) \sin ^2\left(\theta _2\right)\nonumber\\
    &+&72 y^2 \cos \left(2 \phi _2\right) \sin ^2\left(\theta _1\right) \sin ^2\left(\theta _2\right)
	+192 \cos \left(2 \phi _2\right) \sin ^2\left(\theta _1\right) \sin ^2\left(\theta _2\right)+18 x^2 \cos \left(4 \theta _2\right)+3 y^2 \cos \left(4 \theta _2\right)+24 \cos \left(4 \theta _2\right)
	\nonumber\\
    &+&\cos \left(2 \theta _1\right) [2 x^2+27 y^2+12 P_T^2 \left(2 x^2+3 y^2+8\right) \cos \left(2 \phi _1\right) \cos \left(2 \phi _2\right) \sin ^2\left(\theta _2\right)
	+9 \left(6 x^2+y^2+8\right) \cos \left(4 \theta _2\right)\nonumber\\
    &+&12 \cos \left(2 \theta _2\right) [-2 x^2+5 y^2
	+P_T^2 \left(6 x^2+y^2+8\right) \cos \left(2 \phi _1\right) \cos \left(2 \phi _2\right) \sin ^2\left(\theta _2\right)+8]+24]
	-48 x^2 \cos \left(\phi _2\right) \sin \left(2 \theta _1\right) \sin \left(2 \theta _2\right)\nonumber\\
    &+&72 y^2 \cos \left(\phi _2\right) \sin \left(2 \theta _1\right) \sin \left(2 \theta _2\right)
	+96 \cos \left(\phi _2\right) \sin \left(2 \theta _1\right) \sin \left(2 \theta _2\right)+72 x^2 \cos \left(\phi _2\right) \sin \left(2 \theta _1\right) \sin \left(4 \theta _2\right)
	\nonumber\\
    &+&12 y^2 \cos \left(\phi _2\right) \sin \left(2 \theta _1\right) \sin \left(4 \theta _2\right)+96 \cos \left(\phi _2\right) \sin \left(2 \theta _1\right) \sin \left(4 \theta _2\right)
	+12 \cos \left(2 \theta _2\right) \{\cos \left(2 \phi _1\right) [2 \left(-2 x^2+5 y^2+8\right) \sin ^2\left(\theta _1\right)
	\nonumber\\
    &+&3 \left(6 x^2+y^2+8\right) \cos \left(2 \phi _2\right) \sin ^2\left(\theta _2\right)] P_T^2-6 x^2+7 y^2
	+2 \left(6 x^2+y^2+8\right) \sin ^2\left(\theta _2\right) (2 \cos \left(\theta _1\right) \sin \left(2 \phi _1\right) \sin \left(2 \phi _2\right) P_T^2\nonumber\\
    &+&\cos \left(2 \phi _2\right) \sin ^2\left(\theta _1\right))+8\}
	-312\} a_{1,0}^2 b_{1,0}^2,\nonumber\\
	r^0_0 r^2_1 &=& \frac{3}{224} \{7 \sqrt{\frac{3}{2}} y \cos \left(\Delta _1-\Delta _2\right) \{-32 P_T^2 \cos \left(\theta _1\right) \sin \left(\theta _2\right) \sin \left(2 \phi _1\right) \sin \left(2 \phi _2\right) \cos ^3\left(\theta _2\right)
	\nonumber\\
    &+&20 P_T^2 \cos \left(2 \phi _1\right) \sin ^2\left(\theta _1\right) \sin \left(2 \theta _2\right)-4 \cos \left(2 \phi _2\right) \sin ^2\left(\theta _1\right) \sin \left(2 \theta _2\right)+2 [5 \cos \left(2 \theta _1\right)+\cos \left(2 \theta _2\right)
	+7] \sin \left(2 \theta _2\right)
    \nonumber\\
    &-&6 P_T^2 \cos \left(2 \phi _1\right) \cos \left(2 \phi _2\right) \sin \left(2 \theta _2\right)-2 P_T^2 \cos \left(2 \theta _1\right) \cos \left(2 \phi _1\right) \cos \left(2 \phi _2\right) \sin \left(2 \theta _2\right)
	+6 P_T^2 \cos \left(2 \phi _1\right) \sin ^2\left(\theta _1\right) \sin \left(4 \theta _2\right)\nonumber\\
    &-&2 \cos \left(2 \phi _2\right) \sin ^2\left(\theta _1\right) \sin \left(4 \theta _2\right)+3 \cos \left(2 \theta _1\right) \sin \left(4 \theta _2\right)
	-3 P_T^2 \cos \left(2 \phi _1\right) \cos \left(2 \phi _2\right) \sin \left(4 \theta _2\right)\nonumber\\
    &-&P_T^2 \cos \left(2 \theta _1\right) \cos \left(2 \phi _1\right) \cos \left(2 \phi _2\right) \sin \left(4 \theta _2\right)
	+12 \cos \left(2 \theta _2\right) (2 \sin \left(\theta _1\right) \sin \left(2 \phi _1\right) \sin \left(\phi _2\right) P_T^2 \nonumber\\
    &+&\left(P_T^2 \cos \left(2 \phi _1\right)-1\right) \cos \left(\phi _2\right) \sin \left(2 \theta _1\right))
	+4 \cos \left(4 \theta _2\right) (2 \sin \left(\theta _1\right) \sin \left(2 \phi _1\right) \sin \left(\phi _2\right) P_T^2 \nonumber \\
    &+&\left(P_T^2 \cos \left(2 \phi _1\right)-1\right) \cos \left(\phi _2\right) \sin \left(2 \theta _1\right))\}
	-7 x \cos \Delta _1 \{4 \cos \left(2 \phi _1\right) \sin ^2\left(\theta _1\right) \sin \left(2 \theta _2\right) P_T^2
	\nonumber\\
    &-&2 \cos \left(2 \theta _1\right) \cos \left(2 \phi _1\right) \cos \left(2 \phi _2\right) \sin \left(2 \theta _2\right) P_T^2-6 \cos \left(2 \phi _1\right) \cos \left(2 \phi _2\right) \sin \left(2 \theta _2\right) P_T^2
	-18 \cos \left(2 \phi _1\right) \sin ^2\left(\theta _1\right) \sin \left(4 \theta _2\right) P_T^2\nonumber\\
    &+&3 \cos \left(2 \theta _1\right) \cos \left(2 \phi _1\right) \cos \left(2 \phi _2\right) \sin \left(4 \theta _2\right) P_T^2
	+9 \cos \left(2 \phi _1\right) \cos \left(2 \phi _2\right) \sin \left(4 \theta _2\right) P_T^2+4 \cos \left(\theta _1\right) (3 \sin \left(4 \theta _2\right)\nonumber\\
    &-&2 \sin \left(2 \theta _2\right)) \sin (2 \phi _1) \sin (2 \phi _2) P_T^2
	-4 \cos \left(2 \phi _2\right) \sin ^2\left(\theta _1\right) \sin \left(2 \theta _2\right)+2 \left(\cos \left(2 \theta _1\right)+3\right) \sin \left(2 \theta _2\right)\nonumber\\
    &+&6 \cos \left(2 \phi _2\right) \sin ^2\left(\theta _1\right) \sin \left(4 \theta _2\right)
	-9 \cos \left(2 \theta _1\right) \sin \left(4 \theta _2\right)-3 \sin \left(4 \theta _2\right)+4 \cos \left(2 \theta _2\right) [2 \sin \left(\theta _1\right) \sin \left(2 \phi _1\right) \sin \left(\phi _2\right) P_T^2
	\nonumber\\
    &+&\left(P_T^2 \cos \left(2 \phi _1\right)-1\right) \cos \left(\phi _2\right) \sin \left(2 \theta _1\right)]+12 \cos \left(4 \theta _2\right) [\left(1-P_T^2 \cos \left(2 \phi _1\right)\right) \cos \left(\phi _2\right) \sin \left(2 \theta _1\right)
	\nonumber\\
    &-&2 P_T^2 \sin \left(\theta _1\right) \sin \left(2 \phi _1\right) \sin \left(\phi _2\right)]\}\} a_{1,0}^2 b_{1,0}^2,\nonumber\\
	r^0_0 r^2_2 &=& \frac{3 x y}{32 \sqrt{2}} a_{1,0}^2 b_{1,0}^2 \cos \left(\Delta _2\right) \{2 \sin ^2\left(\theta _1\right) \left(4 \cos \left(2 \theta _2\right)+3 \cos \left(4 \theta _2\right)+9\right) \cos \left(2 \phi _2\right)
	+4 \sin ^2\left(\theta _2\right) (3 \left(\cos \left(2 \theta _2\right)+3\right) \nonumber \\
    &+&\cos \left(2 \theta _1\right) \left(9 \cos \left(2 \theta _2\right)+11\right))
	+P_T^2 \cos \left(2 \phi _1\right) [\left(\cos \left(2 \theta _1\right)+3\right) \left(4 \cos \left(2 \theta _2\right)+3 \cos \left(4 \theta _2\right)+9\right) \cos \left(2 \phi _2\right)
	\nonumber\\
    &+&4 \sin \left(2 \theta _1\right) \left(2 \sin \left(2 \theta _2\right)+3 \sin \left(4 \theta _2\right)\right) \cos \left(\phi _2\right)+8 \sin ^2\left(\theta _1\right) \sin ^2\left(\theta _2\right) \left(9 \cos \left(2 \theta _2\right)+11\right)]
	+4 P_T^2 \cos \left(\theta _1\right) (4 \cos \left(2 \theta _2\right)\nonumber\\
    &+&3 \cos \left(4 \theta _2\right)+9) \sin (2 \phi _1) \sin (2 \phi _2)
	-4 \left(2 \sin \left(2 \theta _2\right)+3 \sin \left(4 \theta _2\right)\right) \left(\sin \left(2 \theta _1\right) \cos \left(\phi _2\right)-2 P_T^2 \sin \left(\theta _1\right) \sin \left(2 \phi _1\right) \sin \left(\phi _2\right)\right)\}.
\end{eqnarray}

\newpage
\section{Monte Carlo Simulation and Fit Results of the angular distributions}\label{fit_results}
\begin{table*}[htbp]
\centering
\caption{Monte Carlo simulation and fit results of the projection of the polar angles in the processes $\psi(2S) \to \gamma \chi_{c0}$,  and $\psi(2S) \to \gamma \chi_{c1,2}$}
\label{fit_results_table}
\begin{tabular}{cccc}
	\hline\hline
	Decay mode &  Default $\alpha$ value &  Fitted  $\alpha$ value & Figure \\
	\hline
	$\psi(2S) \to \gamma \chi_{c0}$ & 1.00 & $1.00 \pm 0.01$ & \ref{c0_o_c1}\\
  $\omega \to \pi^+ \pi^- \pi^0$   & -1.00 &  $-1.00 \pm 0.01$ & \ref{c0_o_c3}\\
	\hline
	$\psi(2S) \to \gamma \chi_{c0}$ &1.00 	& $1.02 \pm 0.01$ & \ref{c0_r_c1}\\
  $\rho^0 \to \pi^+ \pi^-$   & -1.00 & $-1.00 \pm 0.01$  & \ref{c0_r_c3}\\
    \hline
   $\psi(2S) \to \gamma \chi_{c1}$ ($\chi_{c1} \to \gamma \rho$) & -0.33 & $-0.34 \pm 0.01$ & \ref{c1_r_c1}\\
      \hline
   $\psi(2S) \to \gamma \chi_{c1}$ ($\chi_{c1} \to \gamma \omega$) & -0.33 & $-0.33 \pm 0.01$ & \ref{c1_o_c1}\\
    \hline
   $\psi(2S) \to \gamma \chi_{c2}$  ($\chi_{c2} \to \gamma \rho$) & 0.08 & $0.07\pm 0.01$ & \ref{c2_r_c1} \\
    \hline
   $\psi(2S) \to \gamma \chi_{c2}$ ($\chi_{c2} \to \gamma \omega$) &0.08 &$0.07 \pm 0.01$ & \ref{c2_o_c1}\\
	\hline\hline
\end{tabular}
\end{table*}

\begin{table*}[htbp]
\centering
\caption{Monte Carlo simulation and fit results in the decays $\chi_{c1,2} \to \gamma V$}
\label{fit_results_table2}
\resizebox{\textwidth}{!}{
\begin{tabular}{cccc}
	\hline\hline
	Decay mode &  Input value &  Fit value & Figure \\
    \hline
    $\chi_{c0} \to \gamma \rho$  & $P_T=0.24$ & $P_T=0.24 \pm 0.01$&  \ref{c0_r_phi}\\
    \hline
    $\chi_{c0} \to \gamma \omega$    & $P_T=0.24$ & $P_T=0.25 \pm 0.01$  & \ref{c0_o_phi}\\
	\hline
    $\chi_{c1} \to \gamma \rho$  & $x=0.43,~P_T=0.24$ & $x=0.43 \pm 0.01,~ P_T=0.25\pm0.02$& \ref{c1_r_c2},\ref{c1_r_c3},\ref{c1_r_phi}\\
      \hline
 $\chi_{c1} \to \gamma \omega$    & $x=0.57,~P_T=0.24$ & $x=0.56 \pm 0.01, ~P_T=0.23\pm0.02$  & \ref{c1_o_c2},\ref{c1_o_c3},\ref{c1_o_phi}\\
    \hline
    $\chi_{c2} \to \gamma \rho$  & $x=1.55, y=2.06, P_T=0.24$& $x = 1.58 \pm 0.79, y = 2.09 \pm 0.77, P_T=0.29\pm0.07$& \ref{c2_r_c2},\ref{c2_r_c3},\ref{c2_r_phi}\\
    \hline
    $\chi_{c2} \to \gamma \omega$  & $x=0, y=1, P_T=0.24$ &$x=0.00 \pm 0.02, y=1.000 \pm 0.02, P_T=0.22\pm0.01$ & \ref{c2_o_c2},\ref{c2_o_c3},\ref{c2_o_phi} \\
	\hline\hline
\end{tabular}
}
\end{table*}

\begin{figure*}[htbp]
	\centering
	\subfigure[]{
		\includegraphics[scale=0.25]{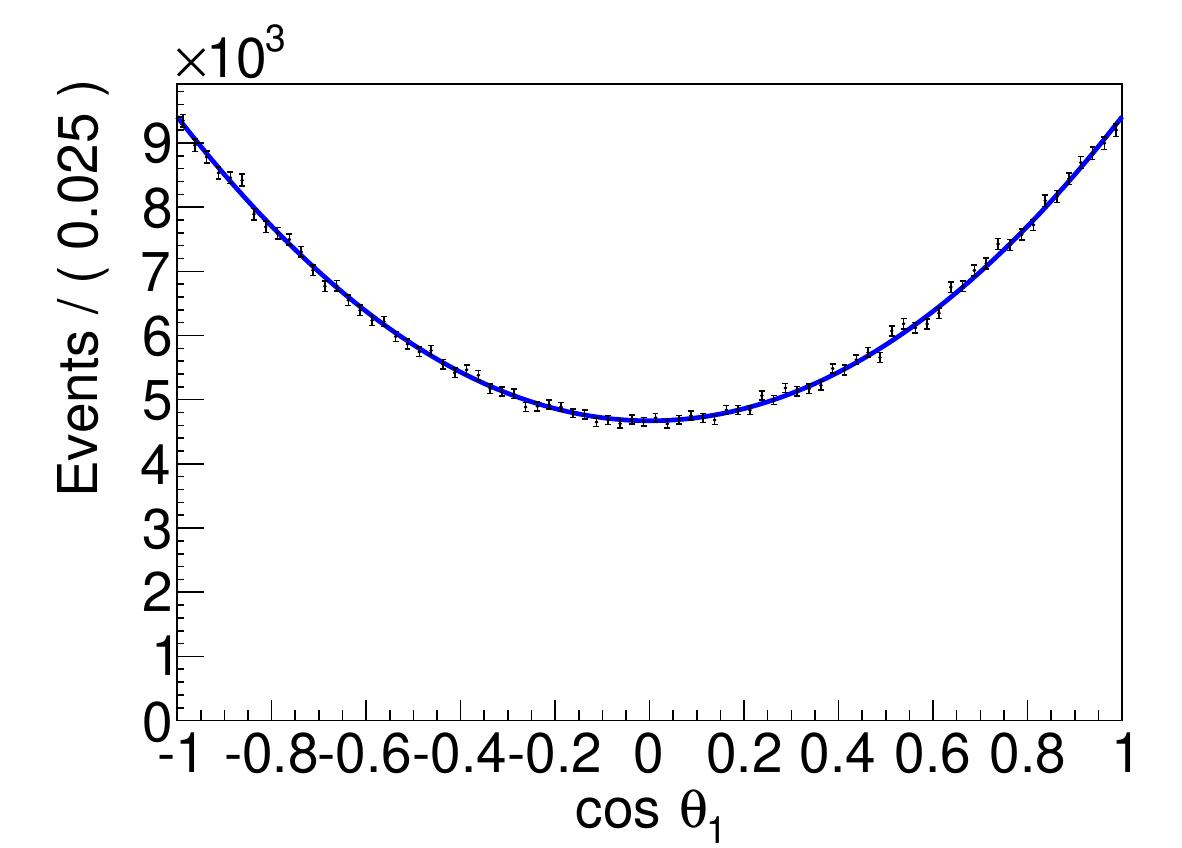}
		\label{c0_o_c1}}
	\subfigure[]{
		\includegraphics[scale=0.25]{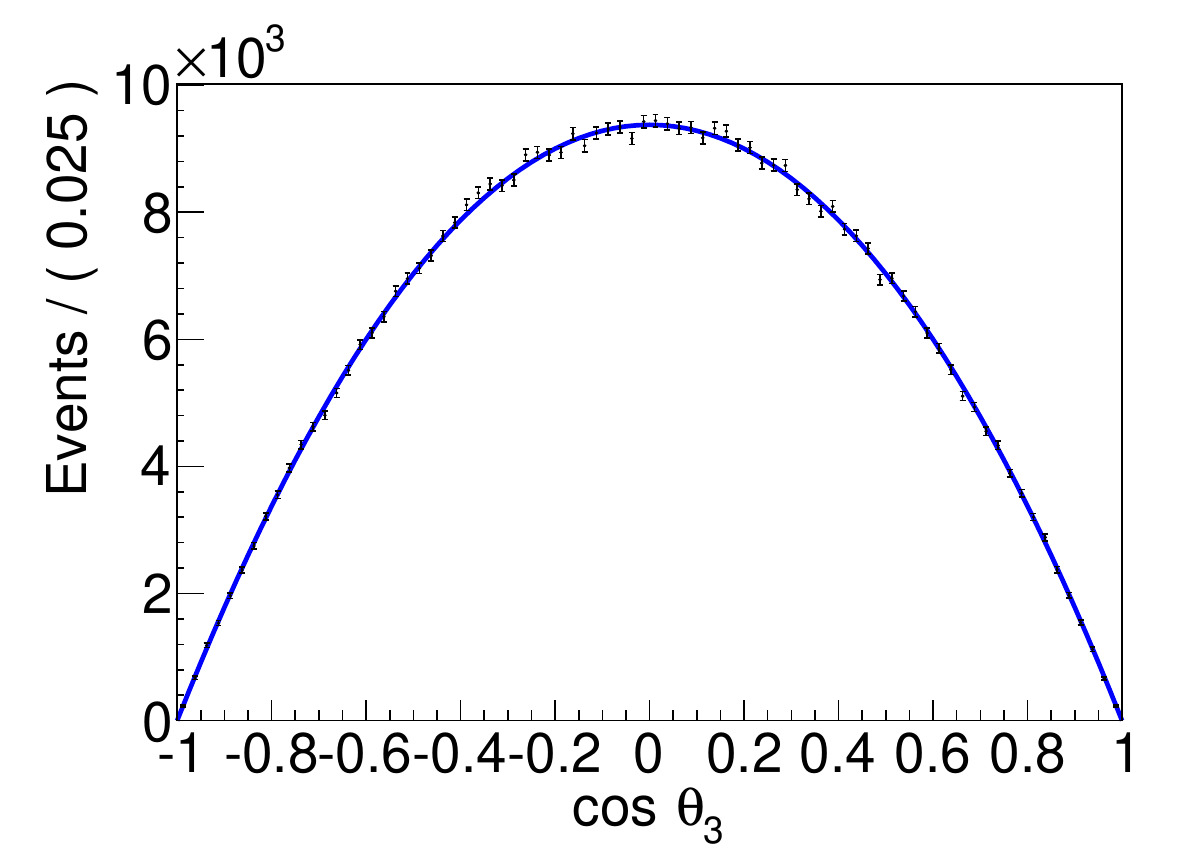}
		\label{c0_o_c3}}
    \subfigure[]{
        \includegraphics[scale=0.25]{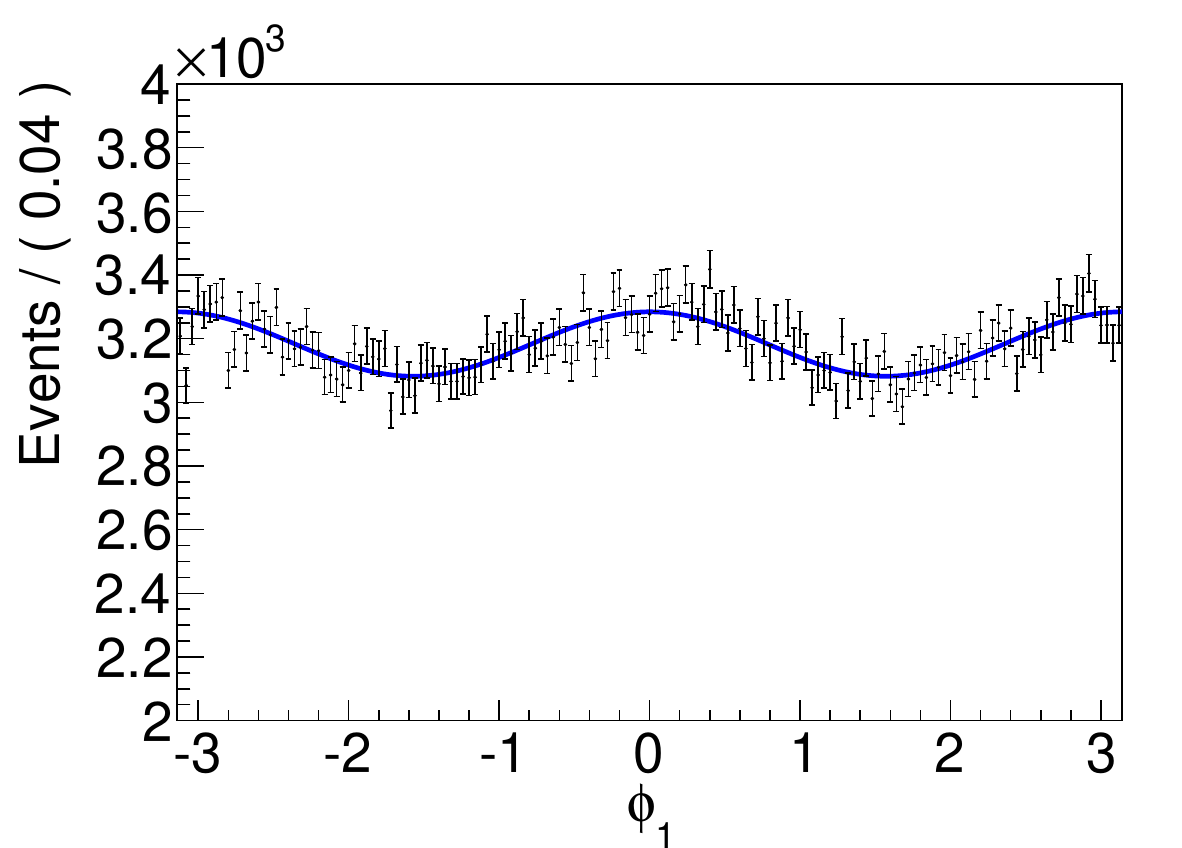}
        \label{c0_o_phi}
    }
	\caption{${\mathrm{d}N}/{\mathrm{d}\cos\theta_1}$,  ${\mathrm{d}N}/{\mathrm{d}\cos\theta_3}$ and ${\mathrm{d}N}/{\mathrm{d}\phi_1}$ distributions versus $\cos\theta_1$, $\cos\theta_3$ and $\phi_1$ in $\psi(2S)\to \gamma \chi_{c0}$ and $\omega \to \pi^+ \pi^- \pi^0$ from $\chi_{c0} \to \gamma \omega$ decays. Dots with error bars are filled with MC events, and the blue solid curve denotes the fit.}
\end{figure*}

\begin{figure*}[htbp]
	\centering
	\subfigure[]{
		\includegraphics[scale=0.25]{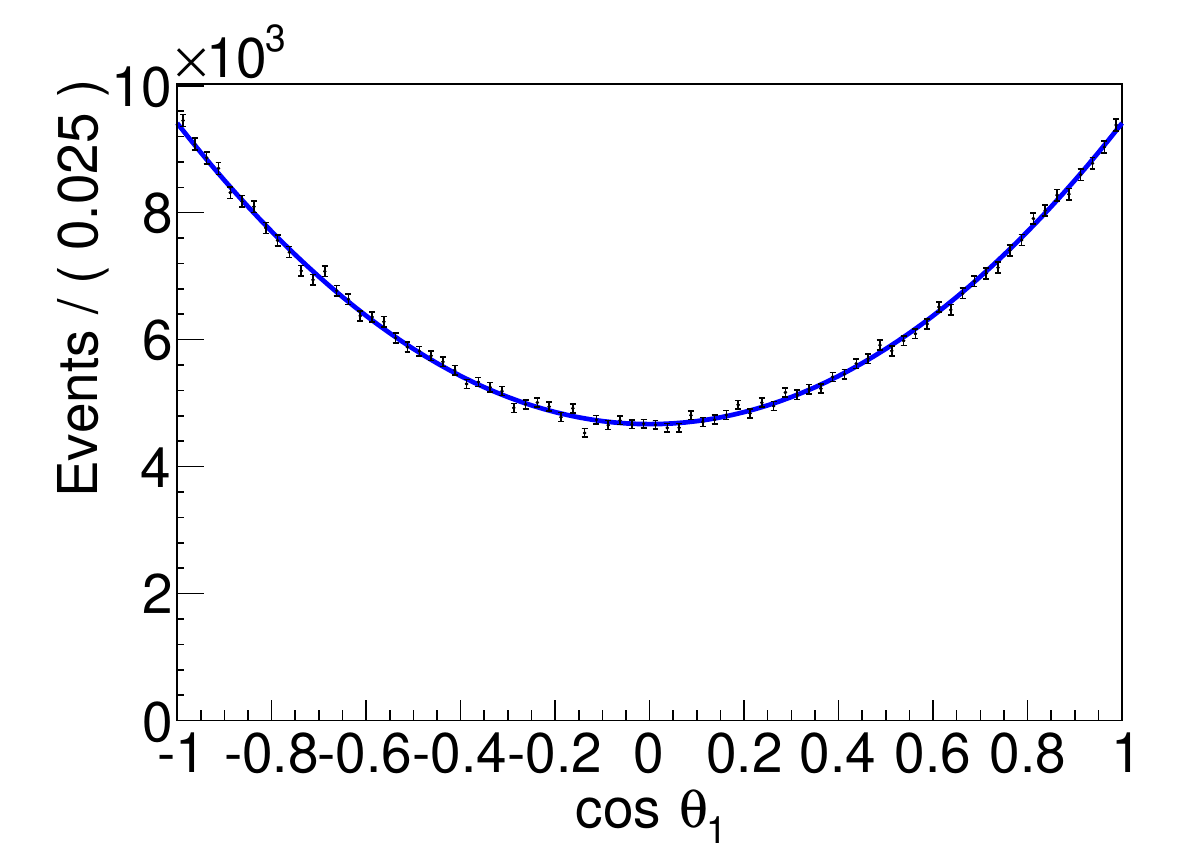}
		\label{c0_r_c1}}
	\subfigure[]{
		\includegraphics[scale=0.25]{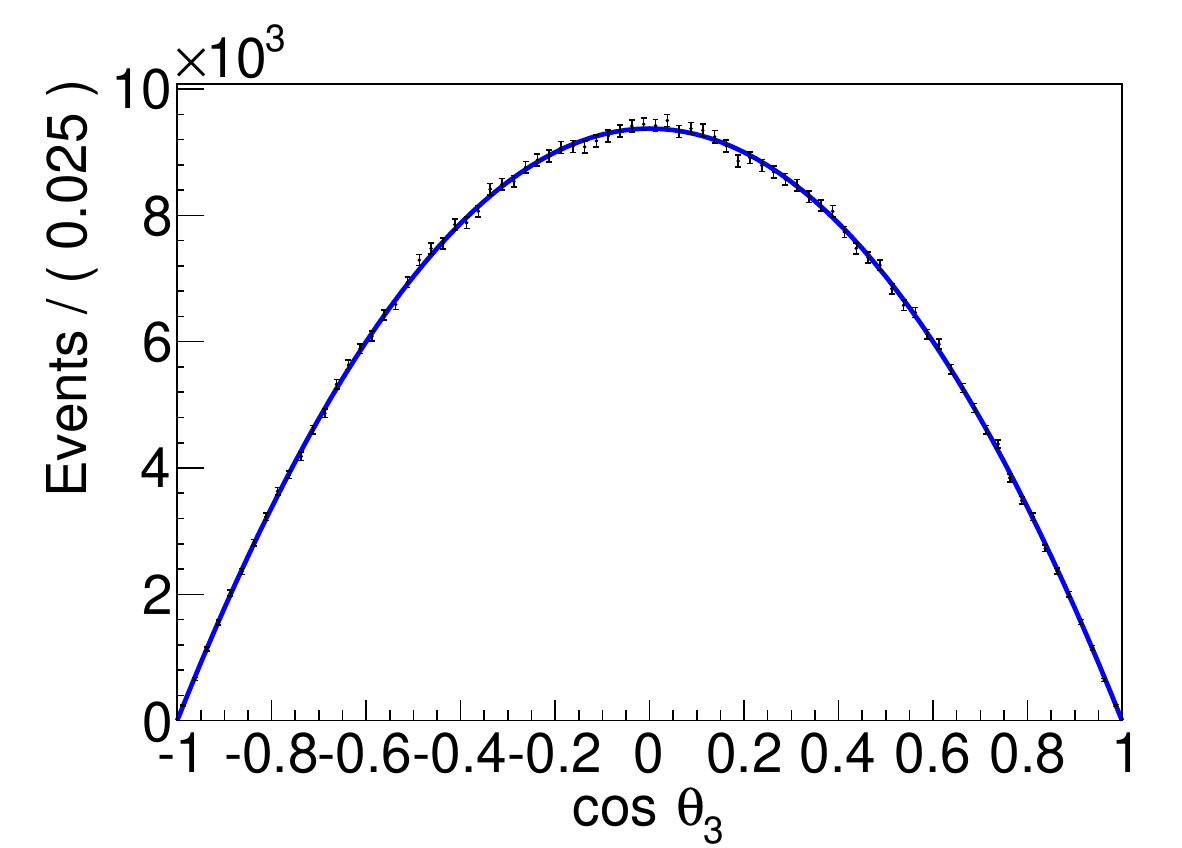}
		\label{c0_r_c3}}
    \subfigure[]{
        \includegraphics[scale=0.25]{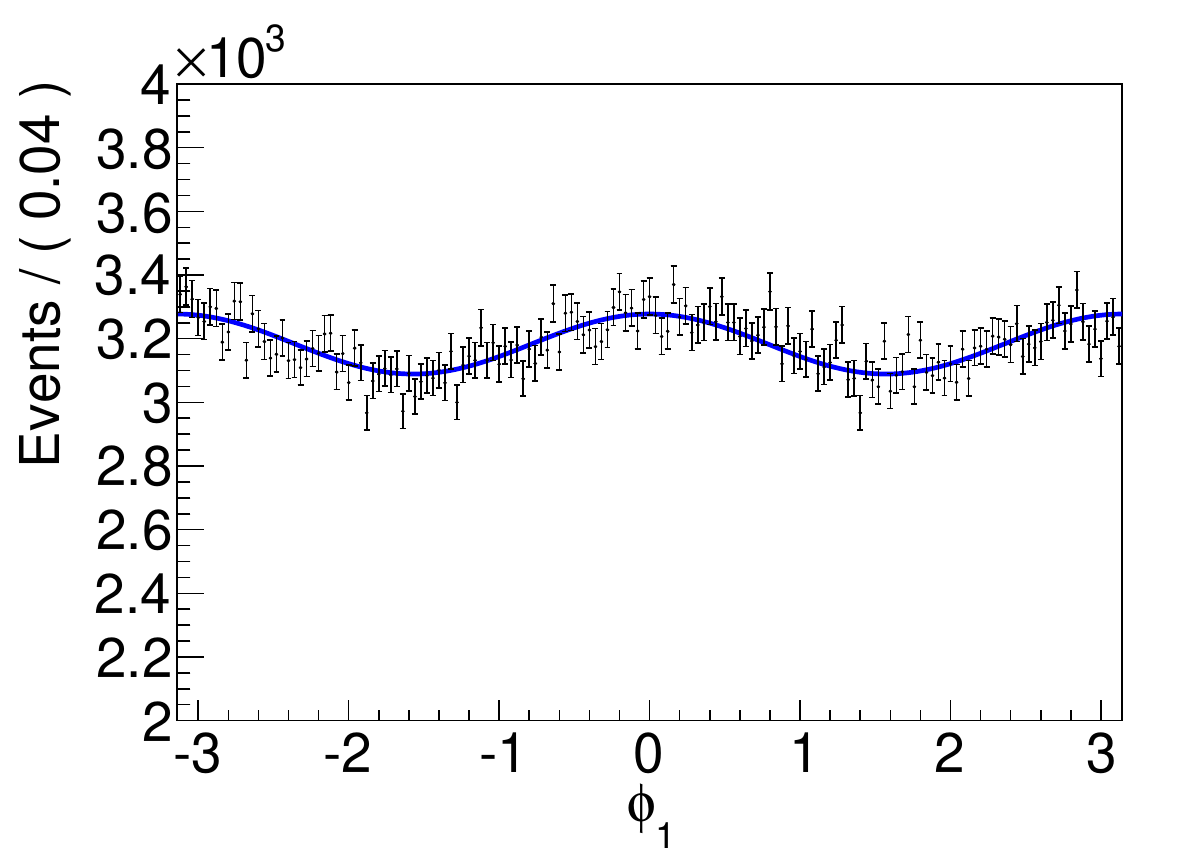}
        \label{c0_r_phi}
    }
	\caption{${\mathrm{d}N}/{\mathrm{d}\cos\theta_1}$,  ${\mathrm{d}N}/{\mathrm{d}\cos\theta_3}$ and ${\mathrm{d}N}/{\mathrm{d}\phi_1}$ distributions versus $\cos\theta_1$, $\cos\theta_3$ and $\phi_1$ in $\psi(2S)\to \gamma \chi_{c0}$ and $\rho^0 \to \pi^+ \pi^-$ from $\chi_{c0} \to \gamma \rho$ decays. Dots with error bars are filled with MC events, and the blue solid curve denotes the fit.}
\end{figure*}

\begin{figure*}[htbp]
	\centering
	\subfigure[]{
		\includegraphics[scale=0.20]{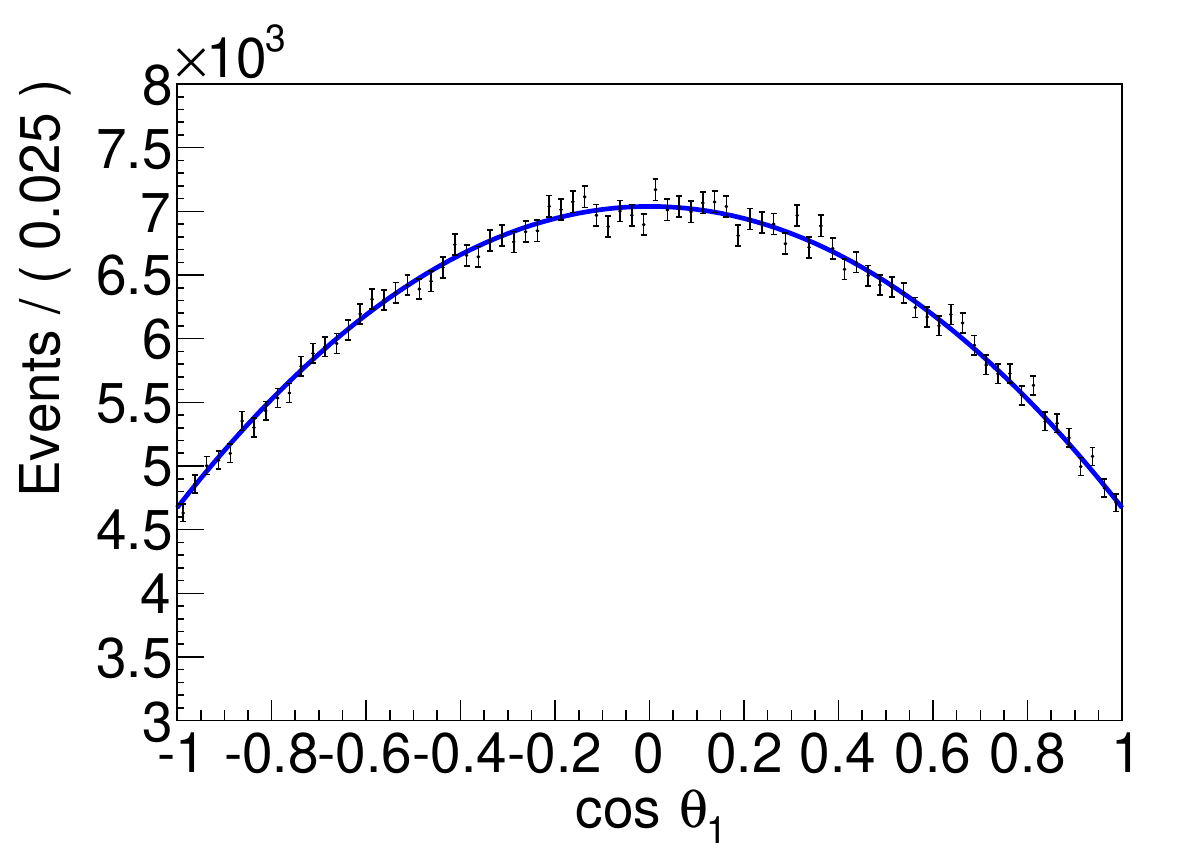}
		\label{c1_r_c1}}
	\subfigure[]{
		\includegraphics[scale=0.20]{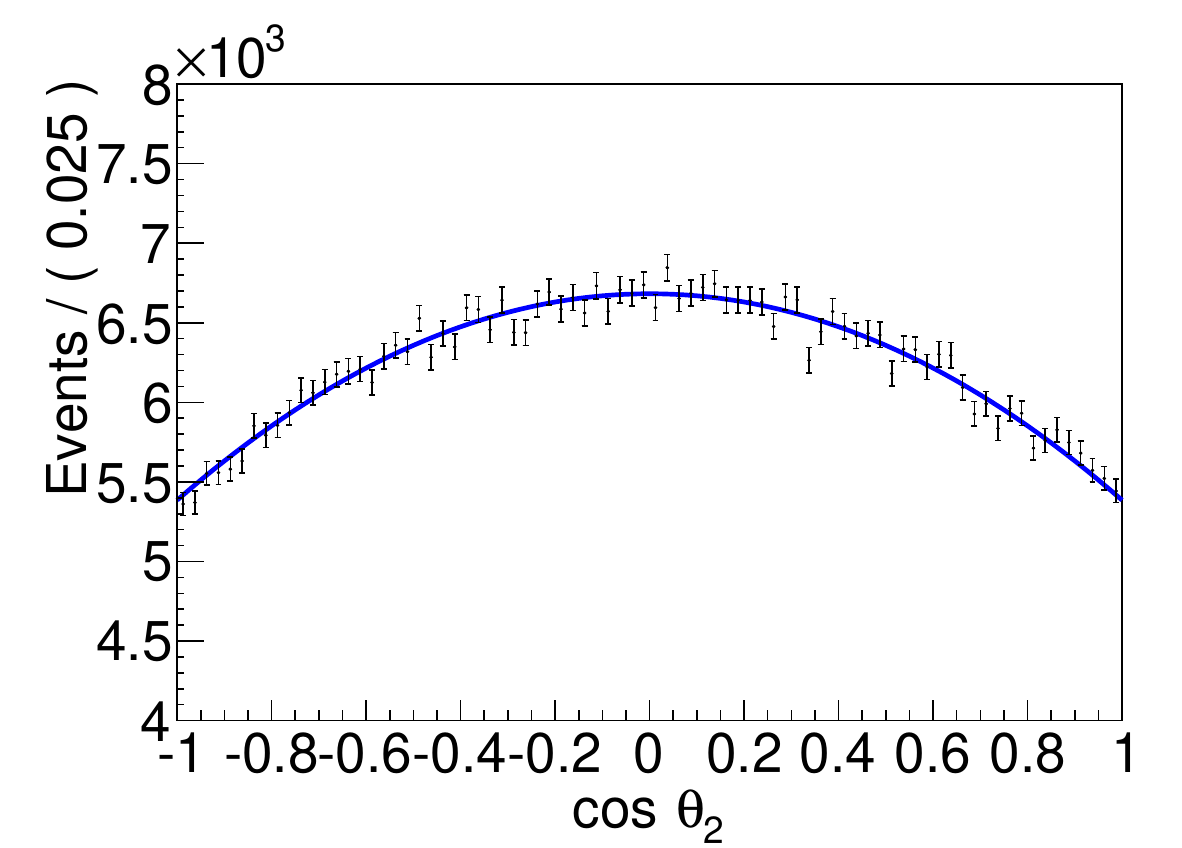}
		\label{c1_r_c2}	}
	\subfigure[]{
		\includegraphics[scale=0.20]{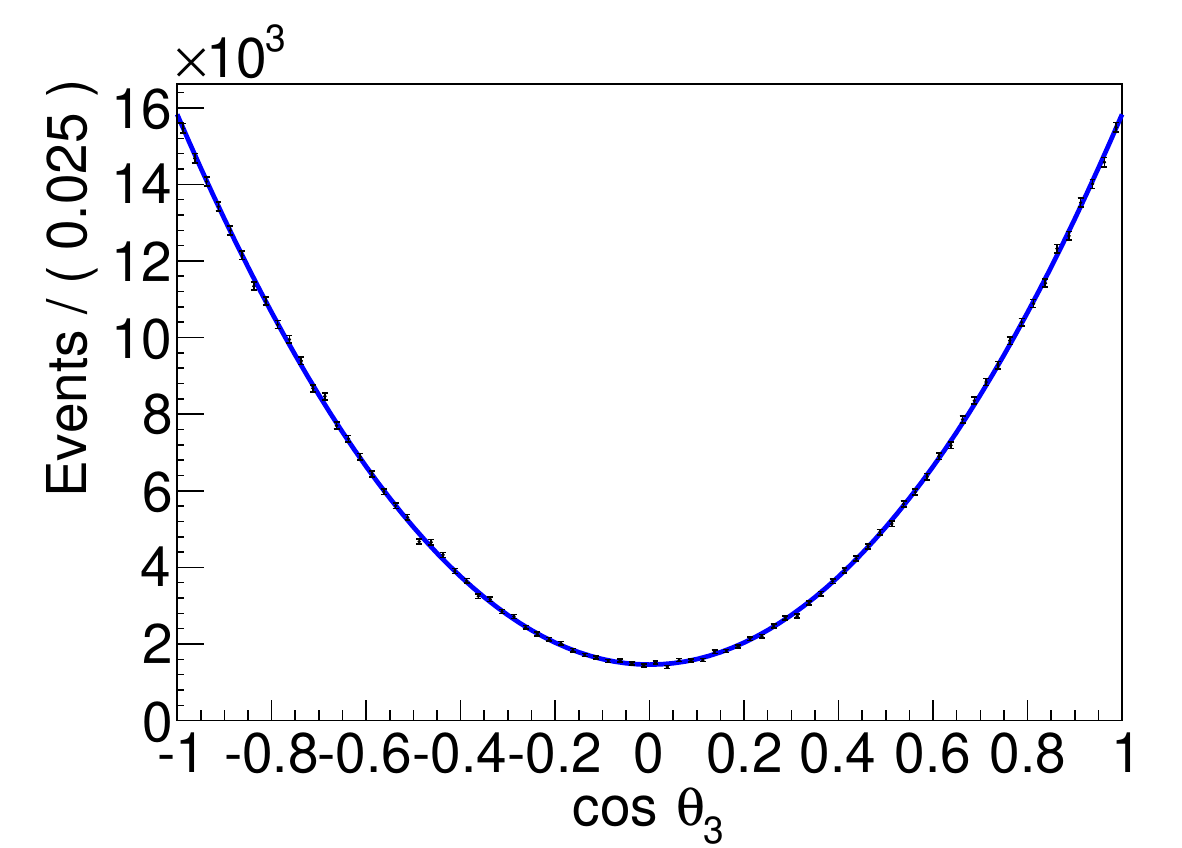}
		\label{c1_r_c3}}
    \subfigure[]{
        \includegraphics[scale=0.20]{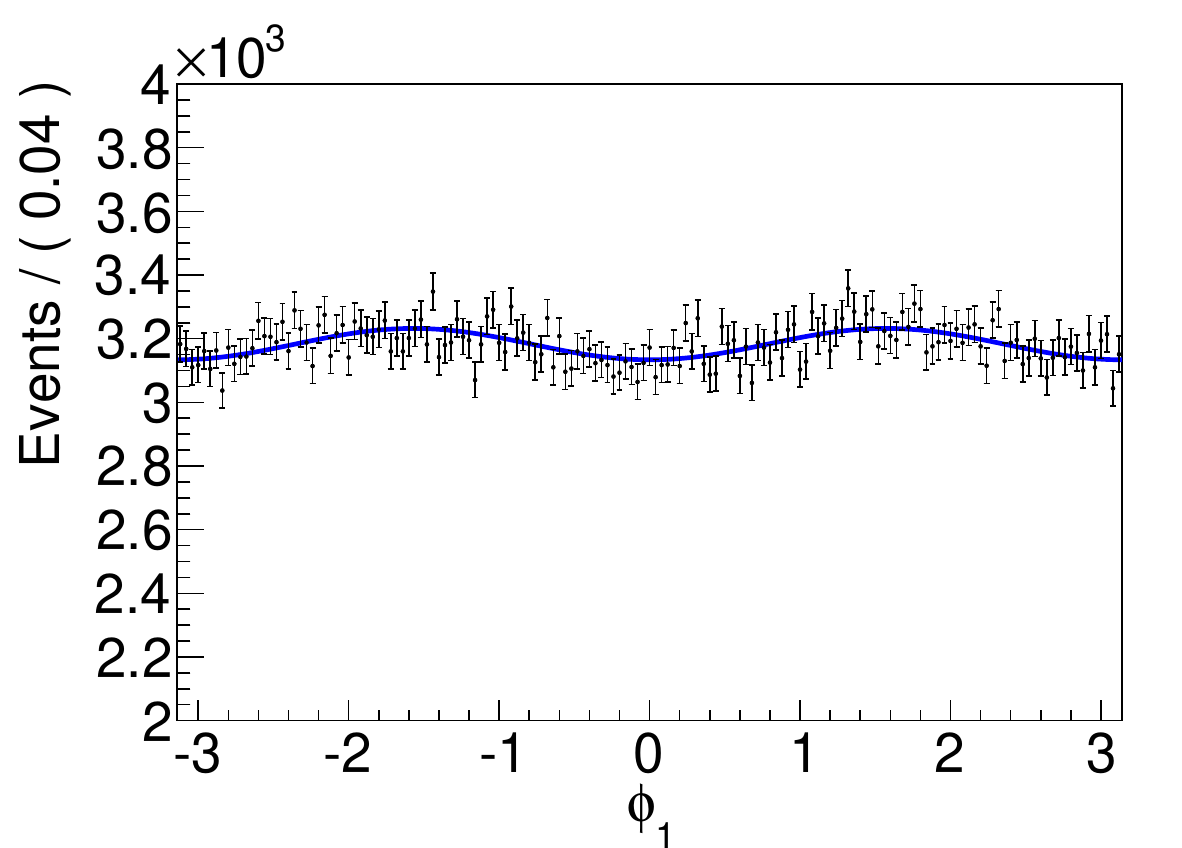}
        \label{c1_r_phi}
    }
	\caption{${\mathrm{d}N}/{\mathrm{d}\cos\theta_i}, i=1,2,3$ and ${\mathrm{d}N}/{\mathrm{d}\phi_1}$ distributions versus $\cos\theta_i$ and $\phi_1$ in $\psi(2S)\to \gamma \chi_{c1}$, $\chi_{c1} \to \gamma \rho$ and $\rho^0 \to \pi^+ \pi^-$ decays. Dots with error bars are filled with MC events, and the blue solid curve denotes the fit.}
\end{figure*}

\begin{figure*}[htbp]
	\centering
	\subfigure[]{
		\includegraphics[scale=0.20]{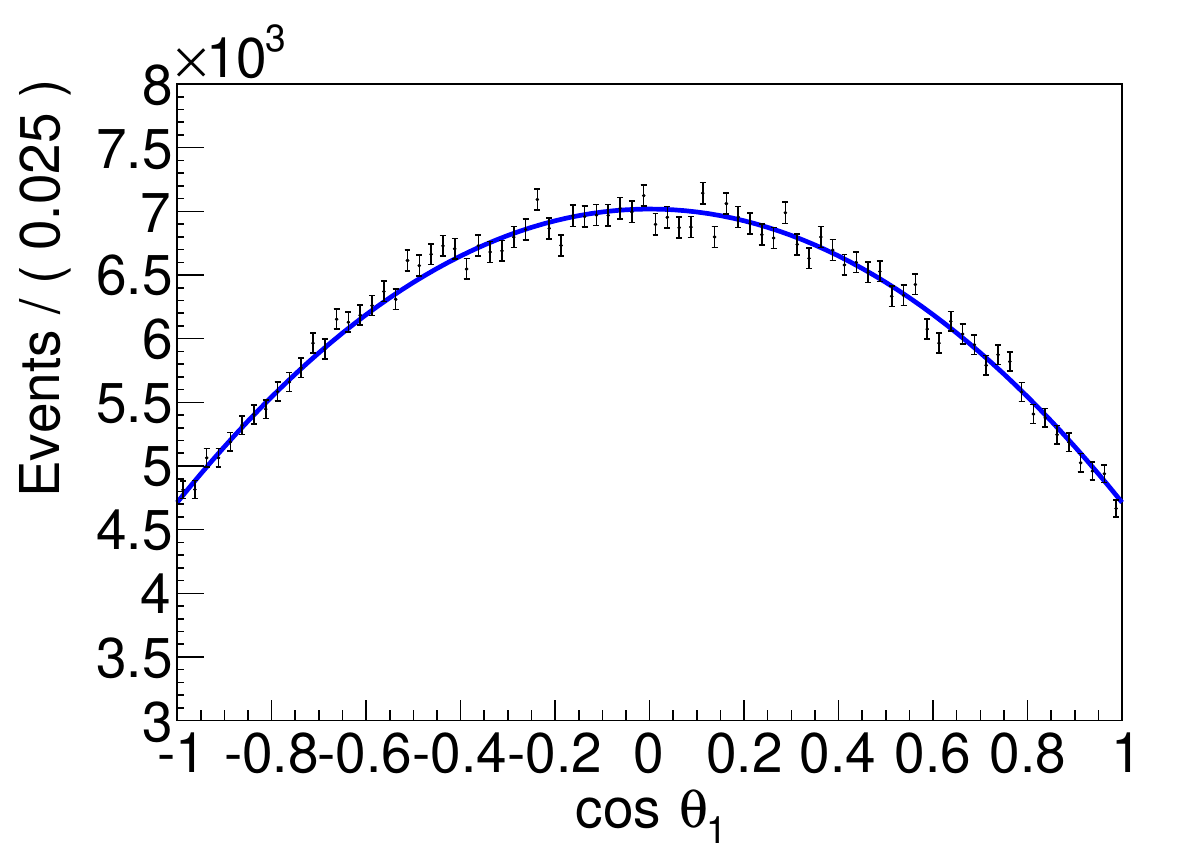}
		\label{c1_o_c1}}
	\subfigure[]{
		\includegraphics[scale=0.20]{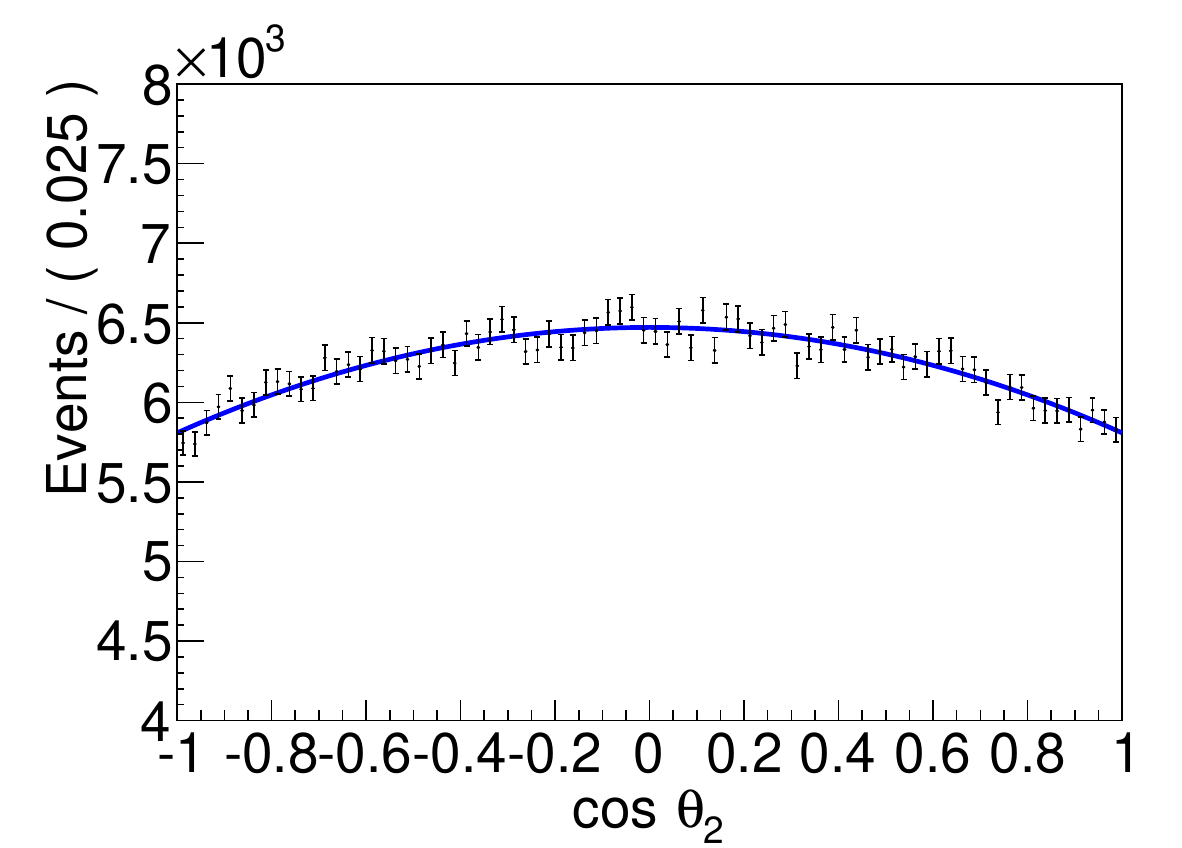}
		\label{c1_o_c2}	}
	\subfigure[]{
		\includegraphics[scale=0.20]{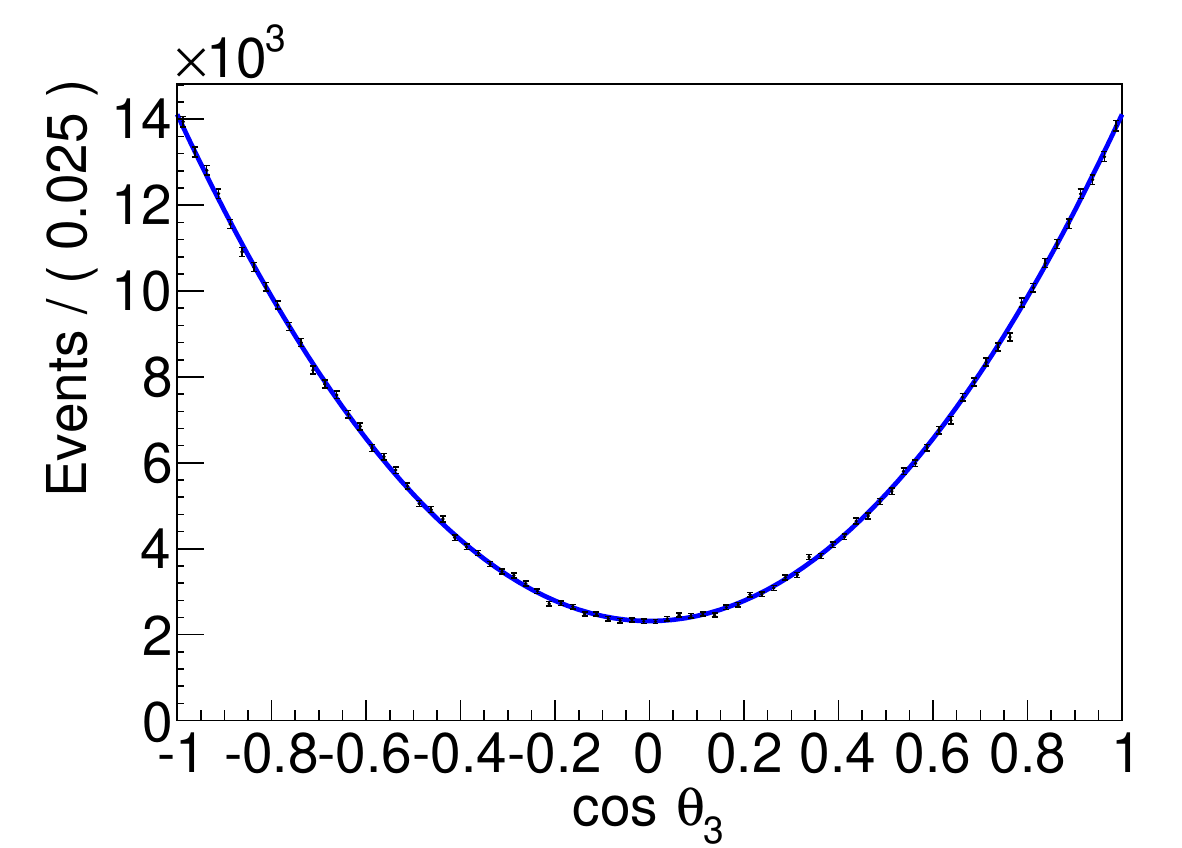}
		\label{c1_o_c3}}
    \subfigure[]{
        \includegraphics[scale=0.20]{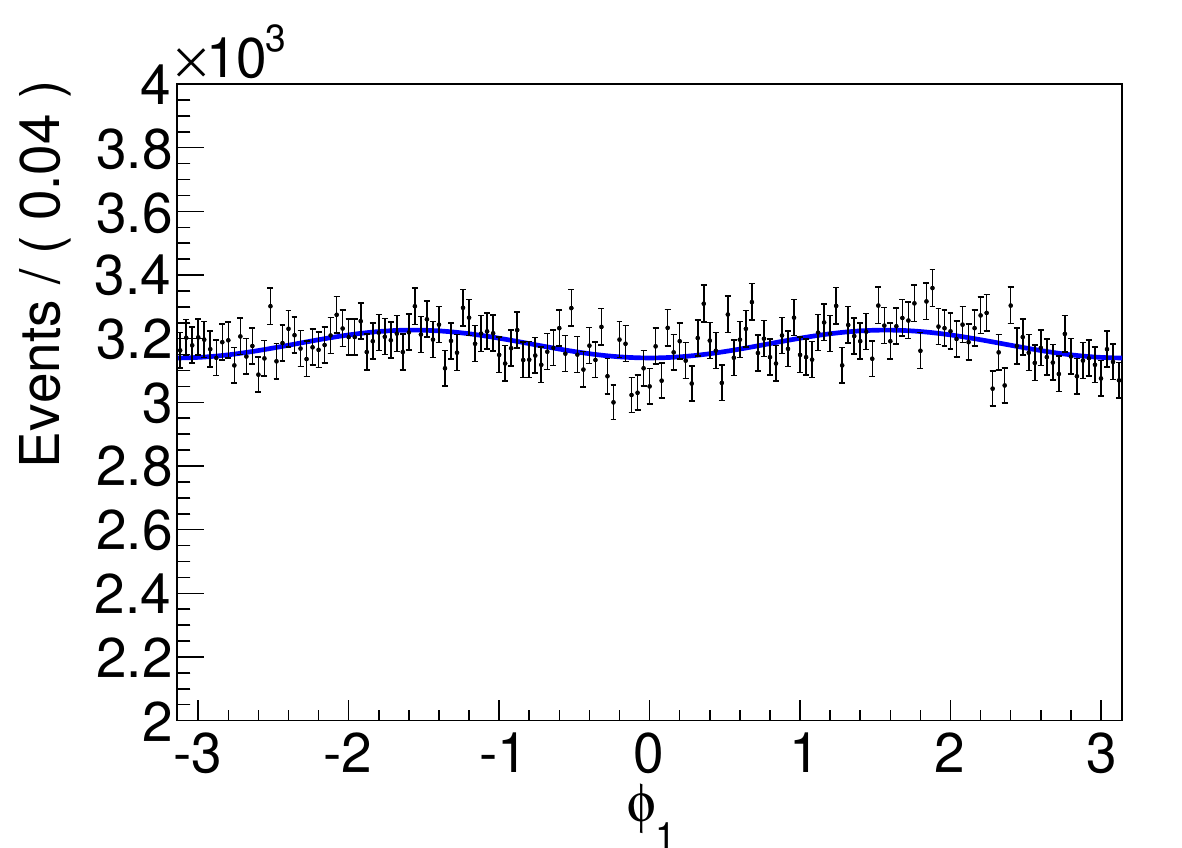}
        \label{c1_o_phi}
    }
	\caption{${\mathrm{d}N}/{\mathrm{d}\cos\theta_i}, i=1,2,3$ and ${\mathrm{d}N}/{\mathrm{d}\phi_1}$ distributions versus $\cos\theta_i$ and $\phi_1$ in $\psi(2S)\to \gamma \chi_{c1}$, $\chi_{c1} \to \gamma \omega$ and $\omega \to \pi^+ \pi^- \pi^0$ decays. Dots with error bars are filled with MC events, and the blue solid curve denotes the fit.}
\end{figure*}

\begin{figure*}[htbp]
	\centering
	\subfigure[]{
		\includegraphics[scale=0.20]{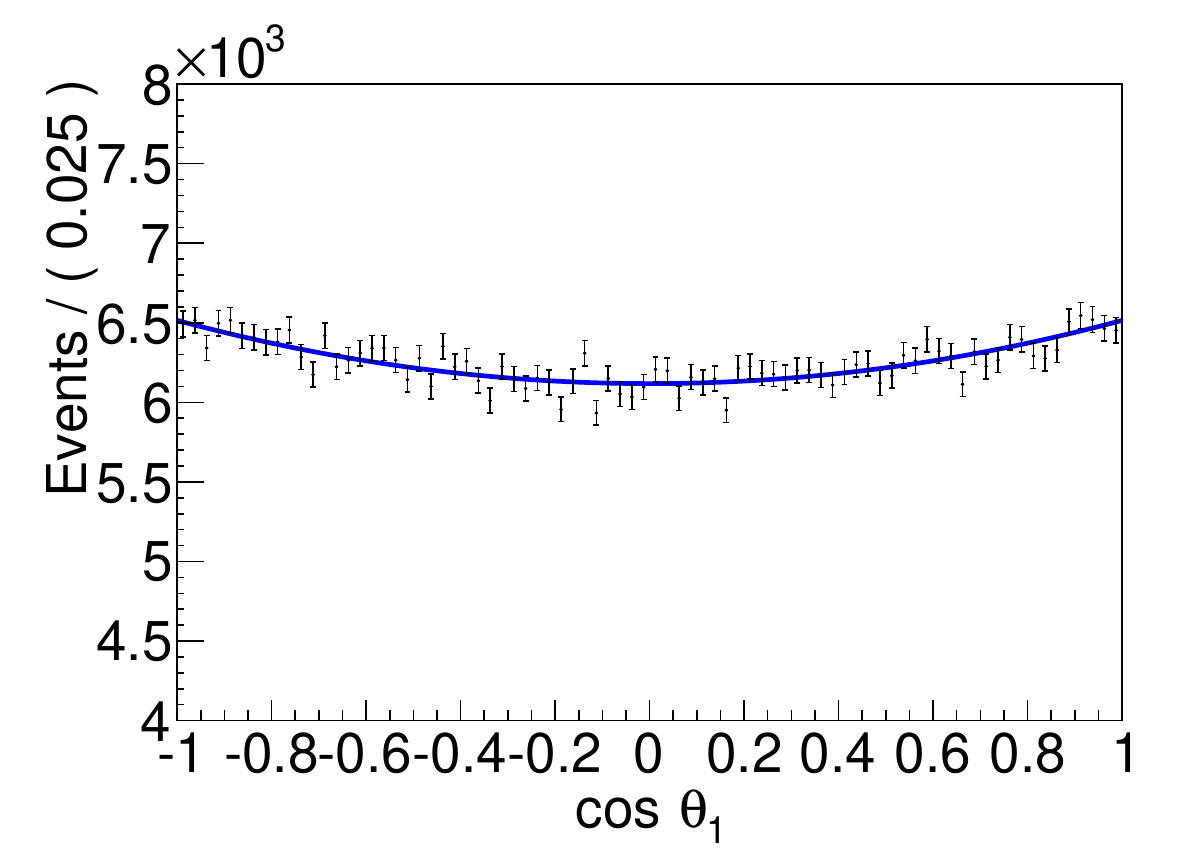}
		\label{c2_r_c1}}
	\subfigure[]{
		\includegraphics[scale=0.20]{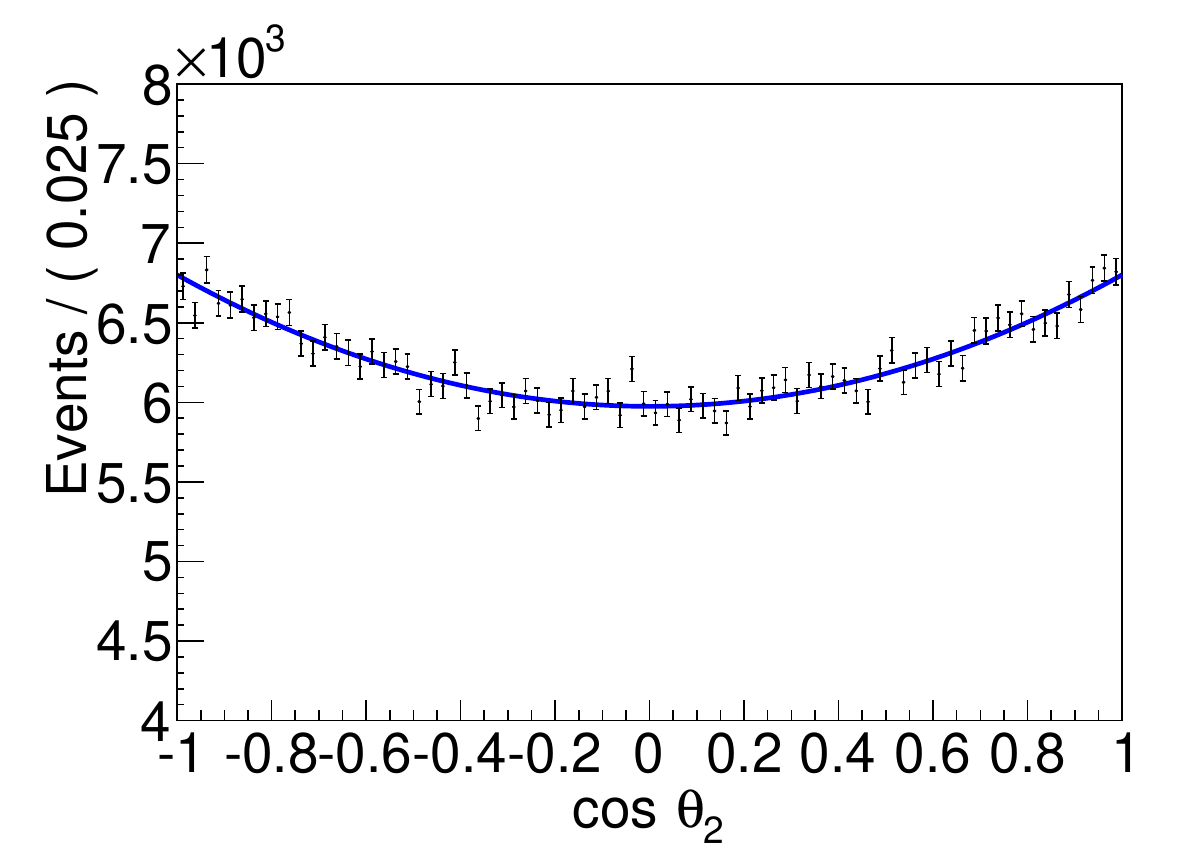}
		\label{c2_r_c2}}
	\subfigure[]{
		\includegraphics[scale=0.20]{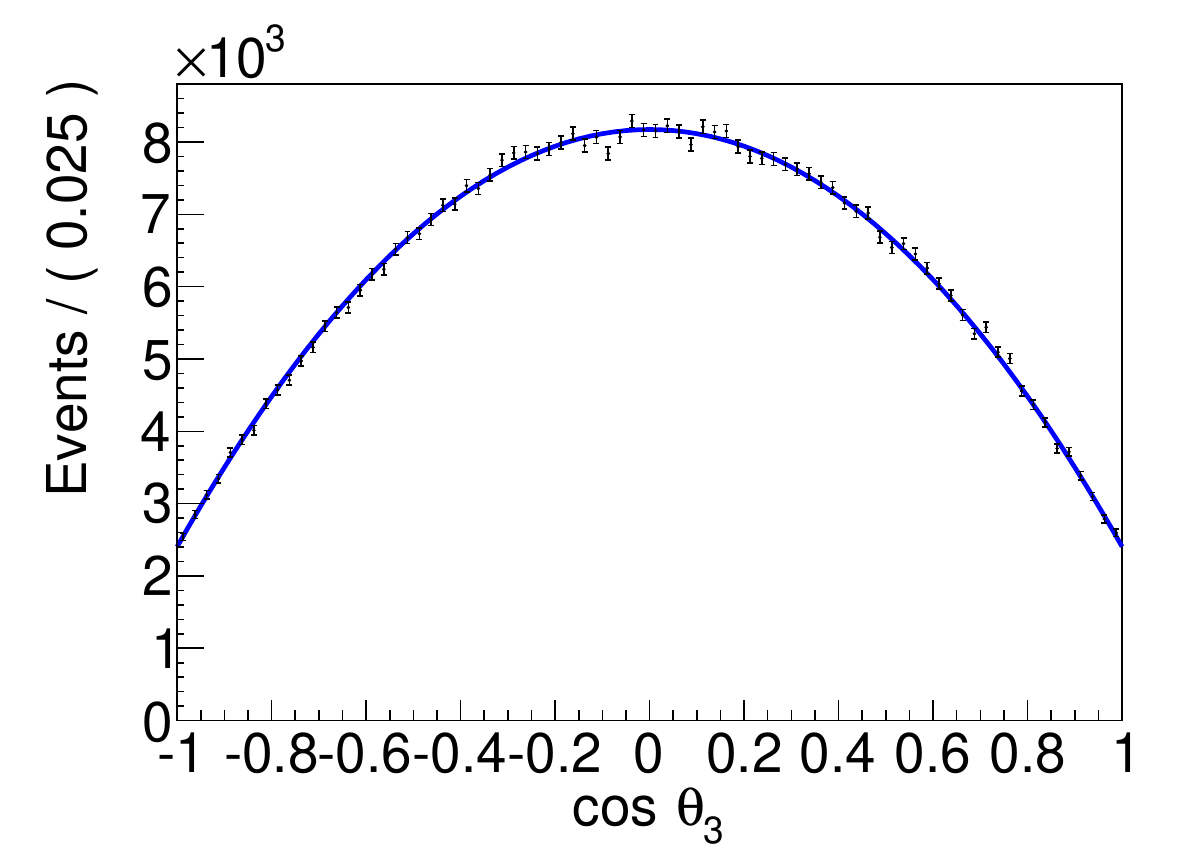}
		\label{c2_r_c3}}
    \subfigure[]{
        \includegraphics[scale=0.20]{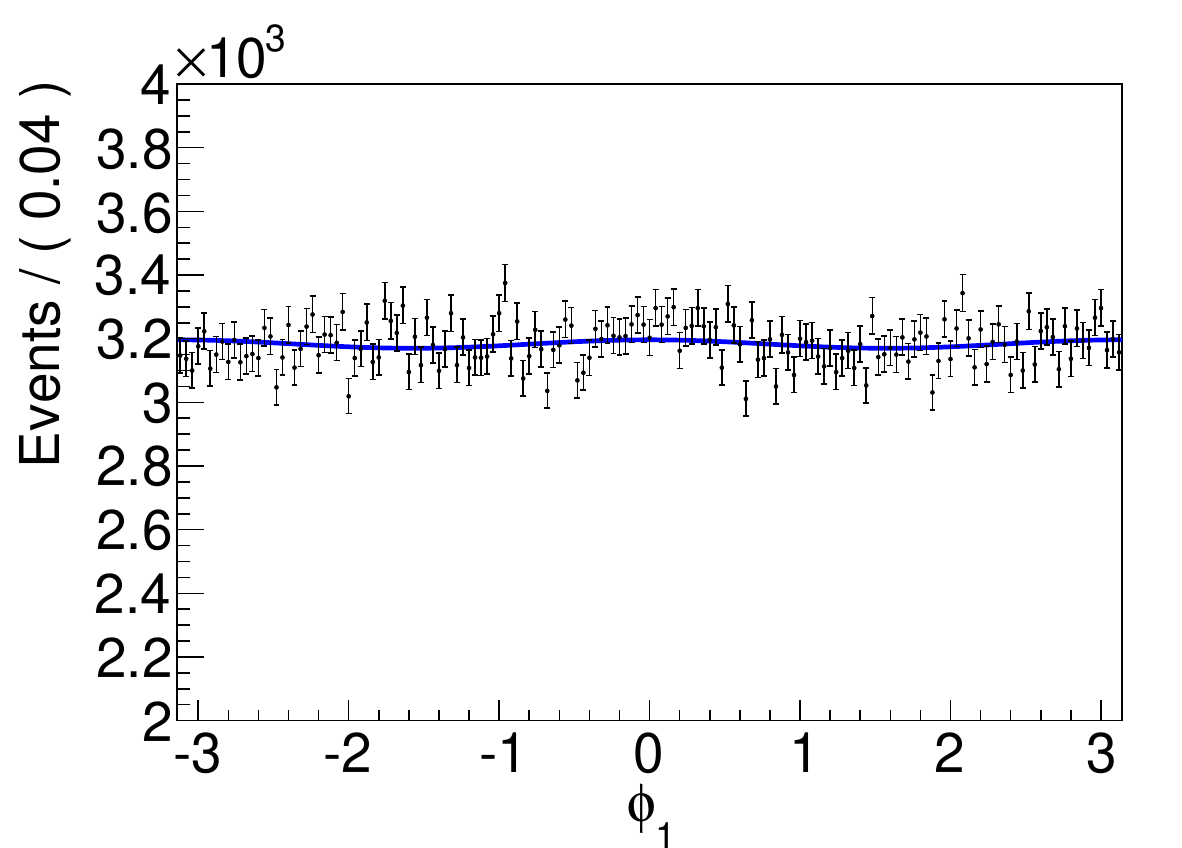}
        \label{c2_r_phi}
    }
	\caption{Fits to the angular distributions of $\cos\theta_i (i=1,2,3)$ and ${\mathrm{d}N}/{\mathrm{d}\phi_1}$  versus $\cos\theta_i$ and $\phi_1$ in $\psi(2S)\to \gamma \chi_{c2}$, $\chi_{c2} \to \gamma \rho$ and $\rho^0 \to \pi^+ \pi^-$ decays. Dots with error bars represent MC events, and the blue solid curve denotes the fit.}
\end{figure*}

\begin{figure*}[htbp]
	\centering
	\subfigure[]{
		\includegraphics[scale=0.20]{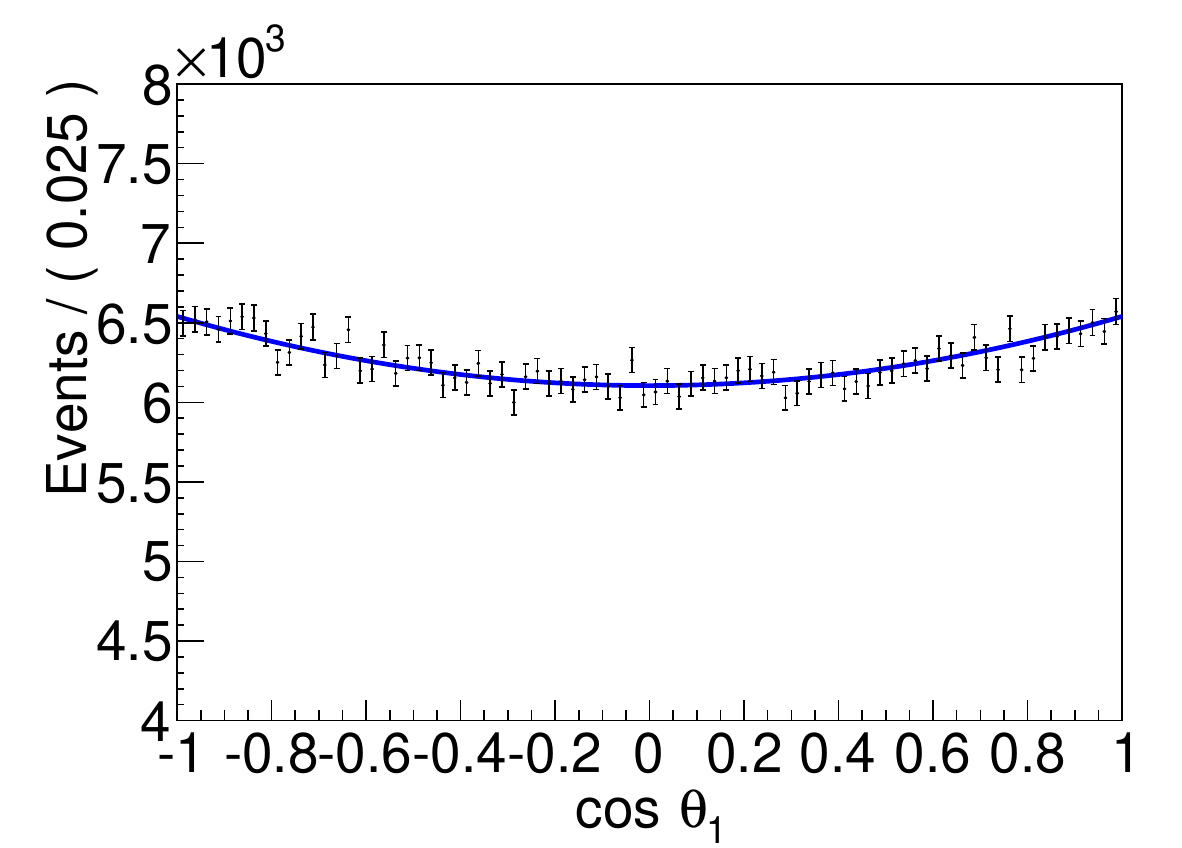}
		\label{c2_o_c1}}
	\subfigure[]{
		\includegraphics[scale=0.20]{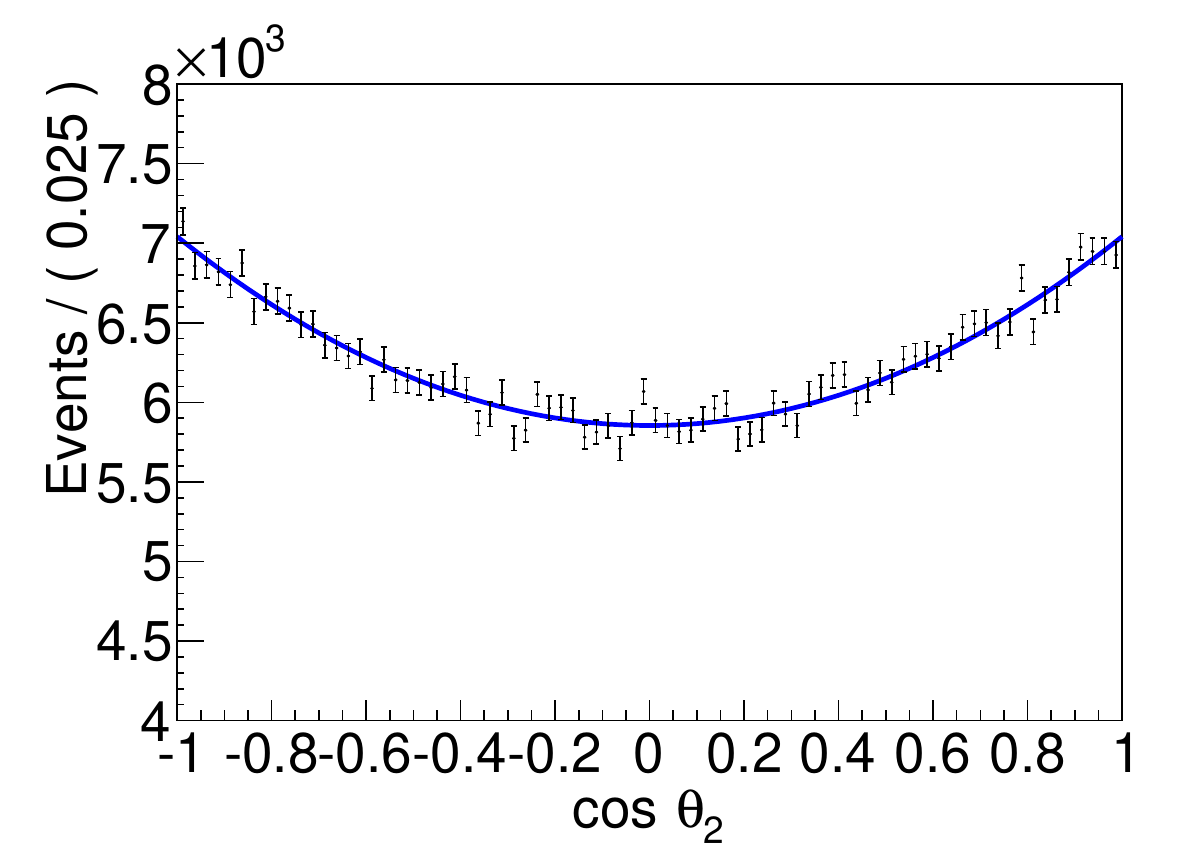}
		\label{c2_o_c2}	}
	\subfigure[]{
		\includegraphics[scale=0.20]{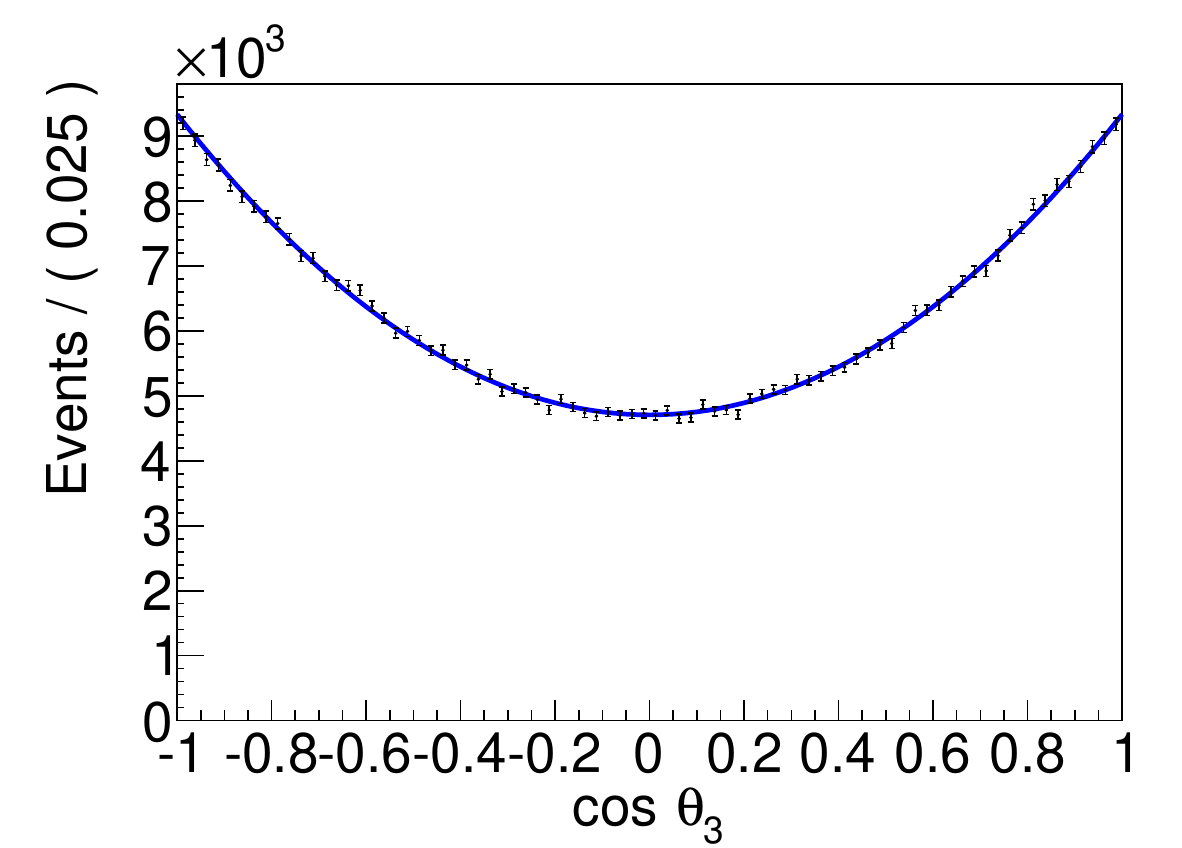}
		\label{c2_o_c3}}
    \subfigure[]{
        \includegraphics[scale=0.20]{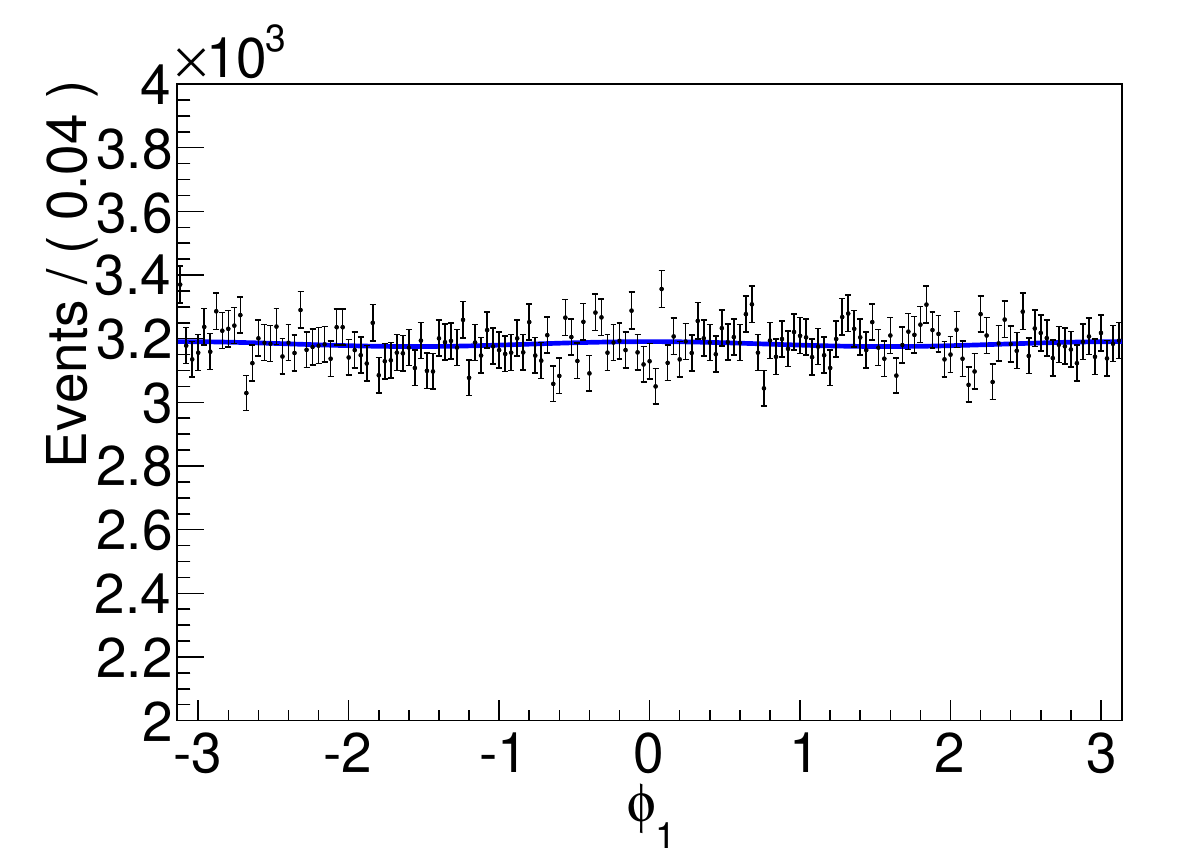}
        \label{c2_o_phi}
    }
	\caption{${\mathrm{d}N}/{\mathrm{d}\cos\theta_i}, i=1,2,3$ and ${\mathrm{d}N}/{\mathrm{d}\phi_1}$ distributions versus $\cos\theta_i$ and $\phi_1$ in $\psi(2S)\to \gamma \chi_{c2}$, $\chi_{c2} \to \gamma \omega$ and $\omega \to \pi^+ \pi^- \pi^0$ decays. Dots with error bars are filled with MC events, and the blue solid curve denotes the fit.}
\end{figure*}


\begin{thebibliography}{99}

\bibitem{Workman:2022ynf}
R.~L.~Workman \textit{et al.} (Particle Data Group),
PTEP \textbf{2022}, 083C01 (2022).

\bibitem{Barnes:2005pb}
T.~Barnes, S.~Godfrey and E.~S.~Swanson,
Phys. Rev. D \textbf{72}, 054026 (2005).

\bibitem{Brambilla:2010cs}
N.~Brambilla, S.~Eidelman, B.~K.~Heltsley \textit{et al.}
Eur. Phys. J. C \textbf{71}, 1534 (2011).

\bibitem{pQCD}
Y.~J.~Gao, Y.~J.~Zhang and K.~T.~Chao,
Chin. Phys. Lett. \textbf{23}, 2376(2006).

\bibitem{QED_NRQCD}
Y.~J.~Gao, Y.~J.~Zhang and K.~T.~Chao,
arXiv:hep-ph/0701009.

\bibitem{CLEO}
J.~V.~Bennett \textit{et al.} (CLEO Collaboration),
Phys. Rev. Lett. \textbf{101}, 151801 (2008).


\bibitem{BESIII:2011ysp}
M.~Ablikim \textit{et al.} (BESIII Collaboration),
Phys. Rev. D \textbf{83}, 112005 (2011).

\bibitem{Chen:2010re}
D.~Y.~Chen, Y.~B.~Dong and X.~Liu,
Eur. Phys. J. C \textbf{70}, 177(2010).

\bibitem{Sokolov:1963zn}
A.~A.~Sokolov and I.~M.~Ternov,
Dokl. Akad. Nauk SSSR \textbf{153}, 1052-1054 (1963)


\bibitem{BESIII:2024lks}
M.~Ablikim \textit{et al.} (BESIII Collaboration),
Chin. Phys. C \textbf{48}(9): 093001 (2024).

\bibitem{Chung:1993da}
S.~U.~Chung,
Phys. Rev. D \textbf{48}, 1225 (1993)
[erratum: Phys. Rev. D \textbf{56}, 4419 (1997)].

\bibitem{Chung:1997jn}
S.~U.~Chung,
Phys. Rev. D \textbf{57}, 431 (1998)

\bibitem{Chung:1971ri}
S.~U.~Chung,
CERN, Geneva, 1969-1970, CERN Yellow Reports: Monographs
doi:10.5170/CERN-1971-008.

\bibitem{AmpF}
P.~C.~Hong, F.~Yan, R.~G.~Ping and T.~Luo,
Chin. Phys. C \textbf{47} (5): 053101 (2023)

\bibitem{Karl:1975jp}
G.~Karl, S.~Meshkov and J.~L.~Rosner,
Phys. Rev. D \textbf{13}, 1203 (1976)

\bibitem{c2gV}
N.~Kivel and M.~Vanderhaeghen,
Phys. Rev. D \textbf{96}, 054007(2017).

\bibitem{Doncel:1973sg}
M.~G.~Doncel, P.~Mery, L.~Michel, P.~Minnaert and K.~C.~Wali,
Phys. Rev. D \textbf{7}, 815 (1973)

\bibitem{a_chi}
H.~Chen and R.~G.~Ping,
Phys. Rev. D \textbf{102}, 016021(2020).

\bibitem{sdm_psi}
X.~Cao, Y.~T.~Liang and R.~G.~Ping,
Phys. Rev. D \textbf{110}, 014035 (2024)

\bibitem{E1}
M.~Ablikim \textit{et al.} (BESIII Collaboration),
Phys. Rev. D \textbf{84}, 092006 (2011).

\bibitem{BESIII:2020nme}
M.~Ablikim \textit{et al.} (BESIII Collaboration),
Chin. Phys. C \textbf{44}(4): 040001 (2020).

\bibitem{Achasov:2023gey}
M.~Achasov, X.~C.~Ai, R.~Aliberti \textit{et al.}
Front. Phys. (Beijing) \textbf{19}, 14701 (2024).


\bibitem{L_function}
T.~Z.~Han, R.~G.~Ping, T.~Luo and G.~Z.~Xu,
Chin. Phys. C \textbf{44}, 013002(2020).

\bibitem{Aushev:2010bq}
T.~Aushev, W.~Bartel, A.~Bondar \textit{et al.}
arXiv:1002.5012 [hep-ex].

\end{thebibliography}
\end{document}